\documentclass[prx,showpacs,superscriptaddress,amssymb,twocolumn,longbibliography]{revtex4-1}

\usepackage{float}
\usepackage{amsmath,amssymb,setspace}
\usepackage{graphicx}
\usepackage{subfigure}  
\usepackage[usenames]{color}
\usepackage{graphicx}
\usepackage{bm, bbm}      
\usepackage{color}
\usepackage{srcltx}
\usepackage{hyperref}
\usepackage{braket}
\hypersetup{
    colorlinks=true,       
    linkcolor=cyan,          
    citecolor=magenta,        
    filecolor=magenta,      
    urlcolor=cyan,           
    runcolor=cyan
}
\usepackage{upgreek}

\usepackage{slashed}

\usepackage{xr}
\usepackage{xcite}

\newcommand{\bes} {\begin{subequations}}
\newcommand{\ees} {\end{subequations}}

\def\>{\rangle}
\def\<{\langle}

\newcommand{\ignore}[1]{}

\newcommand{\TTS}{\mathrm{TTS}}
\newcommand{\TTSopt}[1]{\langle \TTS \rangle^{\ast}_{#1}}

\newcommand{\topt}[1]{t^\ast_{#1}}

\begin{document}
\title{Demonstration of a scaling advantage for a quantum annealer over simulated annealing
}
\author{Tameem Albash}
\affiliation{Department of Physics and Astronomy, University of Southern California, Los Angeles, California 90089, USA}
\affiliation{Center for Quantum Information Science \& Technology, University of Southern California, Los Angeles, California 90089, USA}
\affiliation{Information Sciences Institute, University of Southern California, Marina del Rey, California 90292, USA}
\author{Daniel A. Lidar}
\affiliation{Department of Physics and Astronomy, University of Southern California, Los Angeles, California 90089, USA}
\affiliation{Center for Quantum Information Science \& Technology, University of Southern California, Los Angeles, California 90089, USA}
\affiliation{Department of Chemistry, University of Southern California, Los Angeles, California 90089, USA}
\affiliation{Department of Electrical Engineering, University of Southern California, Los Angeles, California 90089, USA}

\begin{abstract}
The observation of an unequivocal quantum speedup remains an elusive objective for quantum computing. 
A more modest goal is to demonstrate a scaling advantage over a class of classical algorithms for a computational problem running on quantum hardware. 
The D-Wave quantum annealing processors have been at the forefront of experimental attempts to address this goal, given their relatively large numbers of qubits and programmability. A complete determination of the optimal time-to-solution (TTS) using these processors has not been possible to date, preventing definitive conclusions about the presence of a scaling advantage.  The main technical obstacle has been the inability to verify an optimal annealing time within the available range. Here we overcome this obstacle using a class of problem instances constructed by 
systematically combining many-spin frustrated-loops with few-qubit gadgets exhibiting a tunneling event --- a combination that we find to promote the presence of tunneling energy barriers in the relevant semiclassical energy landscape of the full problem --- and we observe an optimal annealing time using a D-Wave 2000Q processor over a range spanning up to more than $2000$ qubits. We identify the gadgets as being responsible for the optimal annealing time, whose existence allows us to perform an optimal TTS benchmarking analysis. We perform a comparison to several classical algorithms, including simulated annealing, spin-vector Monte Carlo, and discrete-time simulated quantum annealing (SQA), and establish the first example of a scaling advantage for an experimental quantum annealer over classical simulated annealing. Namely, we find that the D-Wave device exhibits certifiably better scaling than simulated annealing, with $95\%$ confidence, over the range of problem sizes that we can test. 
However, we do not find evidence for a quantum speedup: SQA exhibits the best scaling for annealing algorithms by a significant margin. This is a finding of independent interest, since we associate SQA's advantage with its ability to transverse energy barriers in the  semiclassical energy landscape by mimicking tunneling. Our construction of instance classes with verifiably optimal annealing times opens up the possibility of generating many new such classes based on a similar principle of 
promoting the presence of energy barriers that can be overcome more efficiently using quantum rather than thermal fluctuations,
paving the way for further definitive assessments of scaling advantages using current and future quantum annealing devices. 
\end{abstract}

\maketitle
\section{Introduction}
The elusive and tantalizing goal of experimentally demonstrating a quantum speedup is being actively pursued using a variety of quantum computing platforms. The holy grail is an exponential speedup, such as expected with Shor's algorithm for factoring integers \cite{Shor:94}, or with the simulation of quantum systems \cite{Feynman1,Lloyd:96,Cirac:2012pi}. This goal is still substantially out of reach given the relatively small scale of current universal quantum computers and quantum simulators ($\sim 20$-$70$ qubits~\cite{google-72q,ibm-50q,intel-49q,Otterbach:2017aa,Zhang:2017aa,Bernien:2017aa}), which prevents the implementation of fault tolerant quantum error correction on a scale that would enable quantum circuits to be executed reliably despite decoherence and noise. However, there is reason for optimism~\cite{Neill:2018aa} that current ``noisy intermediate scale quantum" (NISQ) era \cite{Preskill:2018aa} quantum computers will be capable of demonstrating the important milestone of ``quantum supremacy"~\cite{Harrow:2017aa}, a less ambitious quantum speedup goal than that associated with application-level computational tasks such as factoring or quantum simulation.

The largest quantum information processing devices currently available are quantum annealers, featuring several thousands of noisy qubits and programmable qubit-qubit interactions. Unlike universal quantum computers that operate using quantum gates and the principles of the circuit model~\cite{nielsen2010quantum}, these devices specialize primarily in solving combinatorial optimization problems, and are designed to represent physical implementations of quantum annealing (QA)~\cite{kadowaki_quantum_1998} and the quantum adiabatic algorithm \cite{farhi2001quantum}. While the algorithmic focus in the domain of universal quantum computers has been on demonstrating quantum simulation and quantum supremacy, in QA the primary focus has been on benchmarking the algorithmic performance of quantum annealers against classical algorithms
\cite{q108,speedup,King:2015cs,Hen:2015rt,King:2015zr,DW2000Q,2016arXiv160401746M,Vinci:2016tg,PhysRevX.6.031015,Katzgraber:2015gf,QA-comment}, an effort that has not yet been undertaken with gate model quantum computers. This difference is explained primarily by the relatively large number of qubits available in QA, which enables scaling tests over several orders of magnitude of problem sizes.
Despite the large body of work on benchmarking quantum annealers, conclusive evidence about how their performance scales with problem size has until now been unattainable. 
The primary reason, as we shall discuss in detail, is that it has not been possible to identify an optimal annealing time for any class of problem instances, an obstacle that was first pointed out in Ref.~\cite{speedup}. 

Here, we overcome this obstacle by introducing a new class of problem instances that exhibit an optimal annealing time, and present for the first time a complete algorithmic scaling analysis of a hardware quantum annealer (the D-Wave 2000Q device~\cite{DW2KQ}), up to the largest available problem size of more than $2000$ spins or qubits%
~\footnote{We stress that our observation of an optimal annealing time is not simply due to advances in the quantum annealing hardware; we demonstrate optimal annealing times for these problem instances on the two most recent generations of D-Wave processors.}. 

This advance allows us furthermore to demonstrate the first certifiable observation of an algorithmic scaling advantage obtained using quantum annealing hardware over an important general purpose classical algorithm, namely over simulated annealing with single-spin updates (SA) \cite{kirkpatrick_optimization_1983}. 
Without the identification of an optimal annealing time one can certify a scaling disadvantage for the hardware quantum annealer, but not an advantage~\cite{Hen:2015rt}; this holds for all earlier scaling analyses presented for quantum annealing hardware~\cite{q108,speedup,Hen:2015rt,PhysRevX.6.031015,King:2015zr,King:2015cs,DW2000Q,2016arXiv160401746M,Vinci:2016tg}. 

The advantage of QA over SA we demonstrate holds for a class of problem instances (called ``logical-planted" and defined below) that we constructed by systematically combining a distribution of frustrated cycles of coupled spins over the entire hardware graph and small ``gadgets'' made of a relatively small number of qubits (here we used eight) that have a small quantum gap (on the order of the temperature) and that exhibit a tunneling event that can be established via numerical solution of the Schr\"odinger equation. 
Since both features are flexible, our construction provides a recipe for generalizing our results to a broader class of problem instances. We show that the optimal annealing time arises due to the gadgets, in the sense that instances based only on frustrated loops do not exhibit an optimal annealing time.

To establish the presence of many-qubit tunneling we resort to large scale simulations using the discrete-time simulated quantum annealing (SQA) algorithm \cite{Santoro} which we contrast with the spin-vector Monte Carlo (SVMC) algorithm~\cite{SSSV}.  
Both SQA and SVMC are  transverse-field annealing algorithms, and thus more closely model QA than a temperature-annealing algorithm such as SA does. But while SVMC is purely classical, in the sense that it provides a time-dependent description in terms of unentangled planar rotors,
SQA is a classical algorithm based on path-integral Monte Carlo, which in its continuous-time limit and for sufficiently many spin updates generates samples from the quantum Gibbs state. In particular, at sufficiently low temperatures SQA can mimic tunneling \cite{2015arXiv151008057I,Jiang:2017aa} and describe entangled ground states such as those followed by QA. It is the opposite trends exhibited by SQA and SVMC as a function of the simulation temperature that allows us to argue for the occurrence of many-qubit tunneling. We emphasize that the appropriate energy landscape for tunneling is not the classical energy landscape associated with simulated annealing~\cite{Katzgraber:2015gf} but rather the semiclassical landscape associated with transverse field annealing \cite{Boixo:2014yu,Muthukrishnan:2015ff}.

Our benchmarking analysis reveals that SQA has the best scaling of all the annealing algorithms we tested for the logical-planted instances, in particular outperforming the quantum annealing hardware.  It also outperforms a number of algorithms (described below) designed to specifically exploit features of the ``Chimera" hardware graph of the D-Wave devices~\cite{Choi1,Bunyk:2014hb}. 
The fact that SQA performs so well for the logical-planted instance class is in itself a significant and novel finding about the class of logical-planted problem instances, since one might reasonably expect that as hardware quantum annealers continue to improve, SQA will become a lower bound on the performance of such hardware \cite{Hastings:2013kk}. The reason is that SQA serves as a reasonable classical simulation of a thermally-dominated quantum annealer but of course does not actually physically manifest any of the quantum features (unitary dynamics, coherent tunneling, entanglement) that are expected to come into play in a physical realization of sufficiently coherent QA. 
We show for the logical-planted problem class that, by mimicking tunneling, SQA traverses energy barriers more efficiently as the temperature of the simulation is lowered. We use this to argue that a key reason for the quantum annealer's slowdown relative to SQA is its sub-optimally high temperature~\cite{Albash:2017ab}, which causes it to behave more like SVMC.
Thus, the strong performance of SQA on the logical-planted instance class suggests that this class is a good target or basis for the exploration of an eventual quantum speedup using QA hardware.

We first review and discuss, in Sec.~\ref{sec:TTS}, the time-to-solution metric, and how to establish optimality. Section~\ref{sec:results} presents our results. First, Sec.~\ref{sec:optimal-ta}  establishes the empirical evidence for optimal annealing times for our class of problem instances. 
Then, Sec.~\ref{sec:LQevidence} presents the empirical evidence we have found for a QA scaling advantage over SA, but a disadvantage relative to SQA  and SVMC. In Sec.~\ref{sec:instances}, we introduce and describe the properties of the class of problem instances for which we observe the optimal annealing time and the scaling advantage over SA.  We discuss the implications of our results and provide an outlook in Sec.~\ref{sec:discussion}. Additional technical details and methods are provided in the Appendices.

\section{Optimal time-to-solution}
\label{sec:TTS}

We consider the standard setting where the goal of the optimizer is to find the optimal solution (i.e., the global minimum of the cost function) and one is interested in minimizing the time taken to find the solution at least once. 
There is a tradeoff between finding the solution with a high probability in a single long run of the algorithm, and running the algorithm multiple times with a shorter runtime and (usually) a smaller single-run success probability.  This tradeoff is reflected in the time-to-solution (TTS) metric, which measures the time required to find the ground state at least once with some desired probability $p_\mathrm{d}$ (often taken to be $0.99$):
\begin{equation}
\mathrm{TTS}(t_f) = t_f R(t_f)  \frac{N}{N_{\mathrm{max}}}  \ , \quad R(t_f) =  \frac{\ln (1 - p_\mathrm{d})}{\ln [1 - p_\mathrm{S}(t_f)]} \ .
\label{eq:TTS}
\end{equation}
Here $p_\mathrm{S}(t_f)$ is the success probability of a single-instance run of the algorithm with a runtime $t_f$, and $R(t_f)$ is the required number of runs; success means that the optimal solution was found. The instance size is $N$, and $N_{\max}$ is the size of the largest instance that the device accommodates (typically set by the total number of qubits); the factor $N/N_{\mathrm{max}}$ accounts for maximal parallel utilization of the device.  While $R$ and $N_{\mathrm{max}}/N$ should correspond to whole numbers, we do not round them here since this can result in sharp TTS changes that complicate the extraction of scaling with $N$;  
see Appendix~\ref{app:TTS} for more details.

However, when considering the performance of an algorithm evaluated over an ensemble of randomly chosen instances from the same class, we are typically interested not in the TTS of a single instance but in a given quantile $q$ of the TTS distribution over such instances at a given problem size $N\in[N_{\min},N_{\max}]$.
We denote the $q$-th quantile of the TTS evaluated at $t_f(N)$ by $\langle \TTS(t_f) \rangle_q$, and suppress the $N$ dependence for simplicity.  
Since the goal of optimization is to find the solution as rapidly as possible, there is an optimal $t_f$ value for a given quantile $q$, $\topt{q}$, where $\langle \TTS(t_f) \rangle_q$ is minimized, and we denote 
$\TTSopt{q} \equiv \langle \TTS (\topt{q}) \rangle_q$. While the success probability of individual instances may exhibit many minima (as in the case of coherent evolution when oscillations in the success probability are observed as the annealing time is varied), the quantile of the TTS distribution exhibits only one minimum because the many minima of the individual instances are unlikely to coincide, and there is no ambiguity in the determination of an optimal $t_f$ value.  Finally, it is important to note that since quantiles are solver-dependent, a comparison between different solvers at the same quantile involves different sets of instances. 

We can now state the precise nature of the critical obstacle alluded to above: if $t_q^* <t_{\min}$, where $t_{\min}$ is the smallest possible annealing time on the given quantum annealer, then it becomes impossible to determine $\TTSopt{q}$. As was shown in Refs.~\cite{speedup,Hen:2015rt}, when operating with a suboptimal $t_f$, one can easily be led to false conclusions about the scaling with $N$ of $\langle \TTS(t_f) \rangle_q$ compared to the all-important scaling as captured by $\langle \TTS \rangle^{\ast}_q$, and even be led to conclude that there is a scaling advantage where there is none. 

%
\begin{figure*}[t] %
   \centering
\subfigure[\ TTS for representative instance]{\includegraphics[width=0.32\textwidth]{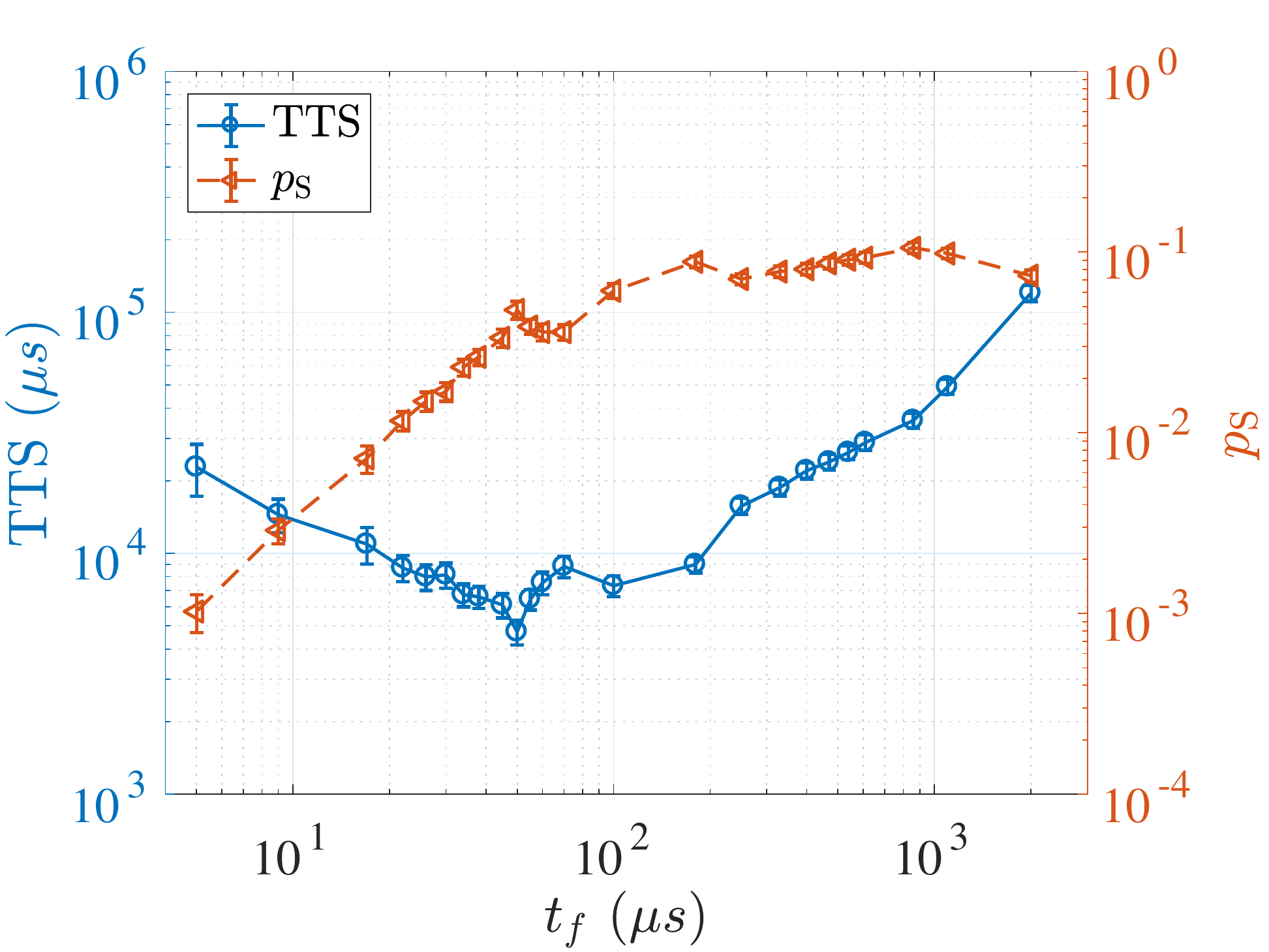} \label{fig:ExampleOptimalAnnealingTime1}} 
\subfigure[\ TTS for ensemble]{\includegraphics[width=0.32\textwidth]{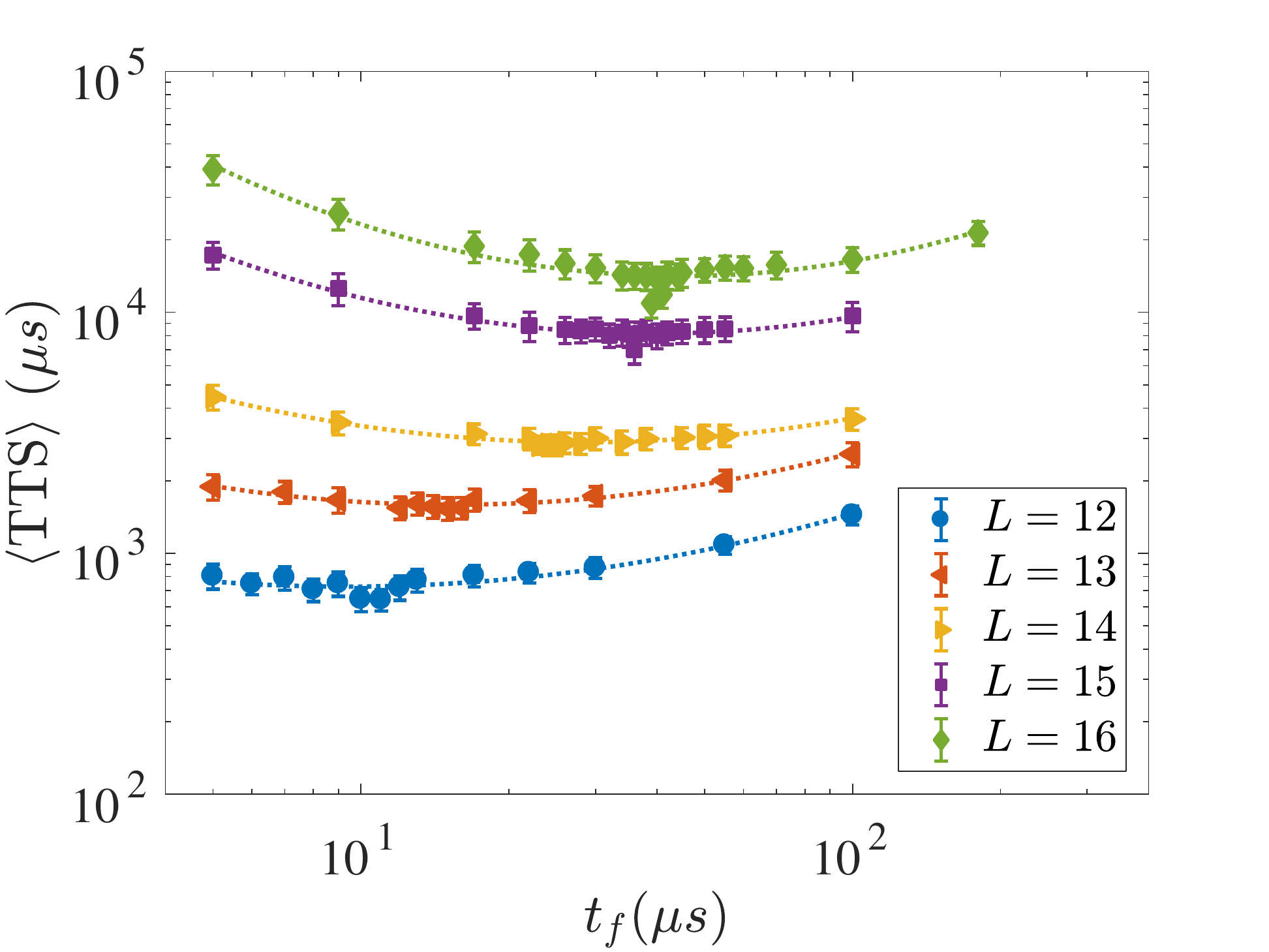} \label{fig:GroupTTS1}}
\subfigure[\ distribution of optimal annealing times]{\includegraphics[width=0.32\textwidth]{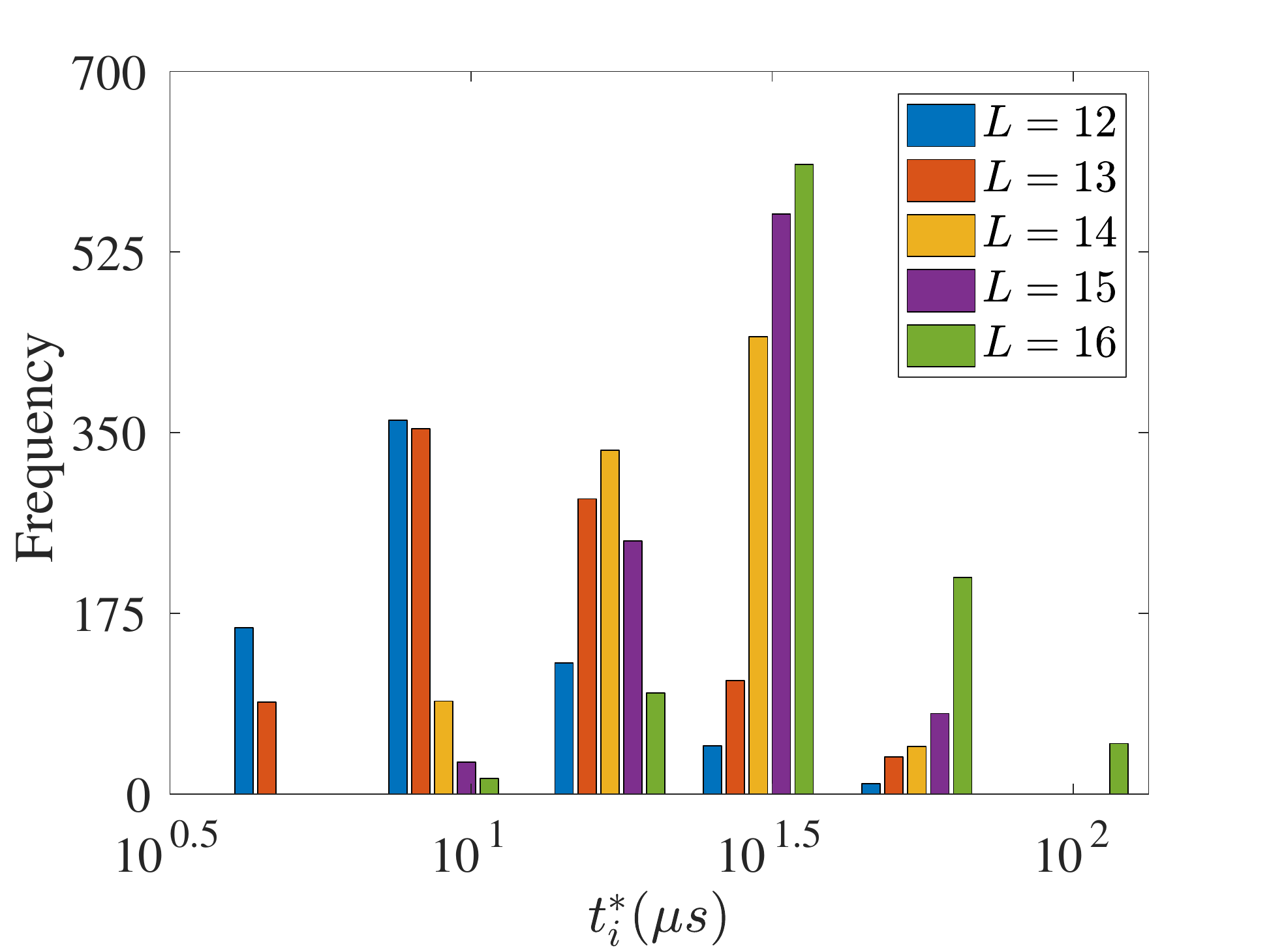} \label{fig:1c}} 
   \caption{\textbf{Optimal annealing time and optimal TTS.} Result shown are for the `logical-planted' instance class. 
(a) TTS (blue solid line) and $p_{\mathrm{S}}$ (red dashed line) for a representative problem instance at size $L=16$. A clear minimum in the TTS is visible at $t^* = 50 \mu$s, thus demonstrating the existence of an optimal annealing time for this particular instance (but not by itself for the problem class). Note that the decreasing or increasing TTS is associated with $p_{\mathrm{S}}$ growing sufficiently fast or too slowly, respectively, with increasing $t_f$.
(b) Median TTS as a function of annealing time for $L\geq 12$, from $1000$ instances. Dotted curves represent best-fit quadratic curves to the data (see Appendix~\ref{sec:InstanceConstruction} for the scaling of $t^*$ with $L$, and Appendix~\ref{app:optimumFitting} for details on the fitting procedure).  The position of the minimum of these curves gives $t^*$.  The position of $\TTSopt \ $ shifts to larger $t_f$ as the system size increases. An optimum could not be established for $L<12$ for this instance class, i.e., it appears that $t^*<5\mu$s when $L<12$. 
(c) The distribution of per-instance optimal annealing times $t_i^*$ for different system sizes, as inferred directly from the positions of the minima as shown in (a).  It is evident that the number of instances with higher optimal annealing times increases along with the system size, in agreement with (b).
} 
\label{fig:OptimalAnnealingTime1}
\end{figure*}

None of the experimental quantum annealing benchmarking studies to date 
\cite{q108,speedup,King:2015cs,Vinci:2016tg,Hen:2015rt,King:2015zr,2016arXiv160401746M,Katzgraber:2015gf,2014Katzgraber,PhysRevX.6.031015,DW2000Q} 
have provided a complete scaling assessment, precisely because it has not been possible to verify that $t_q^* >t_{\min}$ (for any quantile). 
The culprit was the absence of a suitable class of problem instances for which an optimal annealing time could be verified. Here we report on a class of instances that exhibits an optimal annealing time greater than $t_{\min}=5 \mu$s on the D-Wave 2000Q (DW2KQ, fourth generation, for which the largest energy scale is $\sim 50$GHz in $\hbar=1$ units -- see Appendix~\ref{app:DW}) and the D-Wave 2X (DW2X, third generation, for which the largest energy scale is $\sim 40$GHz).  In the main text, we focus on the DW2KQ and provide results from the DW2X in the Appendix.
This allows us to obtain the first complete optimal-TTS scaling results for an experimental quantum annealer, defined as the TTS scaling obtained from certifiably optimal annealing times.

The D-Wave processors used in our study are designed to implement quantum annealing using a transverse field Ising Hamiltonian: %
\begin{equation} \label{eqt:QA}
H(s) = A(s) H_X + B(s) H_{\mathrm{P}}\ , 
\end{equation}
where $s=t/t_f\in[0,1]$, $H_X =  - \sum_{i\in \mathcal{V}} \sigma^x_i$ and $H_{\mathrm{P}} = \sum_{i\in \mathcal{V}} h_i \sigma_i^z + \sum_{(i,j)\in \mathcal{E}} J_{ij} \sigma_i^z \sigma_j^z$ is the Ising, or `problem' Hamiltonian whose ground state we are after. The $\sigma^x_i$ and $\sigma^z_i$ are the Pauli matrices acting on superconducting flux qubits that occupy the vertices $\mathcal{V}$ of a `Chimera' hardware graph $\mathcal{G}$ with edge set $\mathcal{E}$ 
\cite{Choi1,Bunyk:2014hb}
and the local fields $h_i$ and couplings $J_{ij}$ are programmable analog parameters. The system is initialized in or near the ground state
of the initial Hamiltonian $H(0)$, and the annealing schedules $A(s)$ and $B(s)$, which set the energy scale,
are described in Appendix~\ref{app:DW}, along with further technical and operational details, including a schematic of the Chimera graph. The DW2KQ processor comprises $16 \times 16$ unit cells, so we can consider $L \times L$ subgraphs up to $L_{\max}=16$ for our analysis, where each subgraph comprises $L^2$ unit cells, and each complete unit cell comprises $8$ qubits (a small number of unit cells are incomplete, as a total of $21$ out $2048$ qubits are inoperative).

\section{Results}
\label{sec:results}

We start by describing our key results: the evidence for optimal annealing times, and the evidence for a scaling advantage of a physical quantum annealer over SA, along with its scaling disadvantage against the SQA and SVMC algorithms (we review these algorithms and discuss how we implemented and timed them in Appendix~\ref{app:SA_SVMC}).  
We then describe in detail the construction of the class of problem instances exhibiting these properties, and the role of tunneling in explaining them.

\subsection{Evidence for optimal annealing times}
\label{sec:optimal-ta}

We first present the evidence for optimal annealing times in Fig.~\ref{fig:OptimalAnnealingTime1}. 
Figure~\ref{fig:ExampleOptimalAnnealingTime1} shows the TTS for a single representative $L=16$ instance from a class we call `logical-planted' problems. The unambiguous minimum at $t_f=50\mu$s is the optimal annealing time for this instance. 
The presence of a minimum is a robust feature: Fig.~\ref{fig:GroupTTS1} shows that the optimal annealing time feature persists for the median TTS ($\TTSopt{0.5}$), at all sizes $L\in[12,16]$. In all previous benchmarking work, only the rise in $\langle\TTS\rangle $ as a function of $t_f$ was observed, i.e., $t^*$ was always below $t_{\min}$, thus precluding the identification of an optimal annealing time. The increase in the optimal annealing time from $L=12$ to $L=16$ seen in Fig.~\ref{fig:GroupTTS1} can be attributed to the general increase with problem size of the per-instance optimal annealing time as shown in Fig.~\ref{fig:1c}, which shows the distribution of optimal annealing times over all the logical-planted problem instances we tested.

\begin{figure*}
\centering
   \subfigure[\ $q=0.25$]{\includegraphics[width=0.32\textwidth]{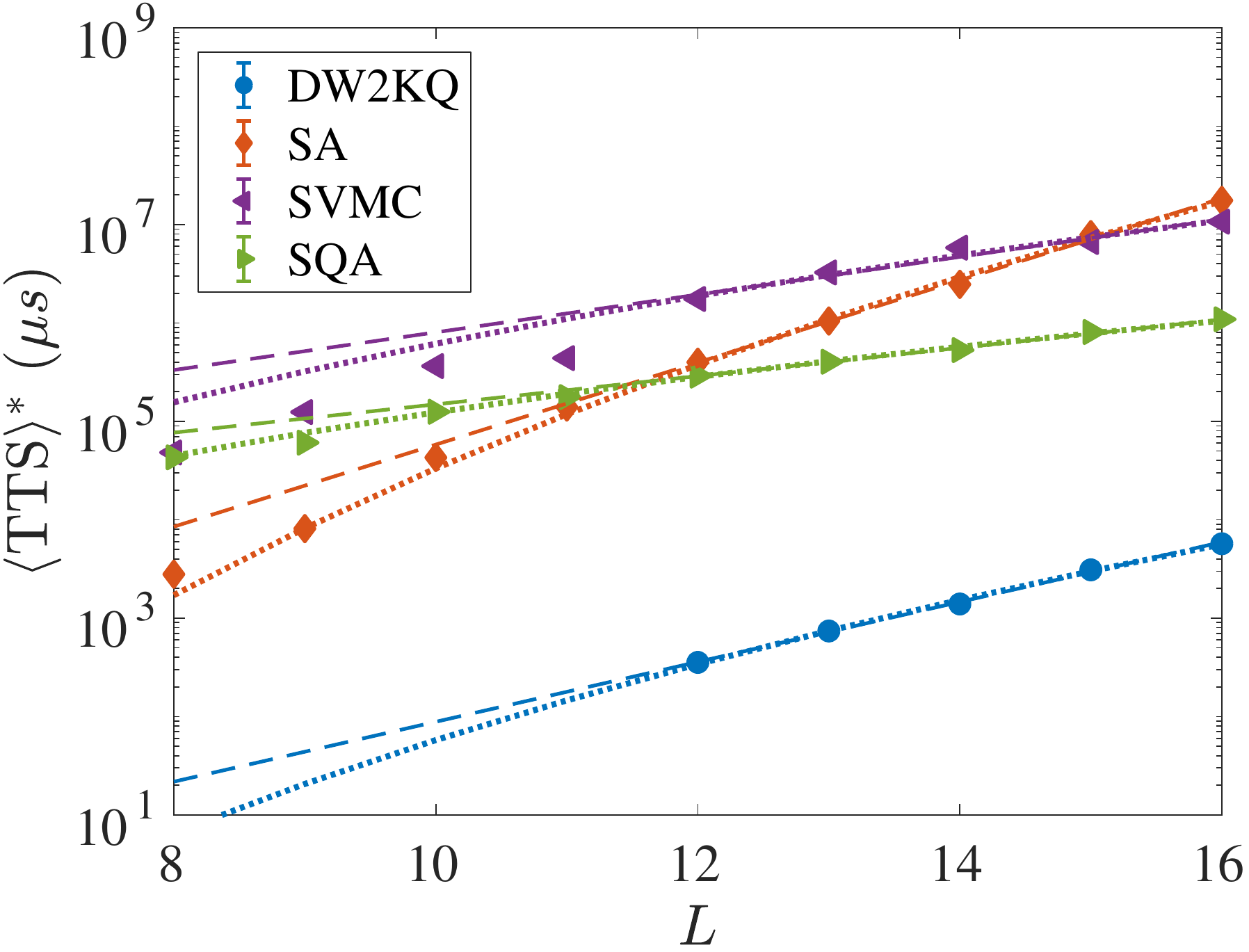} }
\subfigure[\ $q=0.50$]{\includegraphics[width=0.32\textwidth]{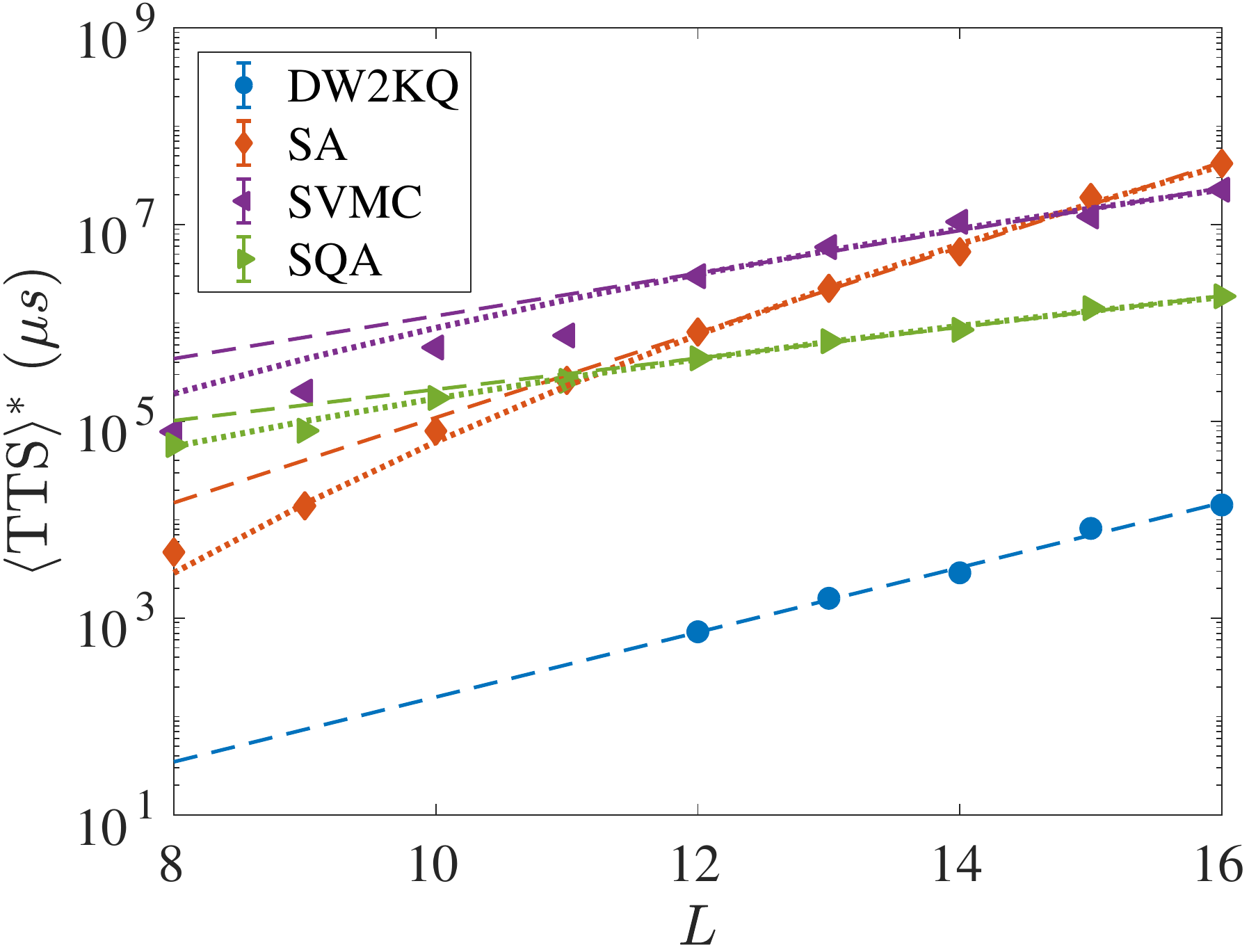} \label{fig:scalingLogical}}
      \subfigure[\ $q=0.75$]{\includegraphics[width=0.32\textwidth]{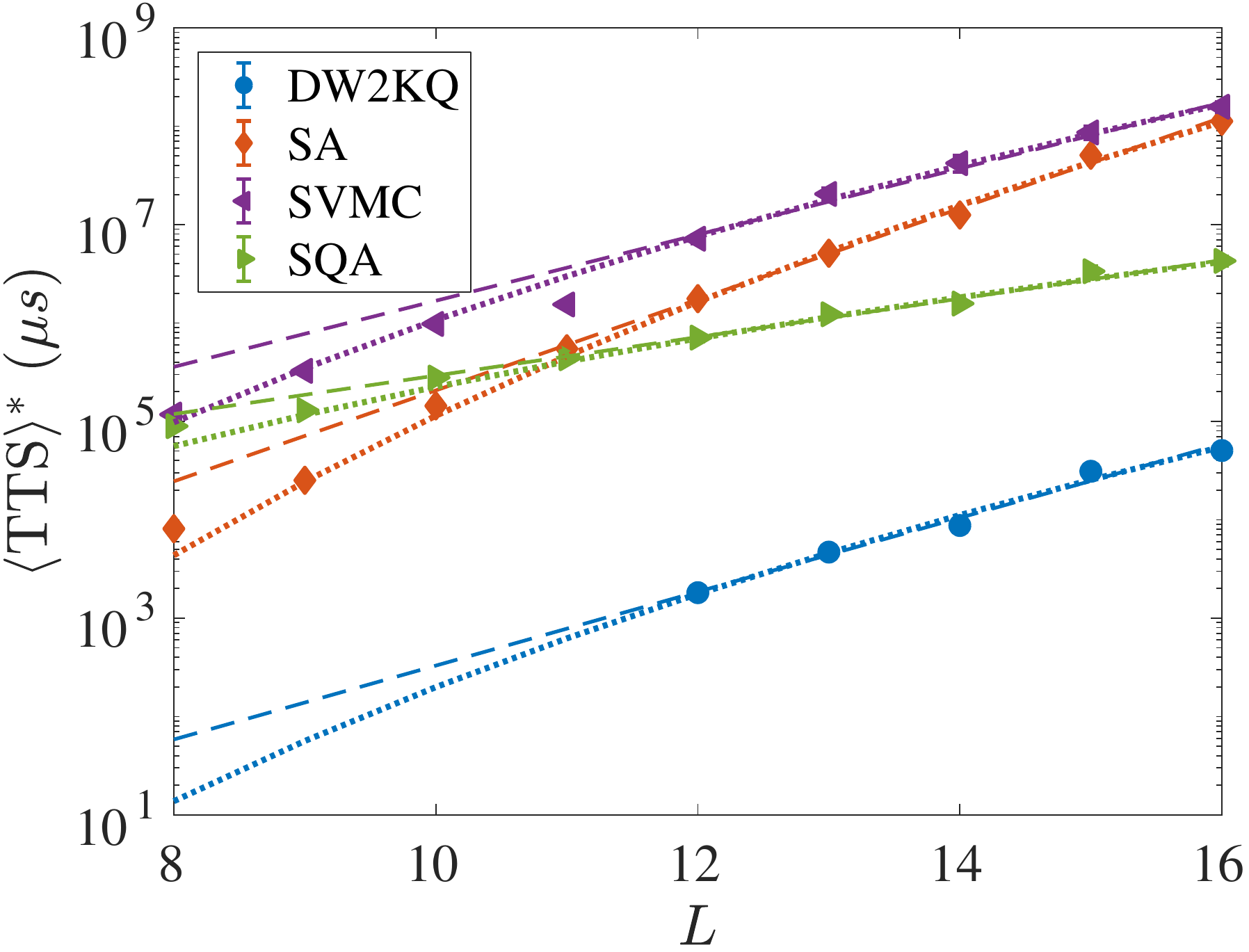} }
\caption{\textbf{Scaling of the optimal TTS with problem size.} Result shown are for the `logical-planted' instance class. The data points represent the DW2KQ (blue circles) and three classical solvers, SA (red diamonds), SVMC (purple left triangles), and SQA (green right triangles). The dashed and dotted curves correspond, respectively, to exponential and polynomial best fits with parameters shown in Fig.~\ref{fig:scalingFitsLogical} (also given in table format in Table~\ref{table:scalingFitsHFS} in the Appendix). Panels (a), (b) and (c) correspond to the $25$th quantile, median, and $75$th quantile, respectively. SVMC and SQA were run with $\beta = 2.5$ here. Additional simulation parameters for SA, SVMC, and SQA are given in Appendix~\ref{app:SA_SVMC}.  
The data symbols obscure the error bars, representing the 95\% confidence interval for each optimal TTS data point (computed from the fit of $\ln\langle \mathrm{TTS} \rangle$ 
 to a quadratic function as explained in Appendix~\ref{app:optimumFitting}).} 
\label{fig:TTSScaling}
\end{figure*}
\begin{figure*}
\centering
   \subfigure[\ fit to $a \exp(b L)$]{\includegraphics[width=0.32\textwidth]{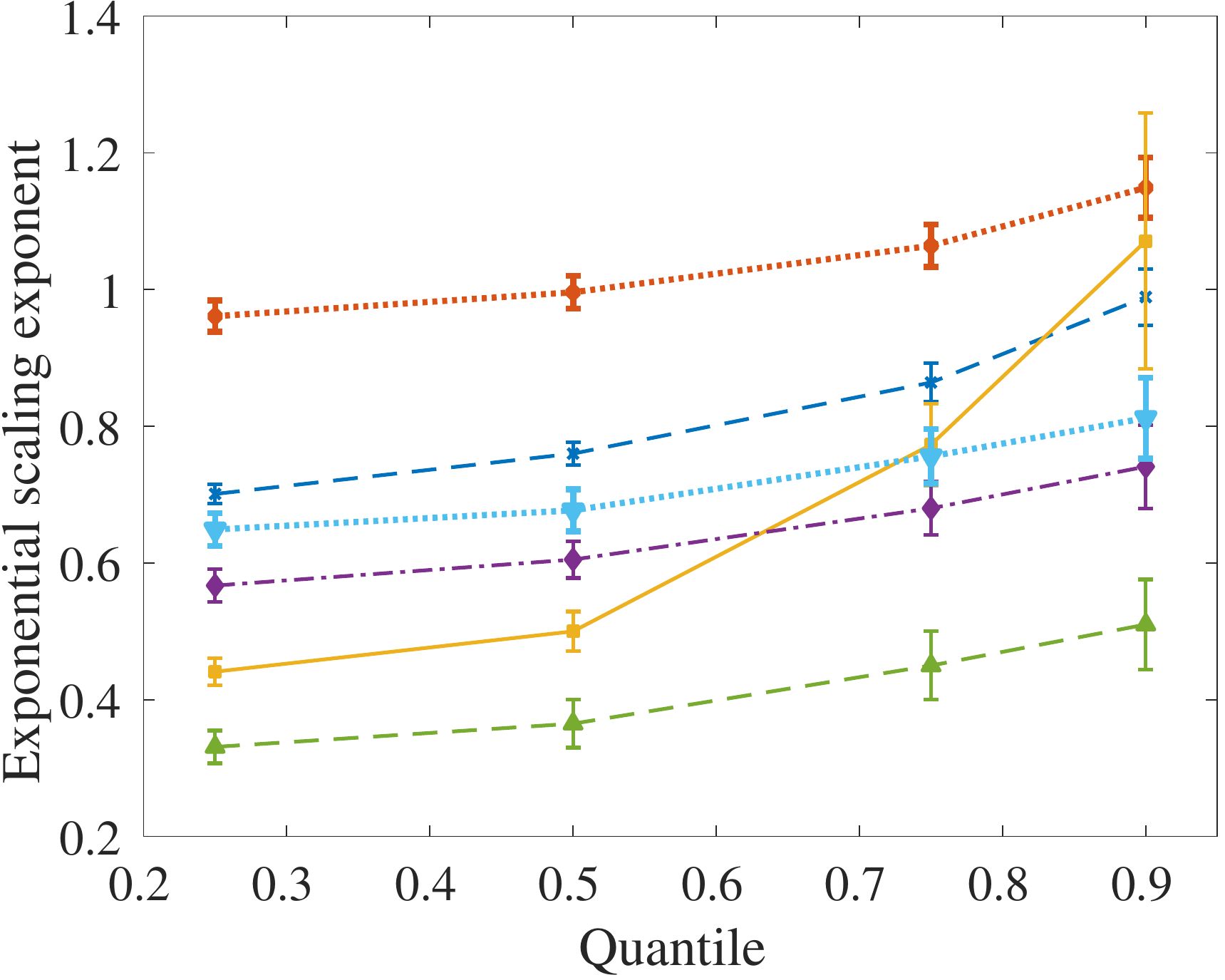} }
      \subfigure[\ fit to $a L^b$]{\includegraphics[width=0.32\textwidth]{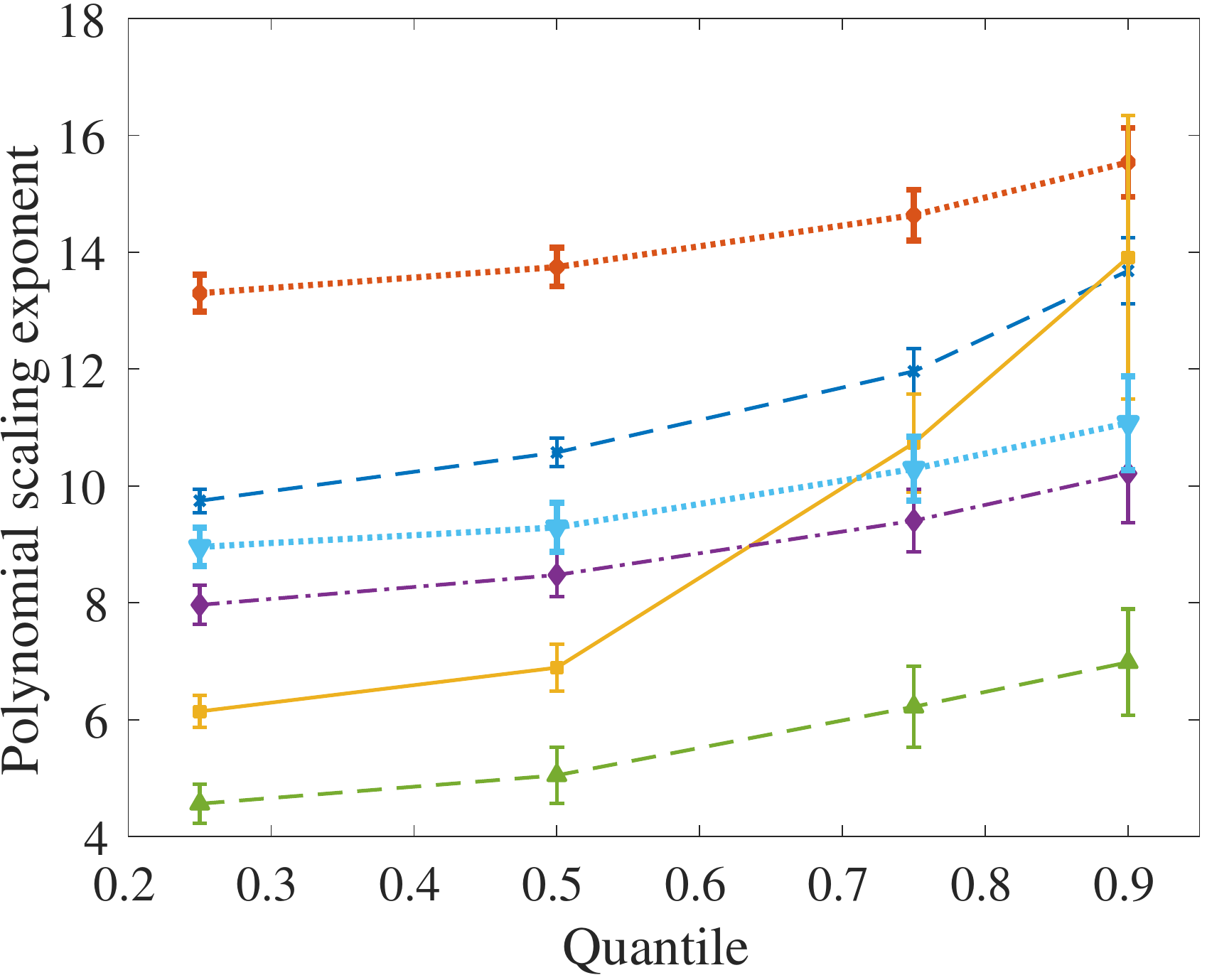} }
        \subfigure{\raisebox{15mm}{ \includegraphics[width=0.125\textwidth]{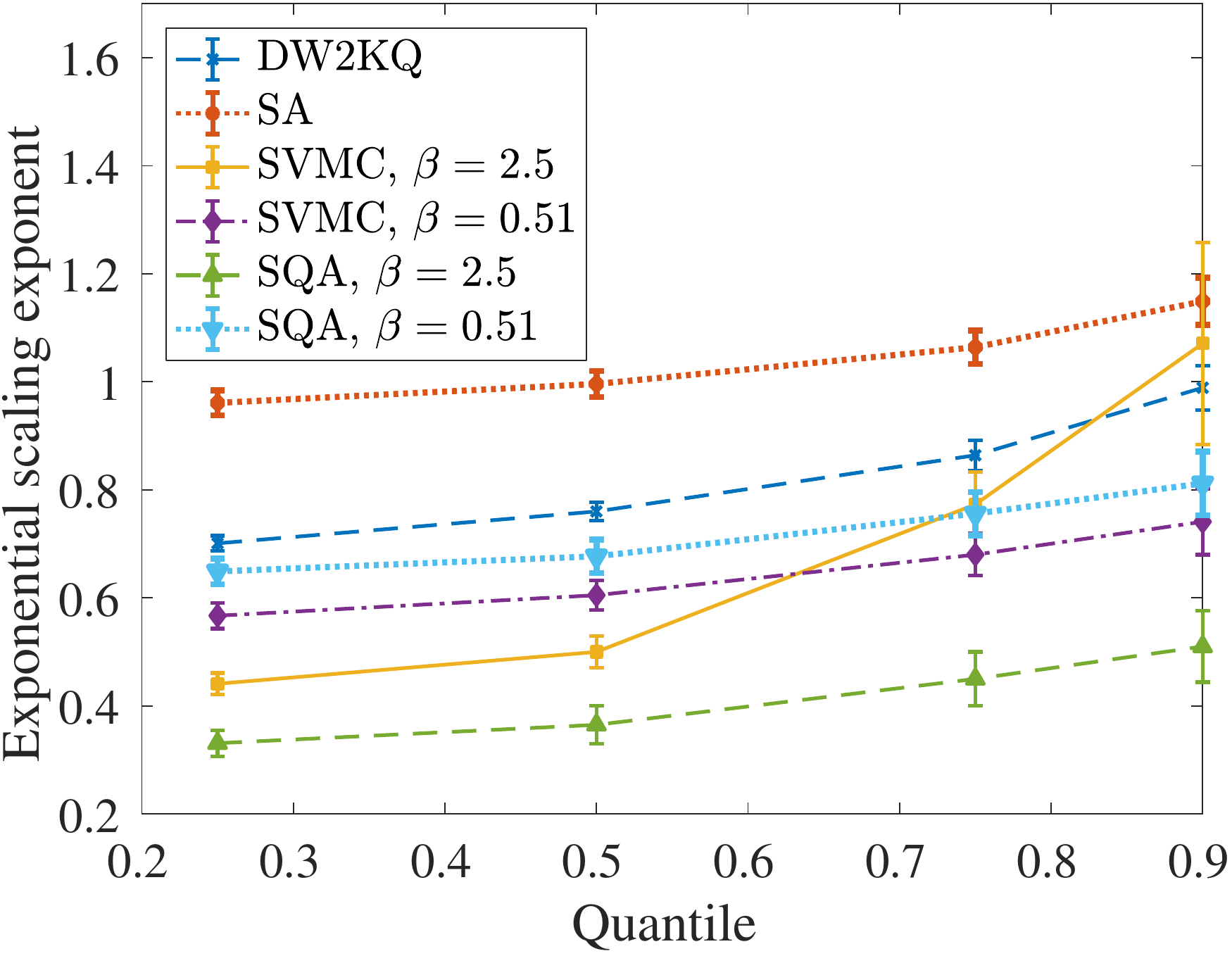} }}
\caption{\textbf{Scaling coefficients for the logical-planted instances.} The data shown are for the coefficient $b$ in fits to (a) $a \exp(b L)$ and (b) $a L^b$  for the logical-planted instances using $L \in [12,16]$ for different quantiles and different solvers.  Results are shown for the the DW2KQ, SA (with a final inverse-temperature of $\beta=5$), SVMC and SQA for two different inverse temperatures.  The value $\beta = 0.51$ corresponds to the operating temperature of the DW2KQ of $15$mK.} 
\label{fig:scalingFitsLogical}
\end{figure*}

\subsection{Evidence for a scaling advantage for QA hardware over simulated annealing, and a disadvantage against SQA and SVMC}
\label{sec:LQevidence}

Having established accessible optimal annealing times ($\geq 5 \mu$s) for the logical-planted instances, we are now ready to present a complete optimal-TTS scaling analysis. Our results for the dependence of $\TTSopt\ $ on problem size are shown in Fig.~\ref{fig:TTSScaling}, where we compare the DW2KQ results to three classical algorithms: 
SA with single-spin updates \cite{kirkpatrick_optimization_1983}, SQA based on discrete-time path-integral quantum Monte Carlo \cite{Santoro}, and SVMC \cite{SSSV}.  
Figure~\ref{fig:scalingFitsLogical} summarizes the performance of each algorithm in terms of the coefficients of exponential and polynomial fits, respectively, for several quantiles and two simulation temperatures for SQA and SVMC
(a hybrid polynomial-exponential fit does not work as well; see Appendix~\ref{app:AlternativeFits}).

\ignore{
\begin{table}[t]
\centering
\begin{tabular}{c c c c c c}
\hline\hline
(a) Solver & $q=0.75$ & $q=0.50$ & $q=0.25$  \\ [0.5ex] 
\hline 
DW2KQ &$0.864 \pm  0.028$ &   $0.760 \pm  0.017$ & $0.701 \pm  0.014$ \\ 
SA &   $1.064 \pm 0.031$ & $0.996 \pm 0.024$ & $0.961 \pm 0.023$\\
SVMC & $0.773 \pm 0.060$ &  $0.500 \pm 0.029$ &$0.441 \pm 0.020$ \\
SQA & $0.450 \pm  0.050$ & $0.365 \pm 0.035 $ & $0.331 \pm 0.024$ \\
[1ex]
\hline
\hline
(b) Solver & $q=0.75$ & $q=0.50$ & $q=0.25$  \\ [0.5ex] 
\hline
DW2KQ &   $11.962 \pm  0.391$ & $10.573 \pm  0.242$ &$9.746 \pm  0.201$  \\ 
SA &   $14.635 \pm 0.433$ & $13.746 \pm 0.331$ & $13.299 \pm 0.316$ \\
SVMC &  $10.735 \pm 0.834$ & $6.890 \pm 0.399$  &  $6.141 \pm 0.273$\\
SQA & $6.221 \pm 0.697$ & $5.047 \pm 0.484 $ & $4.561 \pm 0.331 $ \\
[1ex]
\hline\hline
\end{tabular}
\caption{The coefficient $b$ in fits to (a) $a \exp(b L)$ and (b) $a L^b$  for the logical-planted instances using $L \in [12,16]$; $q$ denotes the quantile. Errors are $95\%$ confidence intervals.}
\label{table:scalingFitsLogical}
\end{table}
}

The results presented in Fig.~\ref{fig:scalingFitsLogical} demonstrate a (95\% confidence) scaling advantage for the DW2KQ over SA in the case of the logical-planted instances, for the entire range of quantiles $[0.25,0.9]$. 
\emph{This represents the first observation of a scaling advantage over SA
on an experimental quantum annealer.} 

However, the SQA algorithm outperforms the DW2KQ in all quantiles and at both the colder inverse temperature of $\beta=2.5$ and the warmer $\beta=0.51$ (which corresponds to the operating temperature of the DW2KQ). The SVMC algorithm outperforms the DW2KQ in all quantiles at the warmer inverse temperature of $\beta=0.51$ and in all quantiles at $\beta=2.5$ except $q=0.9$ where the error bars are too large to make a statistically significant determination.
Thus, the scaling advantage over SA we observe is definitively not an unqualified quantum speedup. 

We note that our results are robust to modifying the SA annealing schedule from a quadratic to a linear function in $\beta$ (see Appendix~\ref{app:SA_DifferentSchedule}), and under a change of the metric to the so-called `quantile-of-ratios speedup'~\cite{speedup} (see Appendix~\ref{app:QofR}).

We also note that of all the solvers featured in Fig.~\ref{fig:scalingFitsLogical}, the scaling of SVMC at $\beta =2.5$ increases fastest from the easiest to the hardest quantile. As we discuss in more detail below, and is clear from Fig.~\ref{fig:scalingFitsLogical}, the SVMC and SQA performance depends strongly on the temperature at which the simulations are run.  Specifically, we find that SVMC performs {better} at higher quantiles at \emph{higher} temperatures, whereas SQA performs {better} at all quantiles at \emph{lower} temperatures.  We attribute this to harder instances involving an energy barrier that SVMC must thermally hop over, while SQA can mimic tunneling through. This also suggests that the DW2KQ performance is severely hindered by its sub-optimally high temperature. To explain this, we next discuss and motivate how we constructed our problem instances.

\subsection{Construction of problem instances with an optimal annealing time}
\label{sec:instances}

Having presented the evidence for optimality and the scaling analysis, we next describe the instance class with these properties.  The two key properties we wish our instances to possess are (1) a guarantee of knowing the ground state energy (a useful feature for benchmarking optimizers at ever-growing problem sizes), and (2) an optimal annealing time on the D-Wave processors.  

\subsubsection{Planted solutions}
In order to guarantee a known ground state energy, we construct `planted' solution instances.  The method builds the problem Hamiltonian as a sum of frustrated loop Hamiltonians $H_{\ell}$, such that
\begin{equation} \label{eqt:Hp1}
H_{\mathrm{P}} = \sum_{\ell} H_{\ell}\ 
\end{equation}
itself is `frustration-free', i.e., the planted solution is the simultaneous ground state of all $H_{\ell}$ terms and hence is the ground state of $H_{\mathrm{P}}$ \cite{Hen:2015rt}.  Without loss of generality, we can always pick the planted solution to be the $\ket{0\cdots 0}$ (all-zero state) configuration, where henceforth the states $\ket{0}$ and $\ket{1}$ denote the eigenstates of the $\sigma^z$ operator with $+1$ and $-1$ eigenvalues, respectively.
We consider planted solutions defined on the logical graph formed by the complete unit cells of the D-Wave hardware graph (i.e., without faulty qubits or couplers; in the case of an ideal Chimera graph, this would form a square lattice) \cite{DW2000Q}.  Frustrated loops are then built on this logical graph, where logical couplings between adjacent unit cells are imposed only when all four physical inter-unit cell couplings are available. 
The intra-unit cell couplers are then all set to be ferromagnetic, guaranteeing that the planted-solution on the hardware graph is the planted-solution on the logical graph with all physical spins in the unit cell set to their corresponding logical spin value.  We refer to these as `logical-planted' instances.  In Appendix~\ref{sec:InstanceConstruction} we introduce `hardware-planted' instances and demonstrate that they also exhibit an optimal annealing time.

\subsubsection{Gadgets}
In order to identify problem instances that exhibit an optimal annealing time, we first recall that previous studies of planted-solution instances on the D-Wave processors \cite{Hen:2015rt,King:2015zr,DW2000Q} found a TTS that rises monotonically as a function of the annealing time. Keeping Fig.~\ref{fig:ExampleOptimalAnnealingTime1} in mind, a decreasing or increasing TTS results from the success probability rising sufficiently fast or too slowly, respectively, with increasing annealing time (we formalize this in Sec.~\ref{sec:gad-resp} below).  
In the case where the system is very weakly coupled to its thermal environment, we can expect a competition between adiabaticity (unitary dynamics) and open system effects such as thermal excitations~\cite{Steffen:2003ys,PhysRevLett.95.250503}, resulting in a peak in the success probability and a minimum in the TTS. Though we note that it is unlikely that the DW2KQ operates entirely in the weak coupling regime (the minimum gap associated with the gadget is already below the temperature energy scale, as shown in Fig.~\ref{fig:GadgetHW}, and we expect the minimum gap of the large instances to be smaller), from this perspective, shifting the minimum in the TTS to larger $t_f$ values corresponds to prolonging the timescale over which adiabaticity dominates over open system effects.  One way to try to accomplish this is by enhancing the role of finite-range tunneling in the dynamics.  Motivated by this insight, and by recent work on the possibility of a computational role of finite multi-qubit tunneling in quantum annealers \cite{Boixo:2014yu,PhysRevX.6.031015}, we introduce a key modification and supplement the planted-solution instance Hamiltonian [Eq.~\eqref{eqt:Hp1}] with terms corresponding to the addition of identical $8$-qubit `gadgets' that exhibit tunneling during their anneal: 

\begin{figure}[b] 
   \centering
   \includegraphics[width=0.9\columnwidth]{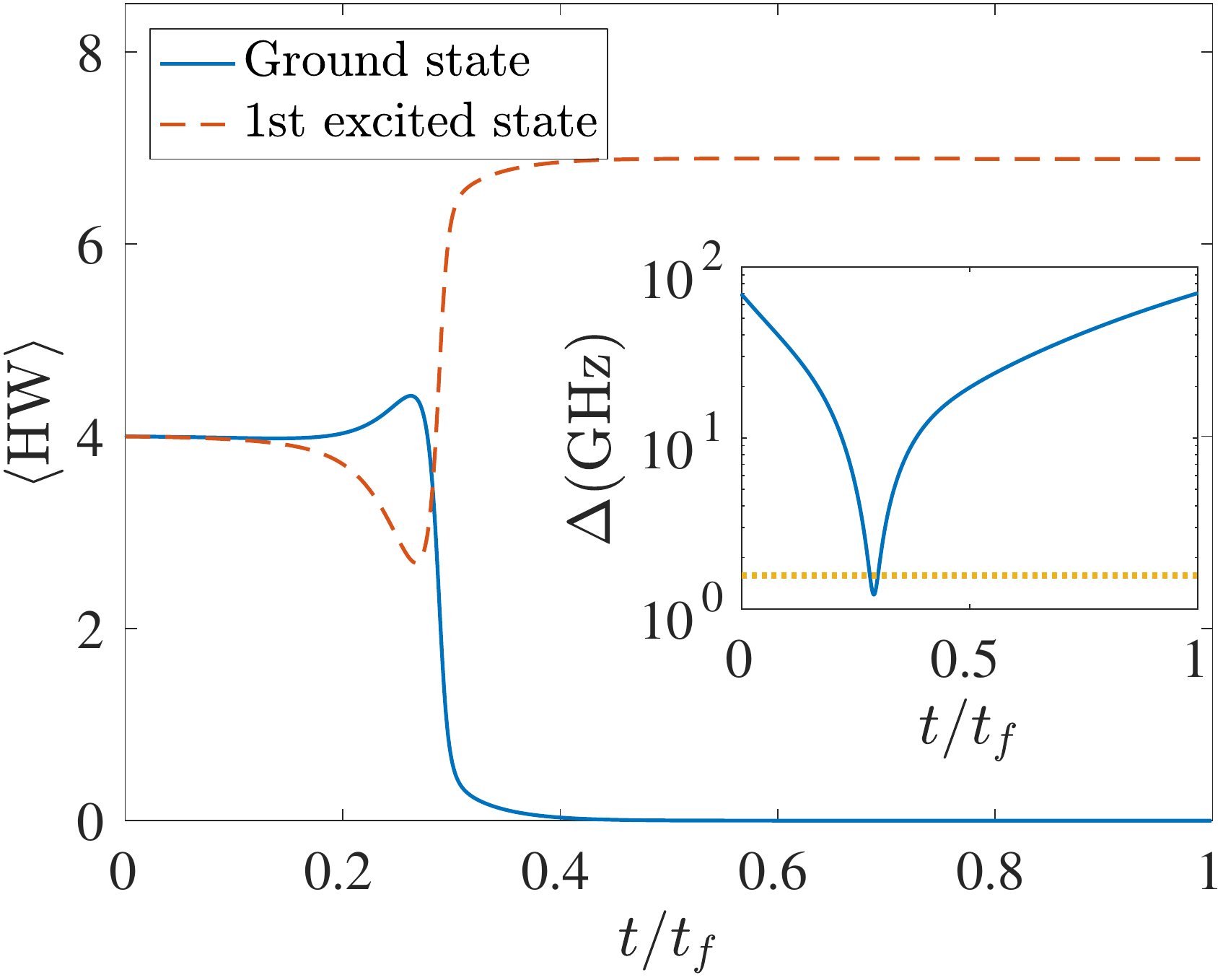} 
   \caption{\textbf{Expectation values of the Hamming weight operator.} Shown are the ground state and first excited state expectation values of $\mathrm{HW} = \frac{1}{2} \sum_{i=1}^n \left( 1 - \sigma_i^z \right)$ for the $8$-qubit gadget using the DW2KQ annealing schedule.  Inset: The ground state energy gap to the first excited state, as calculated using the DW2KQ annealing schedule. 
The dotted line corresponds to the operating temperature of the device.}
   \label{fig:GadgetHW}
\end{figure}
\begin{figure}[b] 
   \centering
   \includegraphics[width=0.5\columnwidth]{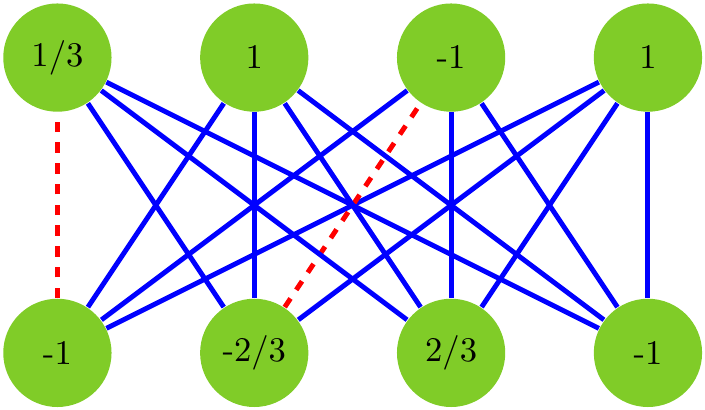} 
   \caption{\textbf{The $8$-qubit gadget used in the instance construction.}  The qubits (green circles) are arranged in a complete bipartite graph.  Blue (red) lines correspond to ferromagnetic (anti-ferromagnetic) Ising couplers with magnitude $1$.  The value of the local fields on the qubits are given inside the circles, with a negative value indicating a spin up preference.}
   \label{fig:8gadget}
\end{figure}
\begin{equation}
H_{\mathrm{P}}' = H_{\mathrm{P}} + \sum_{i\in\mathcal{S}} H_{\mathrm{G}_i}\ .
\end{equation}
Here $H_{\mathrm{G}_i}$ denotes the gadget Hamiltonian in unit cell $i$, and the gadgets are placed into randomly chosen unit cells: $\mathcal{S}$ denotes a randomly chosen subset comprising a fraction $p$ of complete unit cells (we use $p=0.1$).  
The specific $8$-qubit gadget we used fits into the unit cell of the D-Wave processors, and its
%
%
connectivity and parameters are depicted in Fig.~\ref{fig:8gadget}. The ground state of the gadget is the all-zero state of the eight qubits, so that the ground state of the full Hamiltonian remains the all-zero state. The first excited state of the gadget is doubly degenerate with average Hamming weight seven.

\begin{figure*}[t] 
   \centering
  \subfigure[\ gadget only]{ \includegraphics[width=0.3\textwidth]{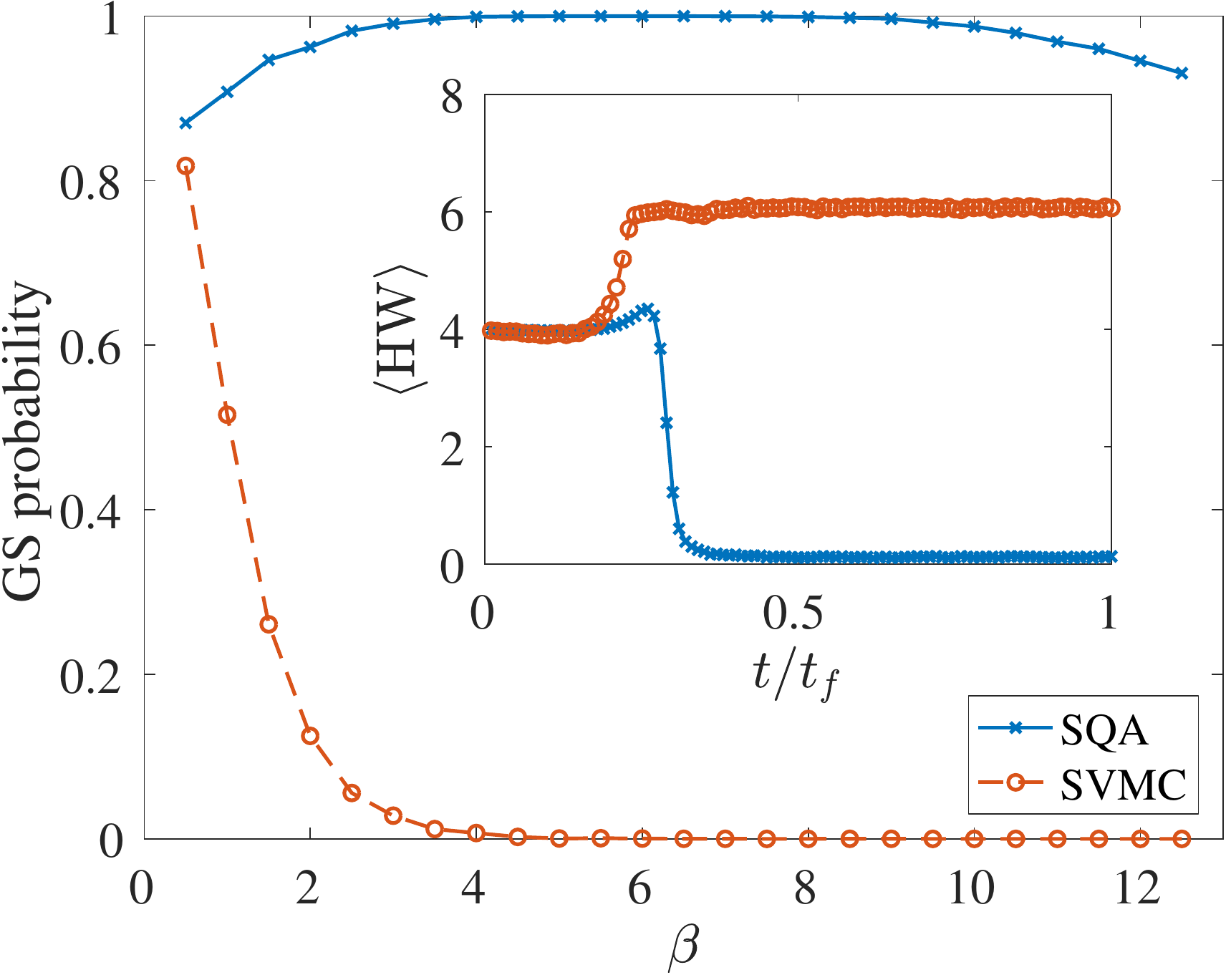} \label{fig:GadgetSQASVMC}} 
  \subfigure[\ instance exhibiting tunneling]{ \includegraphics[width=0.3\textwidth]{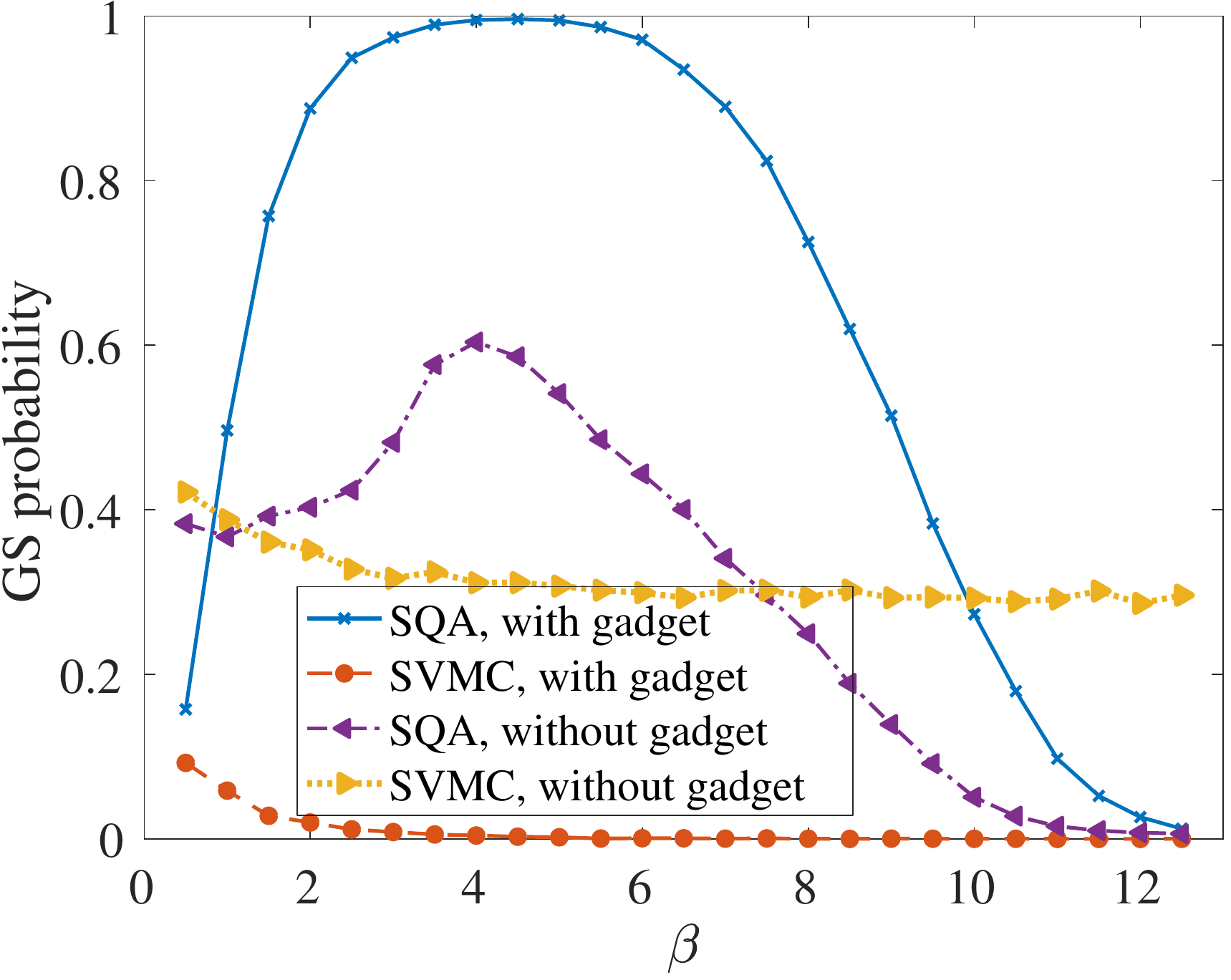} \label{eqt:Instance852}}
  \subfigure[\ instance not exhibiting tunneling]{ \includegraphics[width=0.3\textwidth]{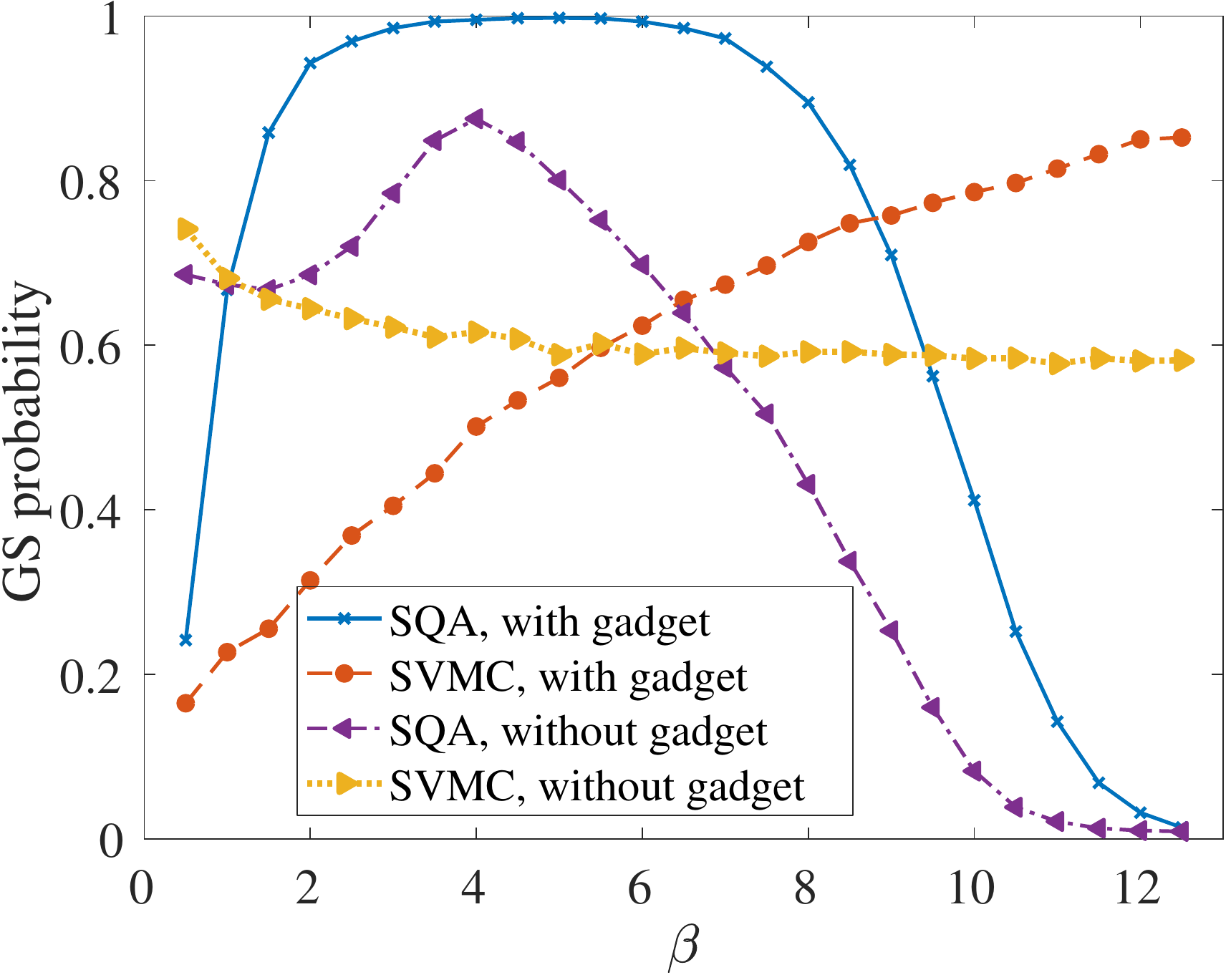} \label{eqt:Instance470} }
   \caption{\textbf{Probability of reaching the ground state at the end of the anneal using SVMC and SQA for different simulation inverse temperatures $\beta$.}  
   (a) Simulation results for the $8$-qubit gadget only. SQA's success probability increases over a wide range of decreasing temperatures, whereas SVMC rapidly deteriorates as the temperature decreases. Inset: Expectation values of the Hamming weight operator $\mathrm{HW} = \frac{1}{2} \sum_{i=1}^n \left( 1 - \sigma_i^z \right)$ for the $8$-qubit gadget for SVMC and SQA using $\beta = 2.5$.  Compare to the behavior of the ground state and first excited state shown in Fig.~\ref{fig:GadgetHW}.  For SVMC, to compute the expectation value of the Hamming weight operator at intermediate $s$ values, we can either project the state to the computational basis (shown) or use the spin-coherent state; the results are almost indistinguishable.  For SQA, we average over the Hamming weight of the configurations in the imaginary-time direction.
   (b) and (c) Probability of reaching the ground state at the end of the anneal using SVMC and SQA for different simulation inverse temperatures $\beta$ using two different instances, with and without the $8$-qubit gadget, at the largest available size $L=16$. (b) An instance that exhibits a clear signature of a tunneling energy barrier when the gadget is introduced. 
   (c) An instance that does not exhibit a signature for a tunneling energy barrier even with the gadget.  In all panels SVMC and SQA simulations used $8$M and $3$M sweeps respectively, and both algorithms use the DW2KQ annealing schedule.  The drop in success probability at large $\beta$ for SQA is because spin updates become less efficient at high $\beta$ and more spin updates are required to maintain the high success probability.}
   \label{fig:L=16SQASVMC}
\end{figure*}
%

%
Generically, one would not expect the annealing properties of $H_{\mathrm{G}}$ to be shared by $H_{\mathrm{P}}'$, but below we show to what extent they are for our instances.

\subsubsection{Tunneling}
We next establish in what sense our gadget exhibits tunneling.  We show in Fig.~\ref{fig:GadgetHW} the expectation value of the Hamming weight operator $\mathrm{HW} = \frac{1}{2} \sum_{i=1}^n \left( 1 - \sigma_i^z \right)$ in the ground state and first excited state, computed by numerically solving the time-dependent Schr\"{o}dinger equation for the evolution of the gadget. This expectation value exhibits a sharp change at the same point in the evolution where the minimum gap occurs (shown in the inset). The ground state reorients itself to the $\ket{0\cdots 0}$ state, while the first excited state reorients to align closely with the $\ket{1\cdots 1}$ state. This already suggests a tunneling transition, but in order to confirm this we wish to establish the presence of an energy barrier in the semiclassical potential that the quantum system must tunnel through during the anneal. Such tunneling transitions have been well-studied in the context of systems with qubit-permutation invariance \cite{Farhi-spike-problem,FarhiAQC:02,Schaller:2007uq,Boixo:2014yu,Muthukrishnan:2015ff}, but less so in the context of systems such as ours without this symmetry.      

The semiclassical potential as derived from the spin-coherent path integral formalism \cite{owerre2015macroscopic} is given by the expectation value of $H(t)$ in  
the spin-coherent state $\ket{\Omega (\vec{\theta}, \vec{\varphi}) }= \otimes_{i=1}^n \left[ \cos\left( \frac{\theta_i}{2} \right) \ket{0}_i + e^{i \varphi_i}  \sin\left( \frac{\theta_i}{2} \right) \ket{1}_i \right]$.
In the context of the transverse field Ising Hamiltonian [Eq.~\eqref{eqt:QA}], the semiclassical potential becomes:

\begin{figure}[t] %
   \centering
\includegraphics[width=0.9\columnwidth]{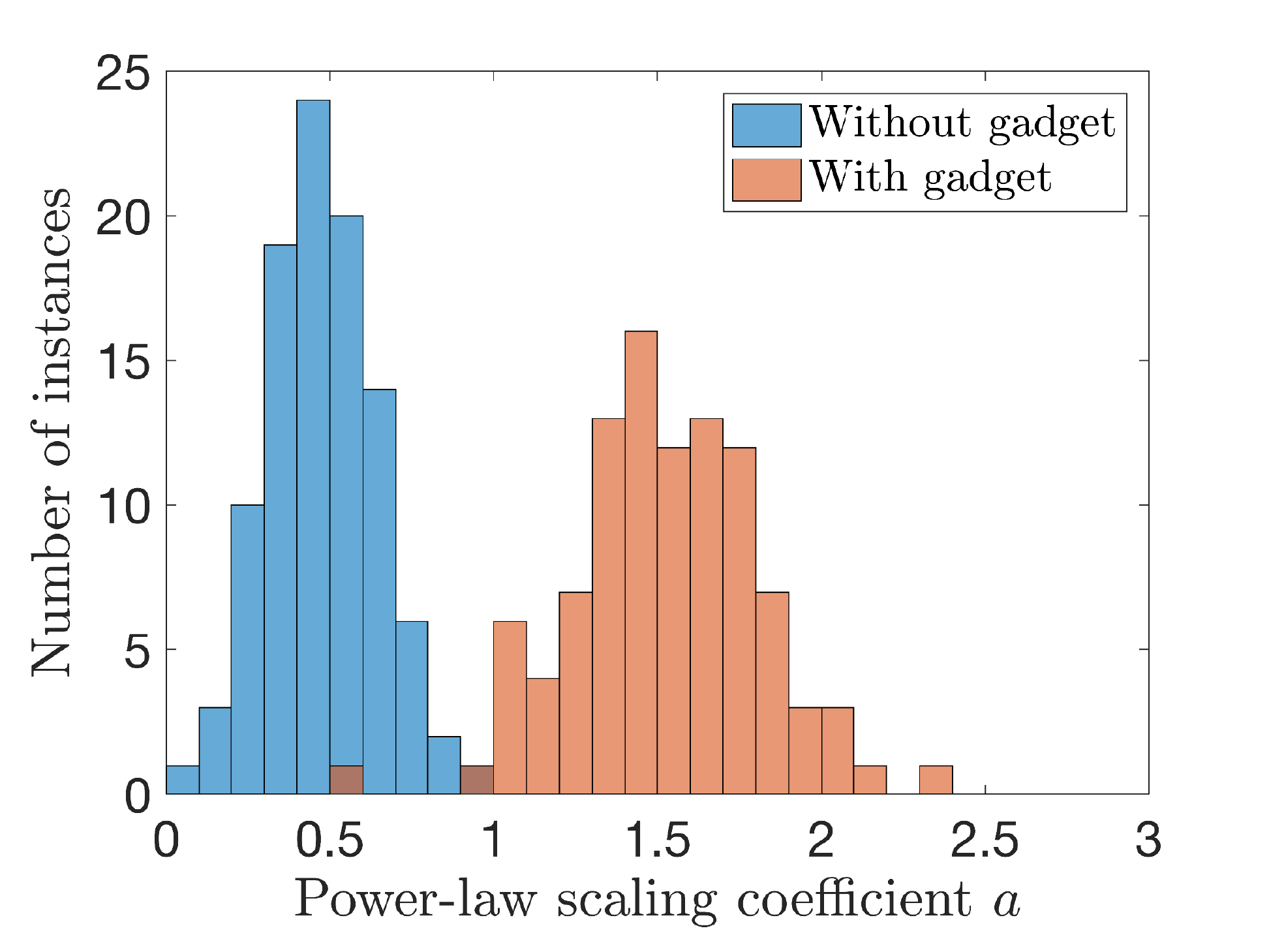}
\caption{\textbf{Empirical scaling behavior with and without the gadget.} Shown is the distribution of the power law scaling coefficient $a$ obtained after fitting $\ln p_{\mathrm{S}}$ to $a \ln t_f + b$ for  $100$ instances at $L=16$, run on the DW2KQ processor. We choose $t_f \in [5, 50 ]\mu$s since this is the range over which the TTS decreases as seen in Fig.~\ref{fig:ExampleOptimalAnnealingTime1}. The instances with the gadget typically exhibit a larger scaling coefficient, which  leads to the observation of an optimal annealing time.
In (a) and (b) error bars represent $95\%$ confidence intervals ($2\sigma$) calculated using $1000$ bootstraps of $100$ gauge transformations \cite{q108}.}
\label{fig:GroupOptimum}
\end{figure}
\begin{eqnarray} \label{eqt:VTIM}
V(\vec{\theta},\vec{\varphi},t) &=& - A(t) \sum_i \sin(\theta_i) \cos(\varphi_i)  \\
&& \hspace{-1cm} + B(t) \left( \sum_{i \in \mathcal{V}} h_i \cos \theta_i + \sum_{(i,j) \in \mathcal{E}} J_{ij} \cos(\theta_i) \cos(\theta_j) \right) \notag \ .
\end{eqnarray}
Equation~\eqref{eqt:VTIM} provides a multi-dimensional energy landscape for the quantum annealing protocol. Unfortunately, due to the absence of any symmetries it is infeasible to exhaustively explore this landscape and identify the actual location of barriers, even under the simplification where $\varphi_i = 0 , \ \forall i$.  
Instead, as proxies for a direct calculation of tunneling transition matrix elements or an instanton analysis~\cite{Jorg:2008aa}, we consider the behavior of the SVMC and SQA algorithms. SVMC performs Metropolis updates on the potential energy landscape, Eq.~\eqref{eqt:VTIM} \cite{SSSV}.  Since this algorithm can only thermally `hop' over energy barriers, we expect its performance to deteriorate with decreasing temperature in the presence of a relevant energy barrier.  On the other hand, a path-integral Monte Carlo based approach like SQA should be able to not only thermally hop over these barriers
but also mimic tunneling through them~\cite{2015arXiv151008057I,Jiang:2017aa,Andriyash:2017aa}, which should 
benefit from a decreasing temperature.  Therefore, we 
expect to be able to identify tunneling energy barrier bottlenecks in the quantum anneal by contrasting the temperature dependence of the performance of SVMC and SQA%
.  

Figure~\ref{fig:GadgetSQASVMC} shows that for our $8$-qubit gadget, SQA and SVMC behave as expected in the presence of a tunneling energy barrier: the success probability of SVMC decreases with decreasing temperature, whereas the success probability of SQA increases with decreasing temperature.  As shown in the inset and comparing to Fig.~\ref{fig:GadgetHW}, we see that SVMC is effectively trapped in the higher excited states, while SQA is able to follow the ground state.

Next, we use the same technique to probe our planted-solution instances with and without the gadget.  We show the behavior of two very different $L=16$ instances in Fig.~\ref{fig:L=16SQASVMC}.  For one of the instances [Fig.~\ref{eqt:Instance852}], the introduction of the gadget adversely affects the performance of SVMC, pushing the success probability to zero for increasing inverse-temperature $\beta$.  For SQA, the improvement in performance as $\beta$ increases is significantly sharper with the gadget.  For the second instance [Fig.~\ref{eqt:Instance470}], we see that while SQA's behavior is almost identical, SVMC exhibits an improving performance with increasing $\beta$ in the presence of the gadget, suggesting an absence of an energy barrier. This analysis demonstrates that the tunneling properties induced by our $8$-qubit gadget can be inherited by the problem instances even at the largest problem size, and that the success probability exhibits a strong temperature dependence. For further details on the behavior of an ensemble of instances see Appendix~\ref{sec:InstanceConstruction}. 

Unfortunately we cannot directly probe tunneling or study the temperature dependence on the D-Wave processors. To the extent that SQA models the behavior of the physical quantum annealer, one may choose to interpret the evidence we have presented above as evidence for the role of tunneling energy barriers induced by the gadgets. 

\subsubsection{The gadget is responsible for the observed optimal annealing time}
\label{sec:gad-resp}
To more directly understand the effect of our gadget on the hardware quantum annealer, it is instructive to contrast the scaling behavior with and without the gadget. Toward that end, we fit 
the empirical success probability $p_{\mathrm{S}}$ to a power law of the form $b (t_f)^a$ (see Appendix~\ref{app:pGSscaling} for the fit quality).  
We show in Fig.~\ref{fig:GroupOptimum} the distribution of the scaling coefficient $a$ for $100$ instances for the logical-planted instances with and without the gadget.  
The two distributions differ substantially: the instances with the gadget exhibit larger coefficients, almost all with a value greater than $1$, resulting in a significantly larger initial rate of increase in $p_{\mathrm{S}}(t_f)$ than for the instances without the gadget. This, in turn, leads to the observed initial decrease in the TTS with increasing annealing time: upon expanding the logarithm in Eq.~\eqref{eq:TTS} for small $p_{\mathrm{S}}$, we find that $\mathrm{TTS}(t_f) \propto t_f/p_\mathrm{S}(t_f) = (t_f)^{1-a}$, so that $\mathrm{TTS}(t_f)$ decreases with $t_f$ provided $a>1$; this is consistent with Fig.~\ref{fig:ExampleOptimalAnnealingTime1}, where the slope of $p_\mathrm{S}(t_f)$ is indeed seen to be initially $>1$, then dropping to $<1$ for larger $t_f$. Since the TTS must eventually increase with the annealing time [ideally, for sufficiently large $t_f$, $R(t_f)\to 1$, at which point $\mathrm{TTS}(t_f)\propto t_f$], this helps to explain why the instances with the gadget exhibit an optimal annealing time. It also suggests a useful heuristic for future studies attempting to identify problem instance classes with an optimal annealing time: the instances should have the property that if $p_{\mathrm{S}}(t_f)=g(t_f)\ll 1$ for some function $g$, then $\mathrm{TTS}(t_f)\propto t_f/g(t_f)$ must be decreasing for some range of $t_f >t_{\min}$ values. This is compatible with any faster-than-linear form for $g$. 
We already alluded earlier to a competition between adiabaticity and thermal excitations as being potentially responsible for an optimal annealing time. Another mechanism, that appears to be more consistent with the fact that the DW2KQ is not operating in the weak coupling regime, is that thermal relaxation is fast for small $t_f$ and then slows down for sufficiently large $t_f$, presumably since the system has already entered the quasistatic regime~\cite{Amin:2015qf}.

\section{Discussion and outlook}
\label{sec:discussion}

The key result of this work is the demonstration of an algorithmic scaling advantage for QA hardware over the SA algorithm for a family of problem instances constructed with frustrated loops and a small gadget that exhibits tunneling.
It is worth emphasizing why the combinatorial optimization and quantum annealing communities have often focused on suboptimal heuristics such as SA (see, e.g., Ref.~\cite{Boixo:2014yu}). SA is not only a very general meta-heuristic, but it is also often been viewed as the 
inspiration for QA, with thermal fluctuations replaced by quantum fluctuations~\cite{kadowaki_quantum_1998}.  
Because of this correspondence between SA and QA, a demonstration of superior performance by QA can presumably be attributed to an advantage of the quantum approach over the thermal approach. The goal is then to leverage this advantage to a broader range of problems. 
However, as we have argued, temperature annealing as in SA is actually quite different from transverse field annealing, so that the analogy between SA and QA needs to be treated with care.
Another concern we face is that while the accuracy threshold theorem provides a theoretical guarantee that for sufficiently low noise levels and through the use of quantum error correction, a finite-size device can be scaled up fault tolerantly \cite{Aharonov:05}, 
in the absence of such an asymptotic guarantee for quantum annealing a finite-size device provides evidence of what can be expected at larger, future sizes, only provided the device temperature, coupling to the environment, and calibration and accuracy errors, can be appropriately scaled down. 

\begin{figure}[t] %
   \centering
   \includegraphics[width=0.8\columnwidth]{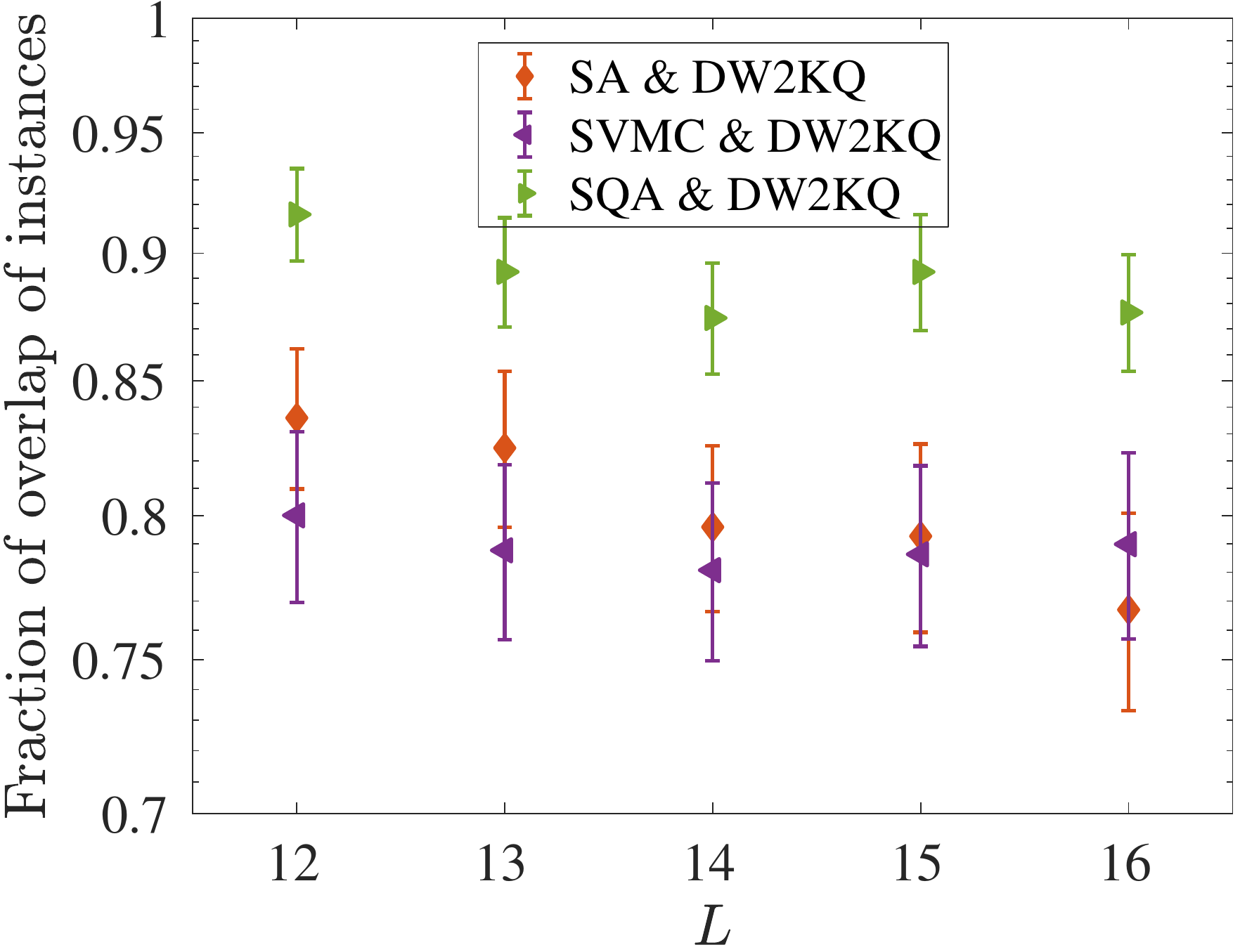} 
   \caption{{\bf Overlap of the instances that fall below the median TTS for the classical solvers and the DW2KQ.}  We calculate the TTS for $1000$ instances with each solver's respective optimal annealing time for the median at a given size $L$, and check which instances fall below the median TTS.  Shown is the (normalized) fraction of the overlap of the instances between the solvers.  Further details are given in Appendix~\ref{app:Overlap}.  
Error bars represent $95\%$ confidence intervals ($2\sigma$) calculated using $1000$ bootstraps of $1000$ instances.}
   \label{fig:Overlap}
\end{figure}

In light of this, what is the significance of our demonstration of a QA scaling advantage over SA? We believe that an important clue lies in the fact that SVMC also exhibits an advantage over SA for these problems. The SA and SVMC algorithms can both be viewed not only as classical analogues of QA, but also as implementing two of its possible classical limits~\cite{kadowaki_quantum_1998,SSSV,Crowley:2016aa}.  While SA performs updates on the classical energy landscape associated with the Ising Hamiltonian, SVMC performs updates on the semiclassical potential associated with the quantum anneal.  A scaling difference between the two, with an advantage for SVMC, suggests that thermal updates on the semiclassical energy landscape is more efficient.  While it is unclear whether the quantum effects in the D-Wave devices that have already been demonstrated on a smaller scale ($N \lesssim 16$) 
\cite{q-sig,q108,Albash:2014if,q-sig2,DWave-entanglement,Dwave,PhysRevX.6.031015,Boixo:2014yu}
remain operative at the much larger scales we have employed in our study, the fact that the DW2KQ also exhibits an advantage over SA suggests that it must be evolving in a landscape that also allows for better scaling. To be specific, this is the landscape associated with transverse field annealing as opposed to temperature annealing.
It is in this sense that quantum effects that are necessarily absent from SA and \emph{might} be present in the quantum annealer, can provide an advantage.  This is especially significant since there is a large overlap between the instances solved at the median quantile by the DW2KQ and all three of the classical algorithms, including SA, as shown in Fig.~\ref{fig:Overlap}. This means that if any quantum effects are responsible for the scaling advantage of the DW2KQ over the SA algorithm, then they are operative in largely the same set of problem instances, so that these instances may define a target class for quantum enhanced optimization.

Likewise, SQA also evolves on the same semiclassical energy landscape as SVMC, but in addition to thermal updates is also capable of mimicking tunneling. The fact that SQA's scaling is far superior to SVMC's, and that this improves as the simulation temperature is lowered, shows that tunneling is effective at enhancing SQA's performance for the logical-planted instances. What does this tell us about the possibility that the relative performance of the algorithms is indicative of quantum effects in the DW2KQ device? It is known that the scaling of quantum Monte Carlo can be as efficient \cite{2015arXiv151008057I,Jiang:2017aa,Mazzola:2017aa} or less efficient \cite{Andriyash:2017aa} than the incoherent tunneling rate scaling of a true quantum annealer.  Therefore, the fact that SQA overwhelmingly outperforms the DW2KQ but that the DW2KQ still outperforms SA, suggests that the device is dominated by classical dynamics with a very small quantum component.  While only speculative at this point, this type of situation might be the best we can hope for in the current generation of highly noisy quantum annealers, without some form of quantum error correction or suppression \cite{jordan2006error,Bookatz:2014uq,Jiang:2015kx,Marvian-Lidar:16,Marvian:2017aa}.

Figure~\ref{fig:scalingFitsLogical} shows how the scaling of both SVMC and SQA is strongly affected when we increase their temperatures to the DW2KQ dilution fridge temperature.  In both cases, for the median and lower percentile, SQA and SVMC's performance is hurt at this higher temperature relative to the colder temperature.  This strongly suggests that the DW2KQ's performance for this class of instances is severely impacted by its temperature, consistent with general expectations~\cite{Albash:2017ab}. However, we also find that  the SVMC algorithm scales better than the DW2KQ. One would expect that if SVMC is the classical limit of the device, then the DW2KQ should perform at least as well as SVMC. One possible explanation for the violation of this expectation is that additional noise sources, such as implementation errors, further degrade the performance of the DW2KQ relative to solvers run on digital classical computers \cite{King:2015zr}.  

We note that SVMC's performance improves at higher percentiles at higher temperatures relative to colder temperatures, which is consistent with the algorithm being able to thermally hop over energy barriers.  This indicates that temperature is another algorithmic parameter that should be optimized separately for each quantile, a point we leave for future studies.

We emphasize that the \emph{discrete}-time SQA algorithm studied here should not be interpreted as a true model of a thermal quantum system.  It has been demonstrated that time-discretization may result in improved residual energy minimization performance over the continuum case \cite{Heim:2014jf}, although this does not necessarily translate into a scaling performance advantage. Nevertheless, the superior performance of the SQA algorithm we have observed is an interesting finding in its own right: we are unaware of another example of an Ising model cost function where SQA with closed boundary conditions bests SA as a ground state solver (Ref.~\cite{PhysRevX.6.031015} reports an example but uses SQA with open boundary conditions). We have attributed this advantage to a more favorable energy landscape with the presence of tunneling barriers than can be traversed efficiently, but to test whether the observed scaling advantage would hold for a thermalizing quantum annealer requires quantum Monte Carlo simulations without Trotter errors 
\cite{sqa2,sandvik:03,Albash:2017a}; 
unfortunately, at the $>2000$ qubits scale we have worked with here, this is computationally prohibitive. The same is true for master equation simulations \cite{ABLZ:12-SI,Boixo:2014yu}, even when implemented using the quantum trajectories method \cite{Yip:2017aa}.

We emphasize that the instances presented here are not necessarily computationally hard, as suggested by the fact that, considering the entire range of sizes we tested, the quality of the polynomial fits is better than that of the exponential fits (see Fig.~\ref{fig:TTSScaling}). In the absence of the gadget, the logical-planted instances are defined on a square lattice and can be solved in polynomial time using the exact minimum-weight perfect-matching algorithm~\cite{Mandra:2017aa}. However, we have confirmed that this algorithm performs poorly once the gadget is included, as expected when local fields are present (see Appendix~\ref{app:HFS}).  Therefore, it is natural that algorithms optimized with respect to the problem structure demonstrate superior performance.  For example, simulated annealing with both single and multi-spin updates (SAC), with the latter being simultaneous updates of all the spins comprising a unit cell (super-spin approximation~\cite{2016arXiv160401746M}), scales significantly better than SA with single spin updates but still does not perform as well as SQA (see  Appendix~\ref{app:HFS}).  
Furthermore, there are many other classical algorithms that do not implement the same algorithmic approach as quantum annealing, such as the Hamze-Freitas-Selby (HFS) \cite{hamze:04,Selby:2014tx} algorithm. The latter exploits the low tree-width of the Chimera connectivity graph and has in all studies to date been the top performer for Chimera-type instances. In contrast, here we find that HFS's scaling performance lies between the DW2KQ and SAC (see Appendix~\ref{app:HFS}).   Another competitive algorithm is parallel tempering with iso-energetic cluster moves~\cite{Houdayer2001,PhysRevLett.115.077201}.
We can expect that a more highly connected hardware graph will prevent algorithms such as HFS 
or SAC from being efficient; which architectures may lend themselves to an unqualified quantum speedup remains an open research question. 

In this work we focused on the task of finding any ground state, and did not address the question of how well quantum annealing can uniformly sample the ground states, commonly referred to as ground state `fair sampling' and a problem that belongs to the complexity class \#P.  It is well-established that quantum annealing with the standard transverse field driver Hamiltonian samples the ground states in a biased manner \cite{Matsuda:2009,q-sig,Zhang2017,PhysRevLett.118.070502}, and our work does not establish to what extent this bias exists for the class of instances we study, nor whether the different algorithms studied exhibit a different bias.  Addressing this question provides another approach for searching for a quantum advantage beyond the standard optimization approach \cite{Albash:2014if,Zhang2017}.

Meanwhile, our hybrid frustration-tunneling based instance construction approach defines a clear path forward by concretely establishing the possibility of generating instance classes with accessible optimal annealing times, amenable to a complete scaling analysis. By ``mining" those instances which exhibit the largest separation between SQA and the top performing alternative classical algorithms, while corroborating that the performance of hardware-based QA is also competitive on the same instances (at a minimum it should certainly continue to beat SA), we expect to be able to identify the features that give rise to a quantum advantage and learn how to amplify the difference. This procedure can be iterated, all the while ensuring that optimal annealing times can be ascertained, in order to amplify the separation. Current QA hardware may simply be too hot and incoherent to exhibit an amplification leading to an unqualified scaling advantage over all classical algorithms for the resulting instance class.  This is especially true for instances amenable to SQA simulations that can reproduce the incoherent tunneling rates of a noisy quantum annealer \cite{2015arXiv151008057I,Jiang:2017aa,Mazzola:2017aa}, in which case more coherent devices will be necessary. Nevertheless the principles we have established here will at the very least provide a means of testing this exciting possibility.

\begin{acknowledgments}
We thank D-Wave Systems Inc. for providing access to the DW2KQ device in Burnaby, Canada. TA thanks James and Andrew King for useful discussions on the GPU implementations of the classical algorithms.  This work was supported under ARO grant number W911NF-12-1-0523, ARO MURI Grant Nos. W911NF-11-1-0268 and W911NF-15-1-0582, and NSF grant number INSPIRE-1551064. This research used resources provided by the USC Center for High Performance Computing and Communications and by the Oak Ridge Leadership Computing Facility, which is a DOE Office of Science User Facility supported under Contract DE-AC05-00OR22725. The research is based upon work partially supported by the Office of
the Director of National Intelligence (ODNI), Intelligence Advanced
Research Projects Activity (IARPA), via the U.S. Army Research Office
contract W911NF-17-C-0050. The views and conclusions contained herein are
those of the authors and should not be interpreted as necessarily
representing the official policies or endorsements, either expressed or
implied, of the ODNI, IARPA, or the U.S. Government. The U.S. Government
is authorized to reproduce and distribute reprints for Governmental
purposes notwithstanding any copyright annotation thereon.
\end{acknowledgments}

\appendix

\section{Time-to-solution and optimality}  
\label{app:TTS}

We provide a derivation of the TTS expression given in Eq.~\eqref{eq:TTS} (see also, e.g., the Supplementary Materials of Ref.~\cite{speedup}).  Let us assume that the probability of observing the ground state energy in any given repetition is $p_{\mathrm{S}}(t_f)$, and we ask how many repetitions $R$ must be performed to observe the ground state energy at least once.  The probability of not observing the ground state energy once in $R$ trials is $(1 - p_S(t_f))^R$.  Therefore, to observe the ground state energy at least once with probability $p_d$ is:
\begin{equation}
1 - p_d = \left( 1 - p_S(t_f) \right)^R \ .
\end{equation}
Solving for $R$ gives the expression in Eq.~\eqref{eq:TTS} in the main text.  Technically, the number of repetitions is defined as $R(t_f) = \protect\lceil \frac{\ln (1 - p_\mathrm{d})}{\ln [1 - p_\mathrm{S}(t_f)]} \protect\rceil$, but we do not include the ceiling operation in the calculation of $R(t_f)$ in this work, since this can result in sharp TTS changes that complicate the extraction of a scaling. Similarly, the ratio $N/N_{\max}$ should be $(\protect\lfloor N_{\max}/N \protect\rfloor)^{-1}$.  The TTS should in principle also include all time-costs accrued by running the algorithm multiple times, such as state initialization and state readout times, as well as multiple programming times if different gauges are used (see Appendix~\ref{app:DW}).  We do not include these here either, because at least on the D-Wave processors, the readout and programming times are several factors larger than the annealing time and hence can effectively mask the scaling behavior. Instead, we restrict $t_f$ to be the runtime between state preparation and readout for all our algorithms.

To see how an analysis of the TTS that does not account for optimal annealing times can lead one astray, consider the following extreme example: suppose $t_f$ is too large at all problem sizes $N\in[N_{\min},N_{\max}]$, such that $R(t_f)=1$ always suffices to find the global minimum; in this case $\protect\langle \protect\TTS(t_f) \protect\rangle \protect\propto N t_f$ for all $N$ (i.e., is constant except for the parallelization factor $N$), which except for trivial problems, must obviously be false.

\section{The D-Wave quantum annealers} 
\label{app:DW}

We used the D-Wave 2000Q (DW2KQ) processor housed at Burnaby, that features $2023$ functional qubits and $5874$ programmable couplers. We also used the DW2X processor housed at USC/ISI, that features $1098$ functional qubits and $3049$ programmable couplers. The minimum annealing times for all D-Wave processors involved in benchmarking studies to date are: $5\mu$s for the D-Wave One, D-Wave Two X, and D-Wave 2000Q, and $20\mu$s for the D-Wave Two. Additional details about the processors are provided below.  For each instance, we ran $100$ random gauges (also known as spin-reversal transforms).  A gauge is the application of a particular bit-flip transformation to the $\sigma^z$ operators in the problem Hamiltonian, i.e., $H_{\mathrm{P}}' \mapsto \prod_{i=1}^N (\sigma_i^x)^{s_i} H_{\mathrm{P}}' \prod_{i=1}^N (\sigma_i^x)^{s_i}$, where each $s_i \in \left\{0,1 \right\}$.  This transformation does not change the eigenvalues of the transverse field Hamiltonian, and it is meant to minimize the effect of local biases and precision errors on the device \cite{q-sig}.  For each gauge we took $n_{\mathrm{reads}}=1000$ readouts, unless constrained by $t_f n_{\mathrm{reads}} < 10^6 \mu$s.  For example for $t_f = 2 $ms, we only took $400$ readouts per gauge.

\begin{figure}[b] %
   \centering
\subfigure[]{\includegraphics[width=0.45\columnwidth]{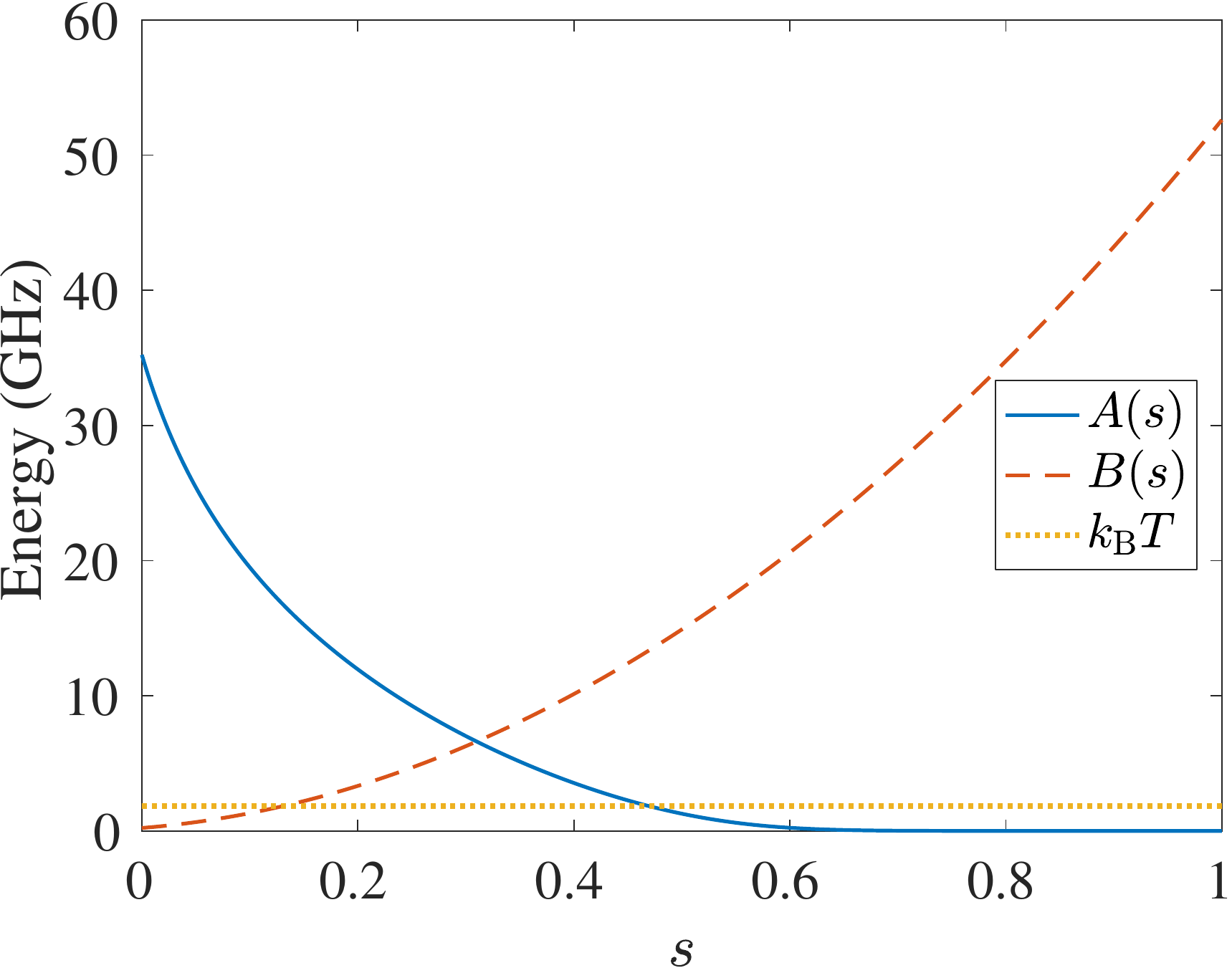}\label{fig:DW2000Qannealing}}
\subfigure[]{\includegraphics[width=0.4575\columnwidth]{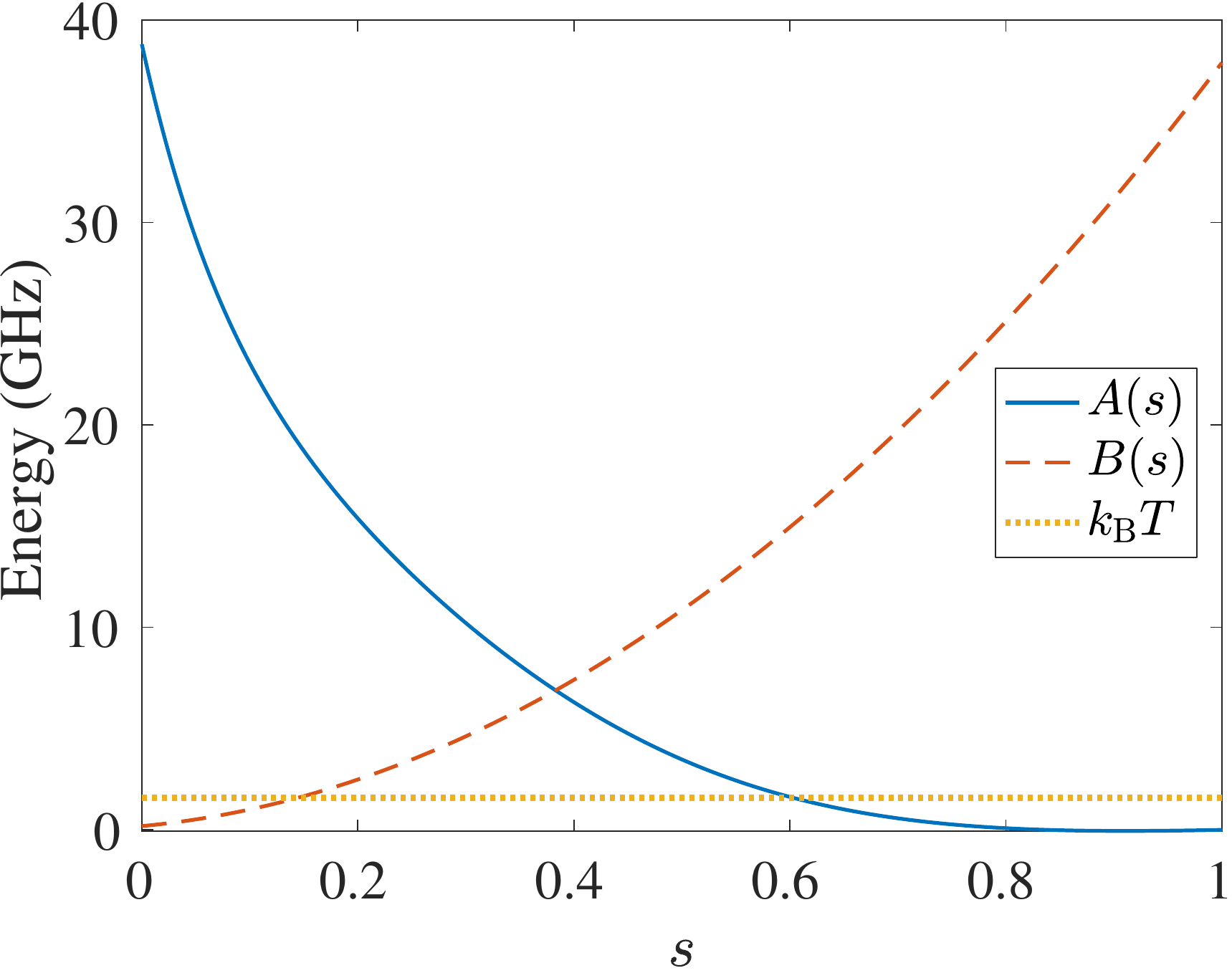}\label{fig:DW2Xannealing}}
\hspace{1cm}
   \caption{\textbf{Annealing schedules for (a) the DW2KQ and (b) the DW2X.}  The units are such that $\hbar = 1$.  As a reference, we include the operating temperatures of the devices, corresponding to $14.1$mK for the DW2KQ and $12.5$mK for the DW2X. Note the different vertical axis scales.}
   \label{fig:AnnealingSchedules}
\end{figure}

\begin{figure*}[t] %
   \centering
\subfigure[]{\includegraphics[width=0.4575\columnwidth]{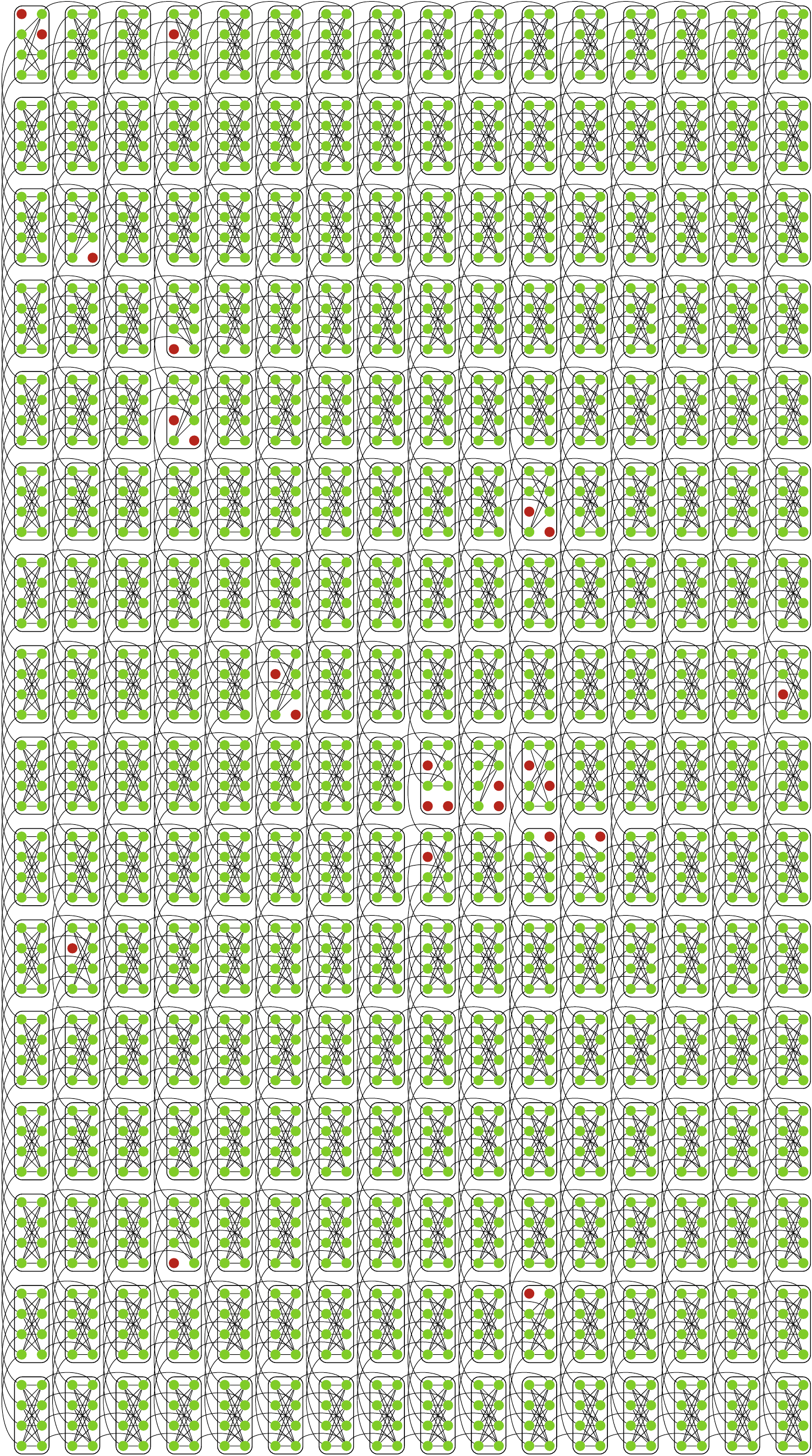}\label{fig:hardwaregraph}}
\hspace{.5cm}
\subfigure[]{\includegraphics[width=0.4575\columnwidth]{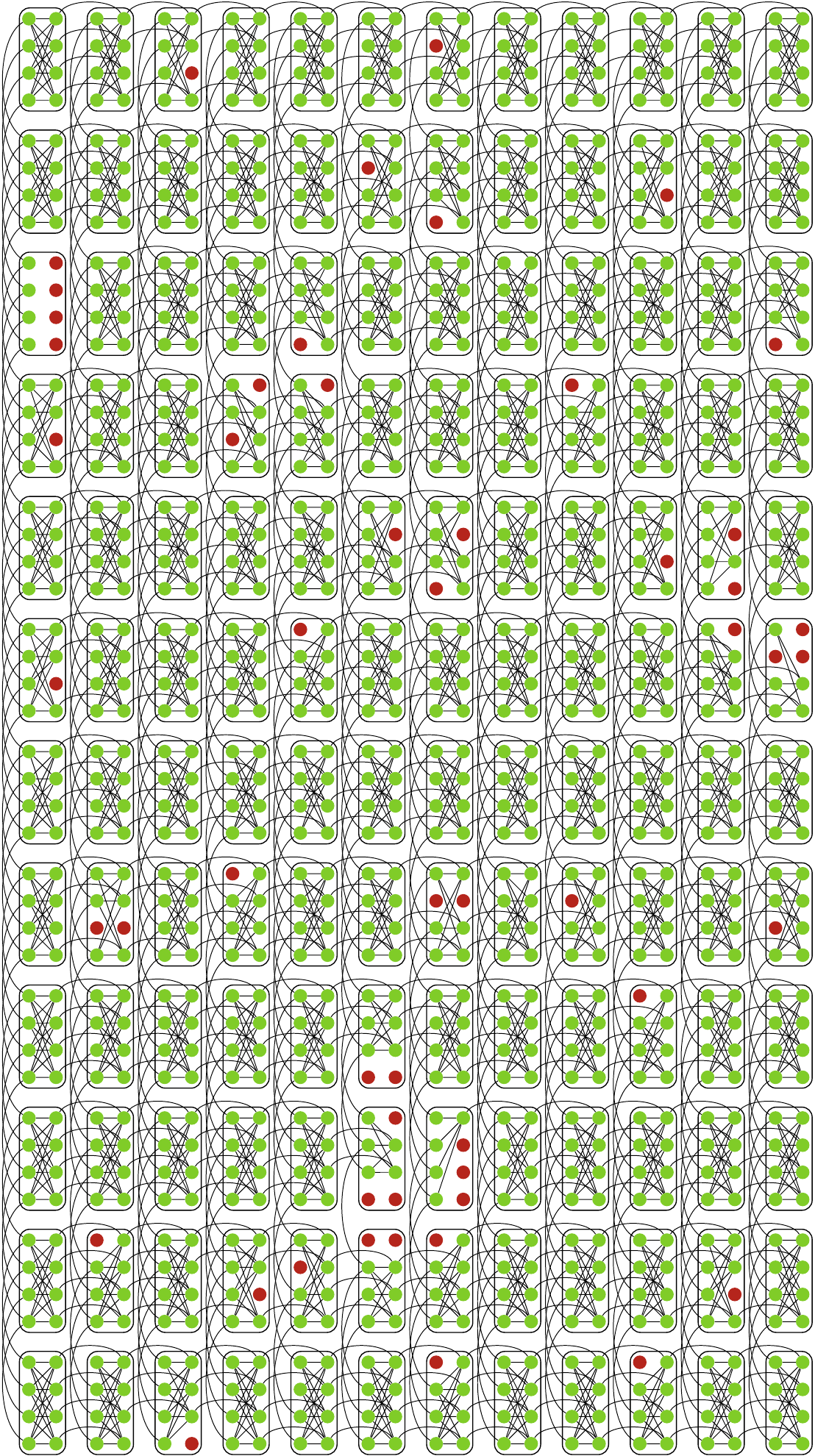}\label{fig:DW2Xhardwaregraph}}
\hspace{.5cm}
\subfigure[]{\includegraphics[width=0.45\columnwidth]{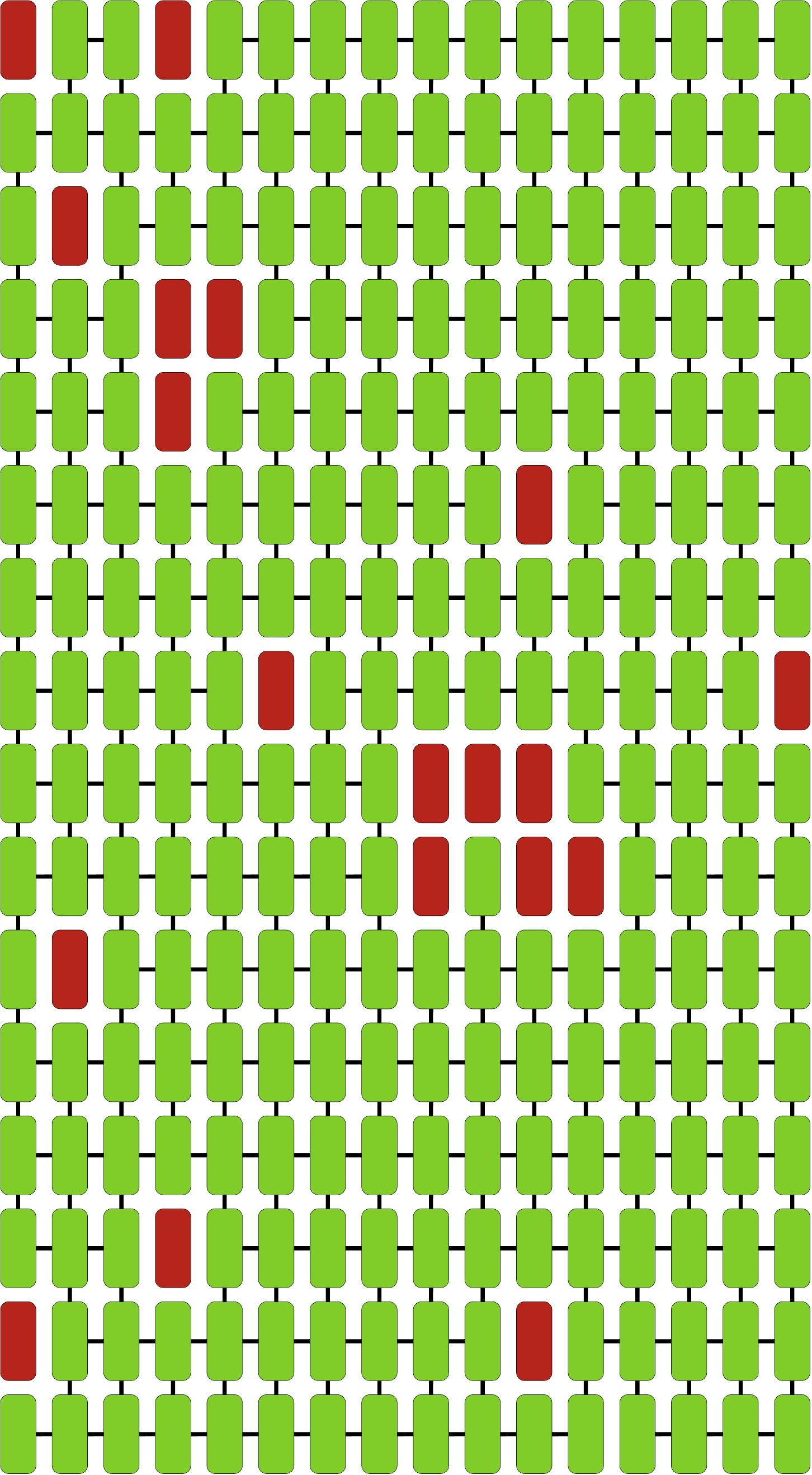}\label{fig:logicalgraph}}
\hspace{.5cm}
\subfigure[]{\includegraphics[width=0.45\columnwidth]{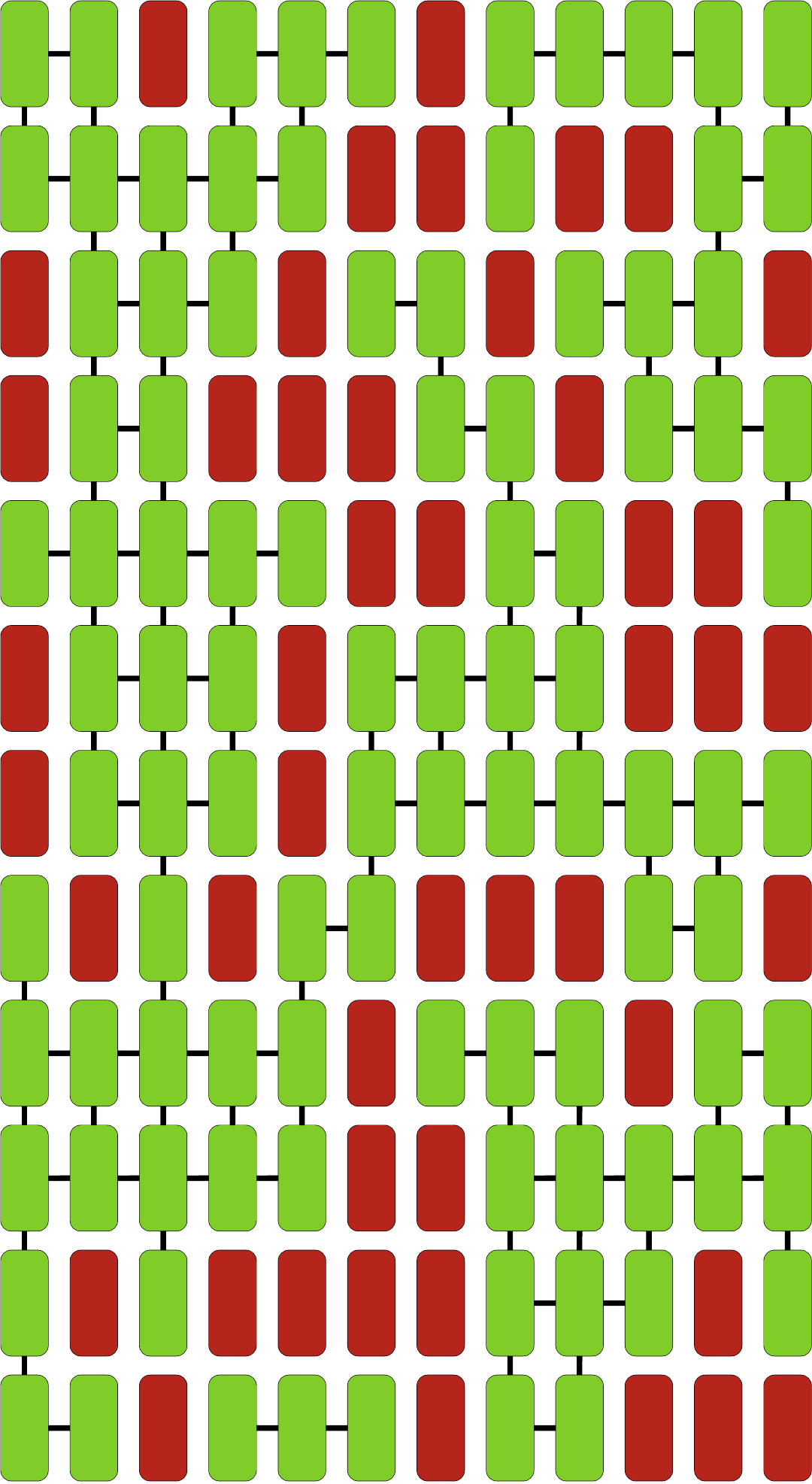}\label{fig:DW2Xlogicalgraph}} 
   \caption{\textbf{Hardware and logical graphs for instance generation.} (a) DW2KQ hardware graph, (b) DW2X hardware graph, (c) DW2KQ logical graph, (d) DW2X logical graph.  For DW2KQ (DW2X) subgraphs of size $L\leq 16$ ($L\leq 12$) were chosen starting from the lower right corner.  (a) and (b):  Available qubits are shown in green, and unavailable qubits are shown in red.  Programmable couplers are shown as black lines connecting qubits. (c) and (d) Complete unit cells are shown in green, and incomplete ones are shown in red.  Logical couplers are shown as black lines between the unit cells. The unit cells are numbered from $0$, starting from the top left corner and moving across rows to the right.}
\label{fig:LargeInstances}
\end{figure*}

The annealing schedules of the D-Wave 2000Q (DW2KQ) processor housed at Burnaby and the DW2X processor housed at USC/ISI devices are shown in Fig.~\ref{fig:AnnealingSchedules}, in units of GHz. These schedules are not measured but computed and reported by D-Wave Systems Inc. based on their flux qubit models.  
  The Chimera hardware graphs of the DW2KQ and DW2X processors we used in this work  are shown in Fig.~\ref{fig:LargeInstances}.

\begin{figure}[t] 
   \centering
   \includegraphics[width=0.8\columnwidth]{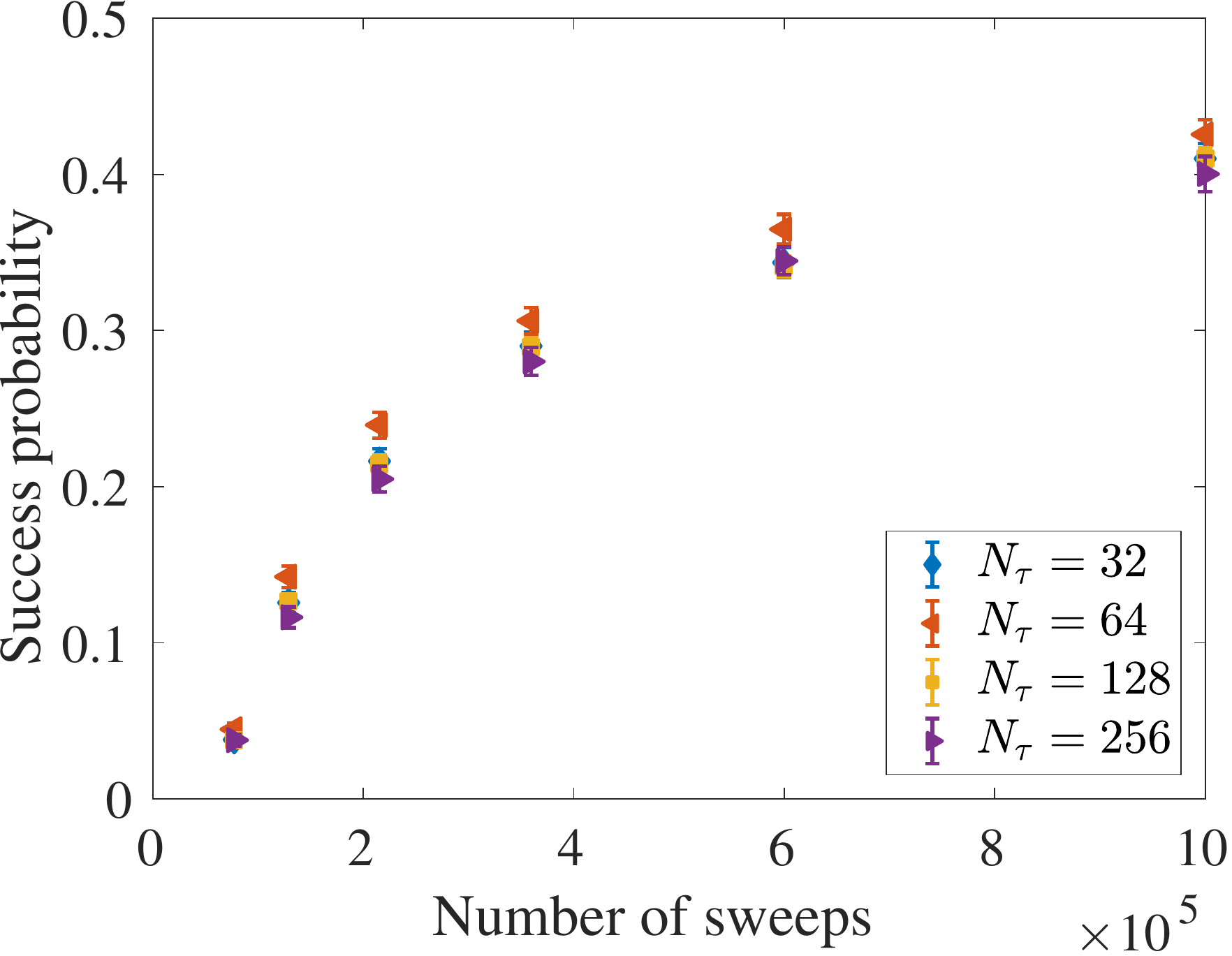} 
   \caption{{\bf Performance of SQA as we vary the number of Trotter slices.} Shown are CPU simulation results for the success probability for a single logical-planted instance, using $32$, $64$, $128$, and $256$ Trotter slices.  The success probability is maximized for $64$ slices. Error bars give the $95\%$ confidence interval generated by performing $1000$ bootstraps.}
   \label{fig:SQATrotter}
\end{figure}
\begin{figure}[t] 
   \centering
 \includegraphics[width=0.8\columnwidth]{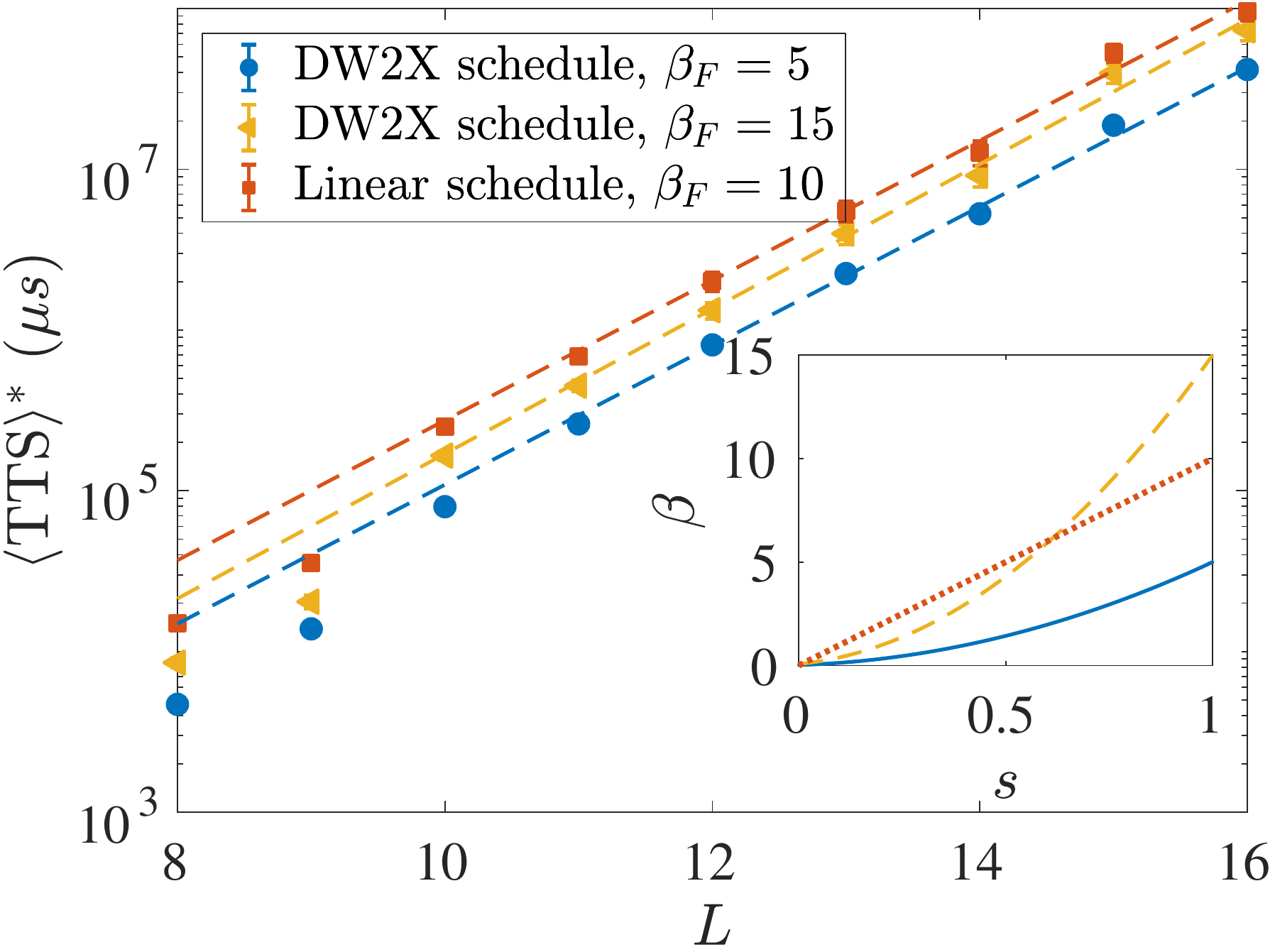} 
   \caption{{\bf Median scaling for SA on the logical-planted instances with three different annealing schedules.}  We compare the median TTS for SA using three different annealing schedules, $0.132 B(s)$ where $B(s)$ is from the DW2X annealing schedule in Fig.~\ref{fig:DW2Xannealing}, $0.396 B(s)$, and a linear schedule.  The dashed lines correspond to the exponential fits $\exp (a + b L)$ with $ a= 12.457 \pm 0.332, b = 0.996 \pm 0.24$ and $ a= 12.489  \pm 0.888 , b = 1.037 \pm 0.064$ for the DW2X schedules and $ a= 13.321 \pm 0.934, b = 1.002 \pm 0.066$ for the linear schedule.  Inset: the annealing schedules in the inverse-temperature $\beta$ as a function of the dimensionless parameter $s$.}
   \label{fig:SAschedule}
   \end{figure}
\begin{table}[t]
\centering
\begin{tabular}{c c c c c c}
\hline\hline
Solver & $a$ & $b$ & $c$  \\ [0.5ex] 
\hline 
DW2KQ &$ -6.953 \pm 1.442$  &    $ 5.017 \pm 1.020$ & $ 0.380\pm 0.090$\\ 
SA &    $ 0.493 \pm 1.102  $ & $9.521 \pm  0.764 $&$-0.193 \pm 0.065  $ \\
SQA & $ 15.893 \pm 5.396 $  & $0.248 \pm 3.933 $ &   $0.508 \pm 0.360 $ \\
SVMC & $3.050 \pm 1.814$ &  $ 9.316 \pm1.241$ & $-0.075 \pm 0.104$ \\
[1ex]
\hline\hline
\end{tabular}
\caption{The coefficient $(a, b, c)$ in fits of $\ln \TTSopt\ $ to $a + b \ln L + c L$ for the hardware-planted instances using $L \in [8,16]$.  Errors are $95\%$ confidence intervals.}
\label{table:scalingFitsHardwareExpPoly}
\end{table}

\begin{figure}[t] 
   \centering
   \includegraphics[width=0.8\columnwidth]{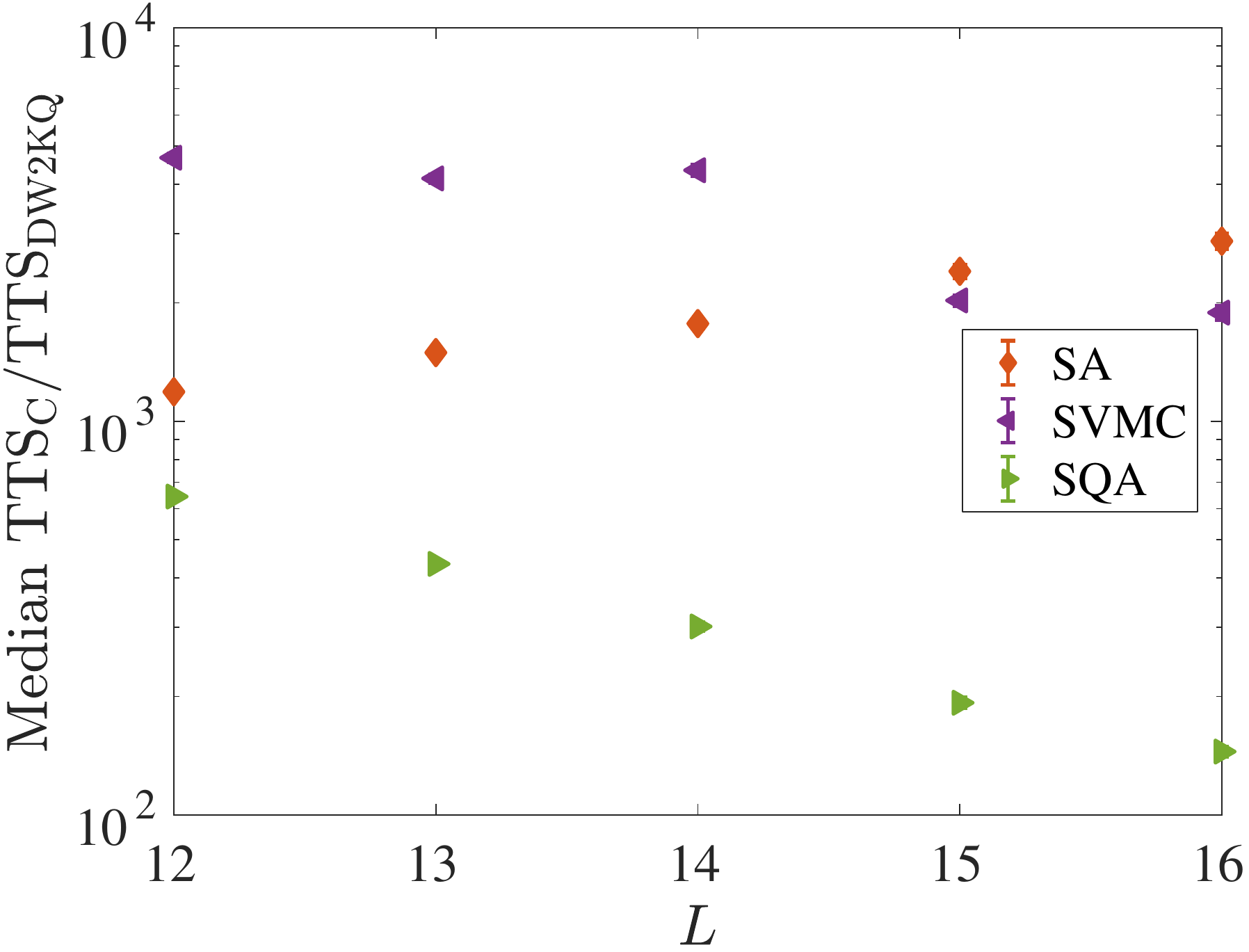} 
   \caption{{\bf Median quantile-of-ratios for the logical-planted instances.}  We show the ratio of the different classical solvers $\mathrm{C} = \left\{\mathrm{SA},\mathrm{SQA},\mathrm{SVMC} \right\}$ to the DW2KQ.  A positive slope, as for SA and SVMC, indicates a scaling advantage for the DW2KQ, while a negative slope, as for SQA, indicates a slowdown.  The data symbols obscure the error bars representing the $95\%$ confidence intervals ($2\sigma$) calculated using $1000$ bootstraps of $1000$ instances.}
   \label{fig:QofR}
\end{figure}
 \begin{figure}[t] %
   \centering
\subfigure[\ DW2X]{\includegraphics[width=0.8\columnwidth]{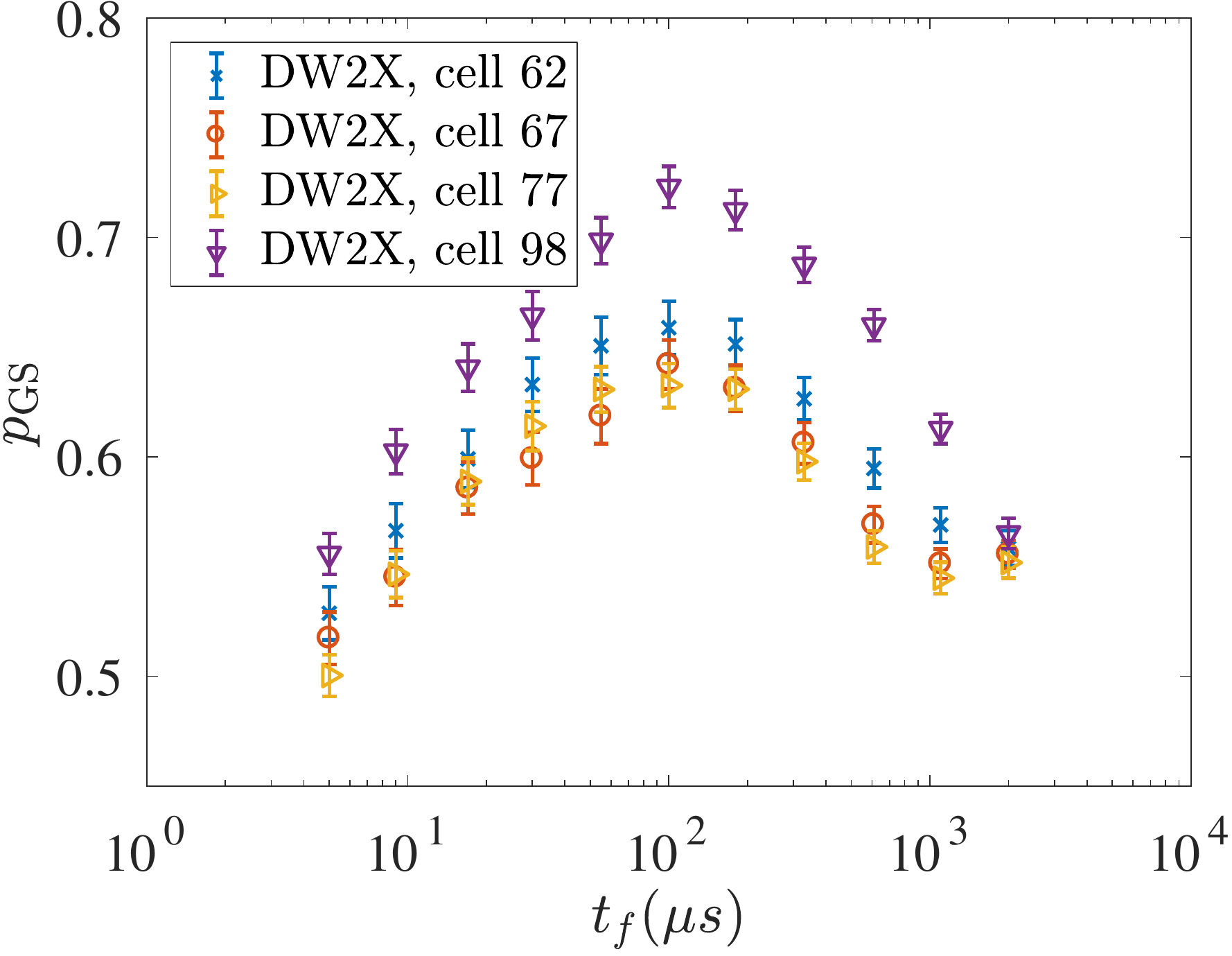} \label{fig:gadgetDW2X}} 
\subfigure[\ DW2KQ]{\includegraphics[width=0.8\columnwidth]{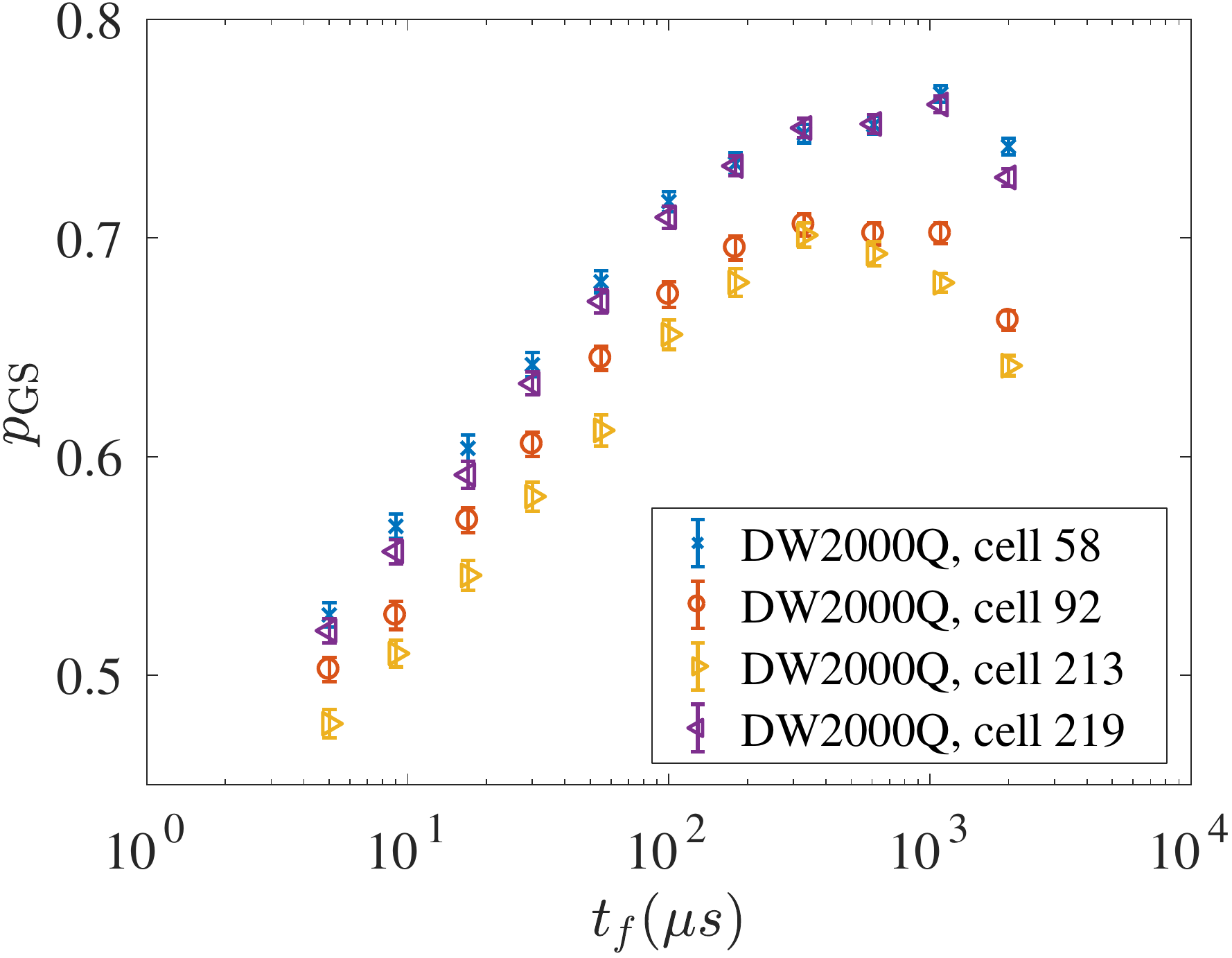} \label{fig:gadgetDW2KQ}}
   \caption{\textbf{Results for the $8$-qubit gadget on the DW2KQ and DW2X.} (a) The gadget on four different unit cells of the DW2X.  (b) The gadget on four different unit cells of the DW2KQ.  On both devices we used $1000$ gauges with $1000$ anneals per gauge.  Error bars on the data points are $2\sigma$ calculated using $1000$ bootstraps of each gauge.}
\label{fig:gadget}
\end{figure}
\begin{figure*}[t] %
   \centering
\subfigure[]{\includegraphics[width=0.66\columnwidth]{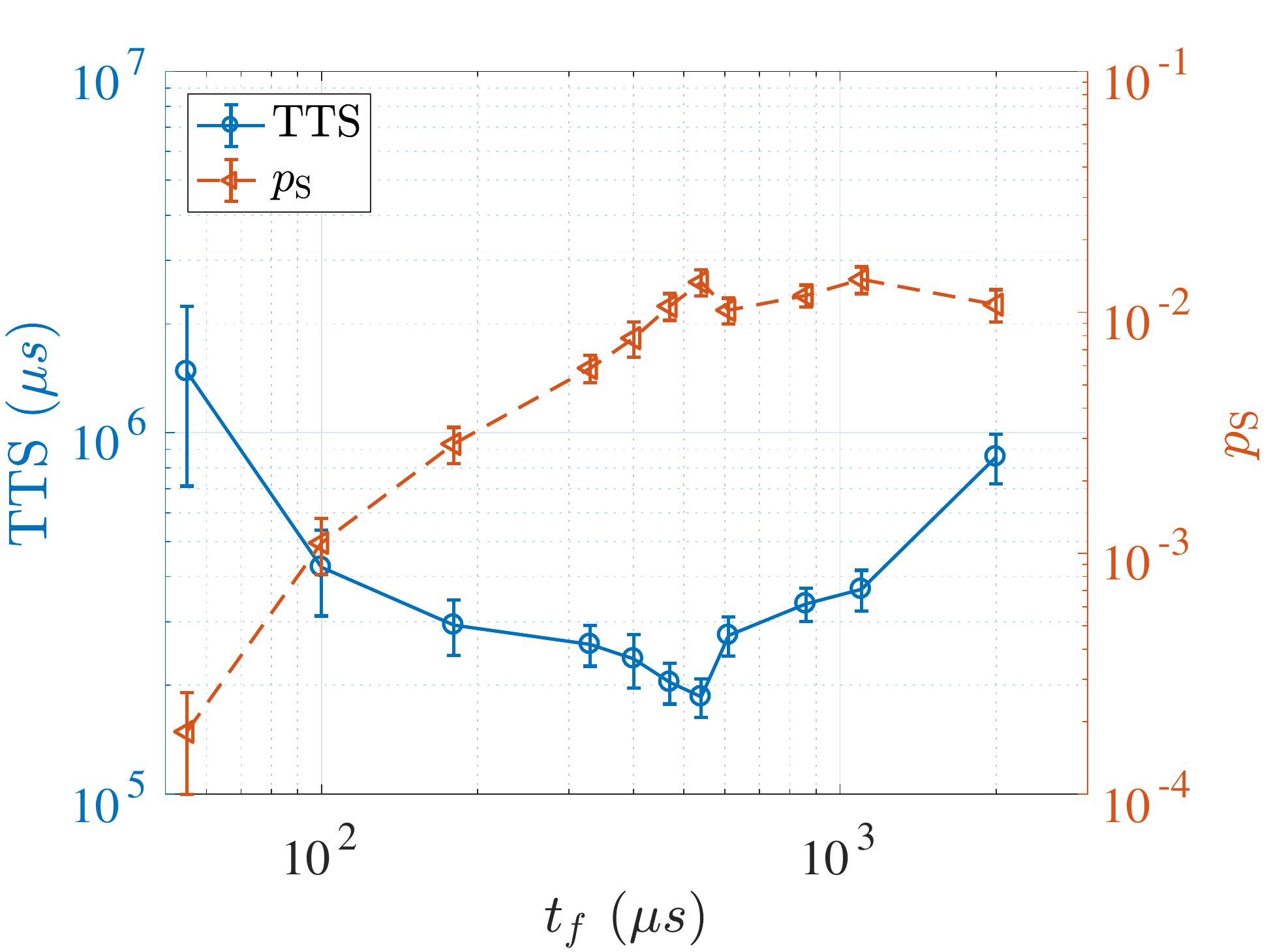} \label{fig:ExampleOptimalAnnealingTime2}} 
\subfigure[]{\includegraphics[width=0.66\columnwidth]{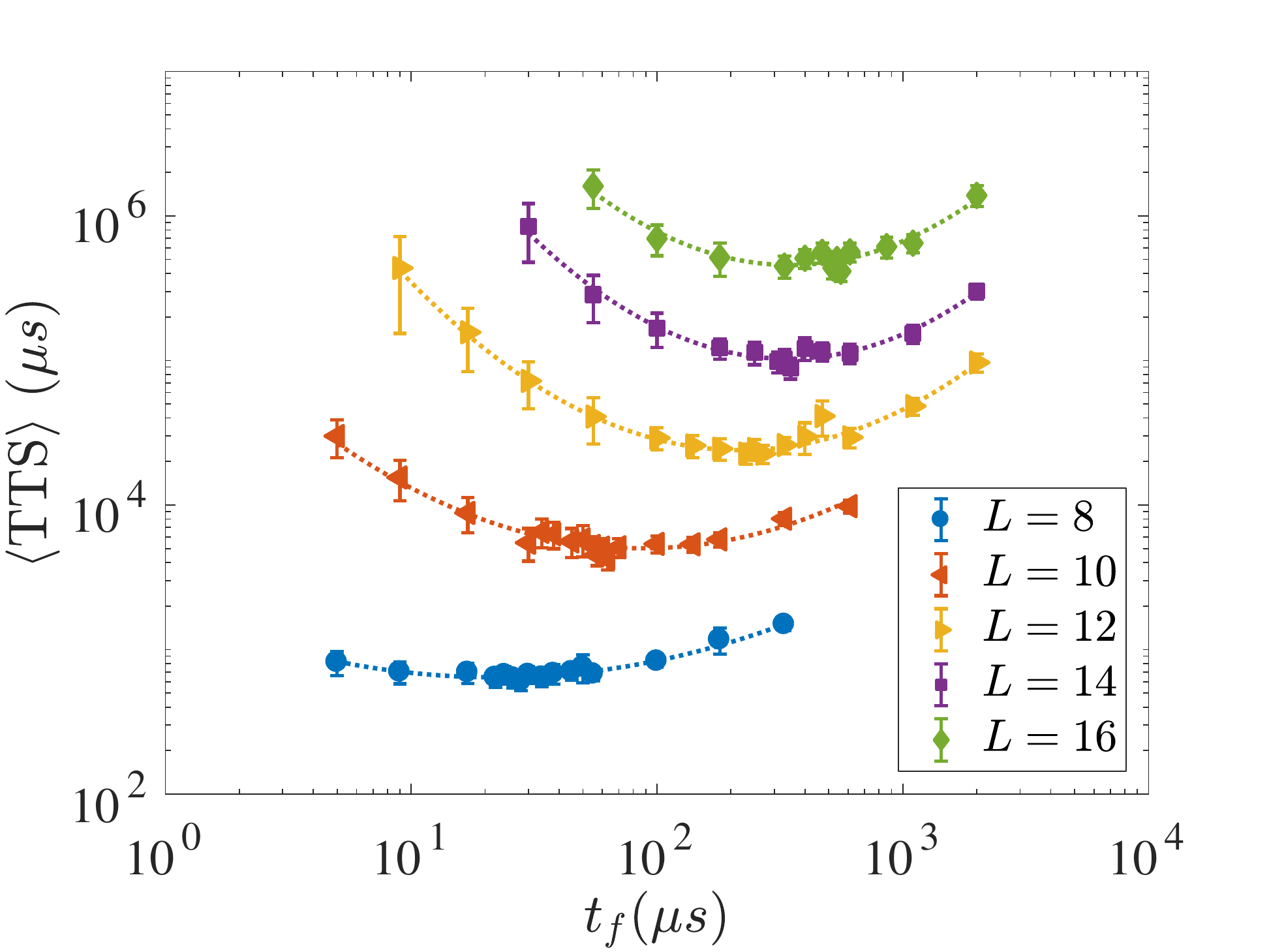} \label{fig:GroupTTS2}}
\subfigure[] {\includegraphics[width=0.66\columnwidth]{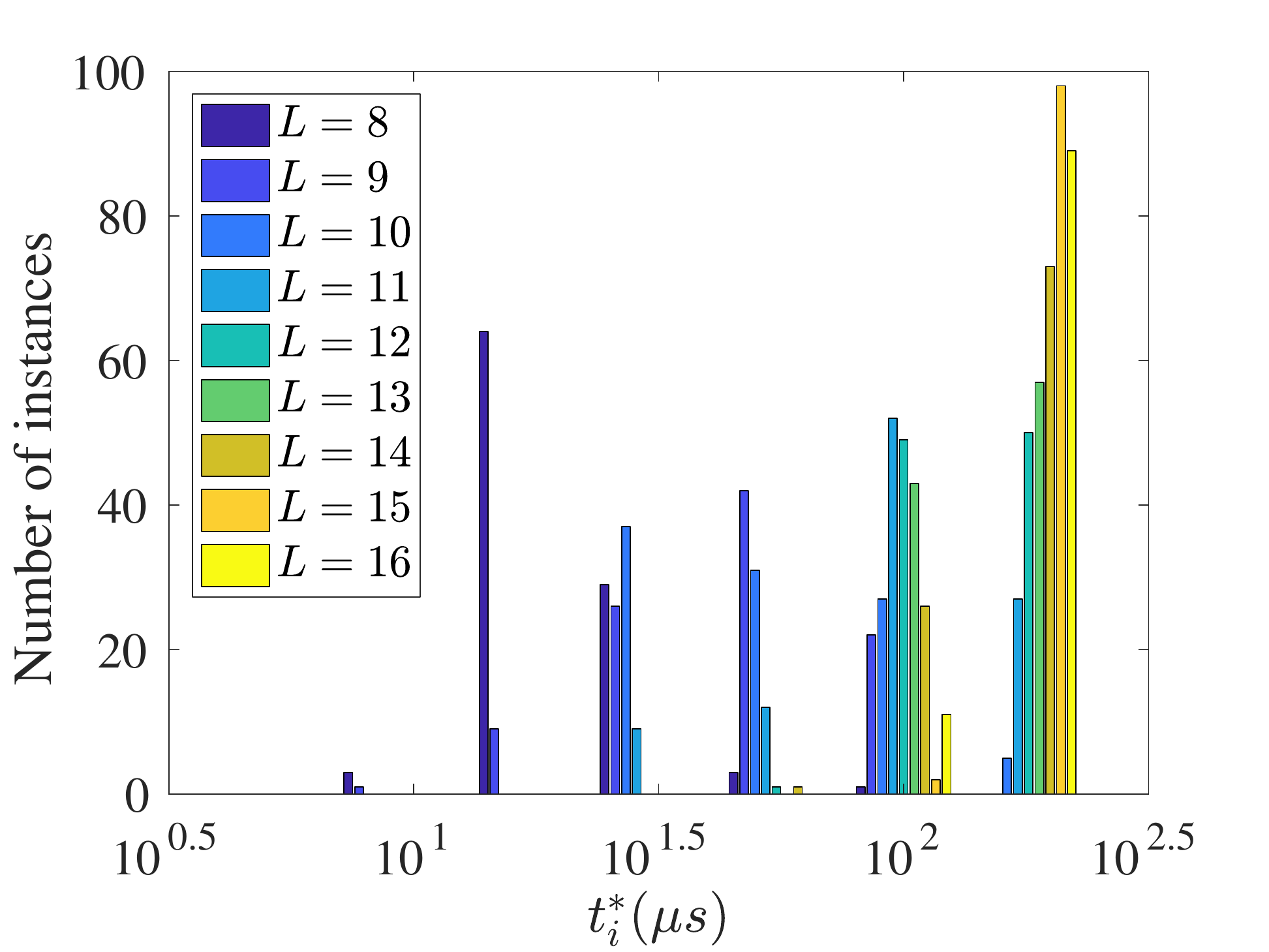} \label{fig:HardwareDistribution}}
\caption{\textbf{Optimal annealing time and optimal TTS for the `hardware-planted' instance class on the DW2KQ.} 
(a) TTS (blue solid line) and $p_{\mathrm{S}}$ (red dashed line) for a representative problem instance at size $L=16$. 
A clear minimum in the TTS is visible at $t^* \approx 550 \mu$s, thus demonstrating the existence of an optimal annealing time for this instance. 
(b) Median TTS as a function of annealing time for $L\geq 8$. Dotted curves represent best-fit quadratic curves to the data (see Appendix~\ref{app:optimumFitting} for details).  The position of $\TTSopt \ $ shifts to larger $t_f$ as the system size increases. An optimum could not be established for $L<8$ for this instance class, i.e., it appears that $t^*<5\mu$s when $L<8$. 
(c) The distribution of instance optimal annealing times for different system sizes, as inferred directly from the positions of the minima as shown in (a).  It is evident that the number of instances with higher optimal annealing times increases along with the system size, in agreement with (b).
In (a) and (b) error bars represent $95\%$ confidence intervals ($2\sigma$) calculated using $1000$ bootstraps of $100$ gauge transformations \cite{q108}.
} 
\label{fig:OptimalAnnealingTime2}
\end{figure*}

\begin{figure}[t] 
   \centering
   \includegraphics[width=0.8\columnwidth]{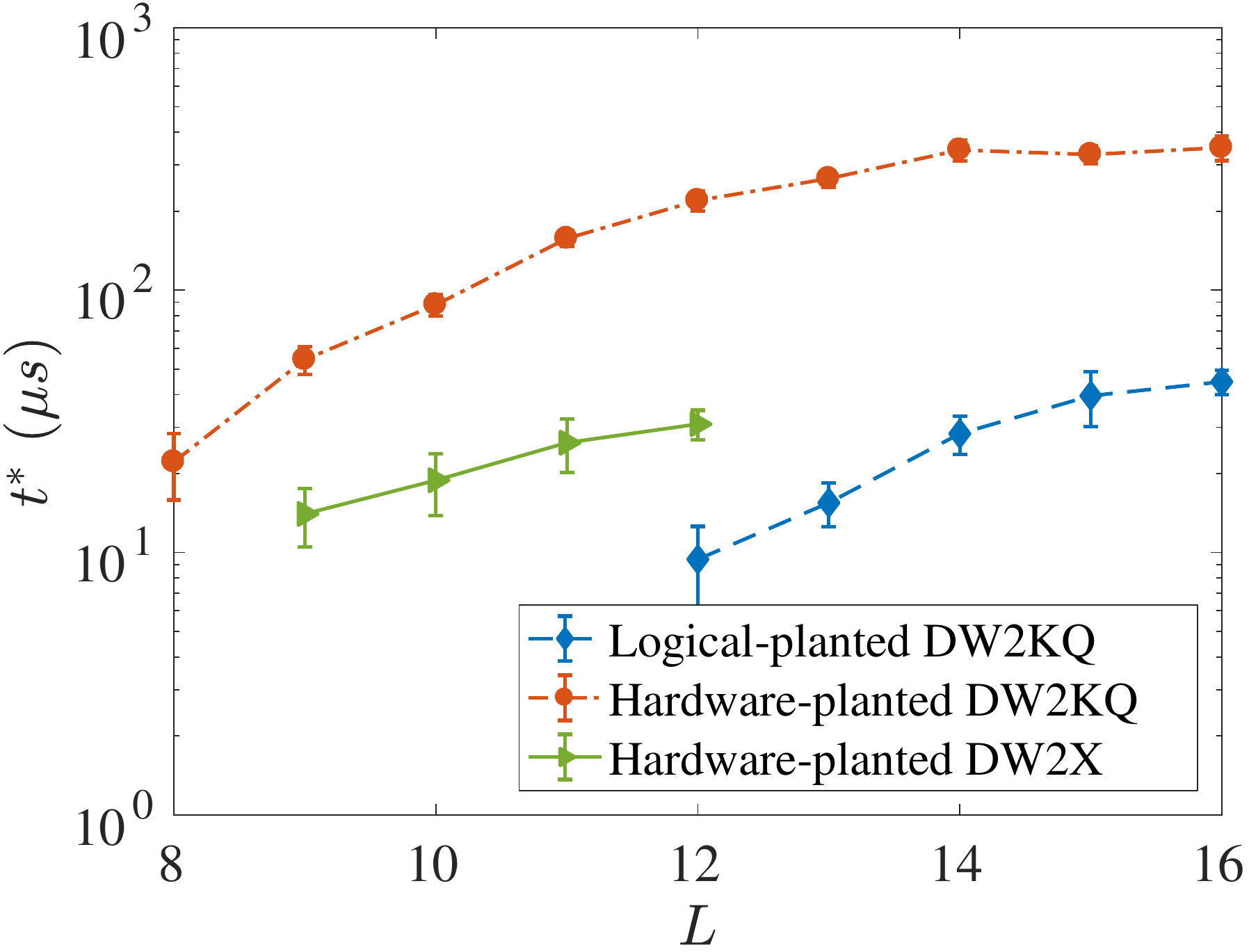} 
   \caption{{\bf Scaling of $t^\ast$ with problem size for the two problem classes.}  Error bars represent the 95\% confidence interval for the location of $t^\ast$ by fitting to a quadratic function as described in Appendix~\ref{app:optimumFitting}. 
   }
      \label{fig:toptScaling}
\end{figure}
\begin{figure*}[tbp] 
   \centering
   \subfigure[\ $q=0.25$]{\includegraphics[width=0.3\textwidth]{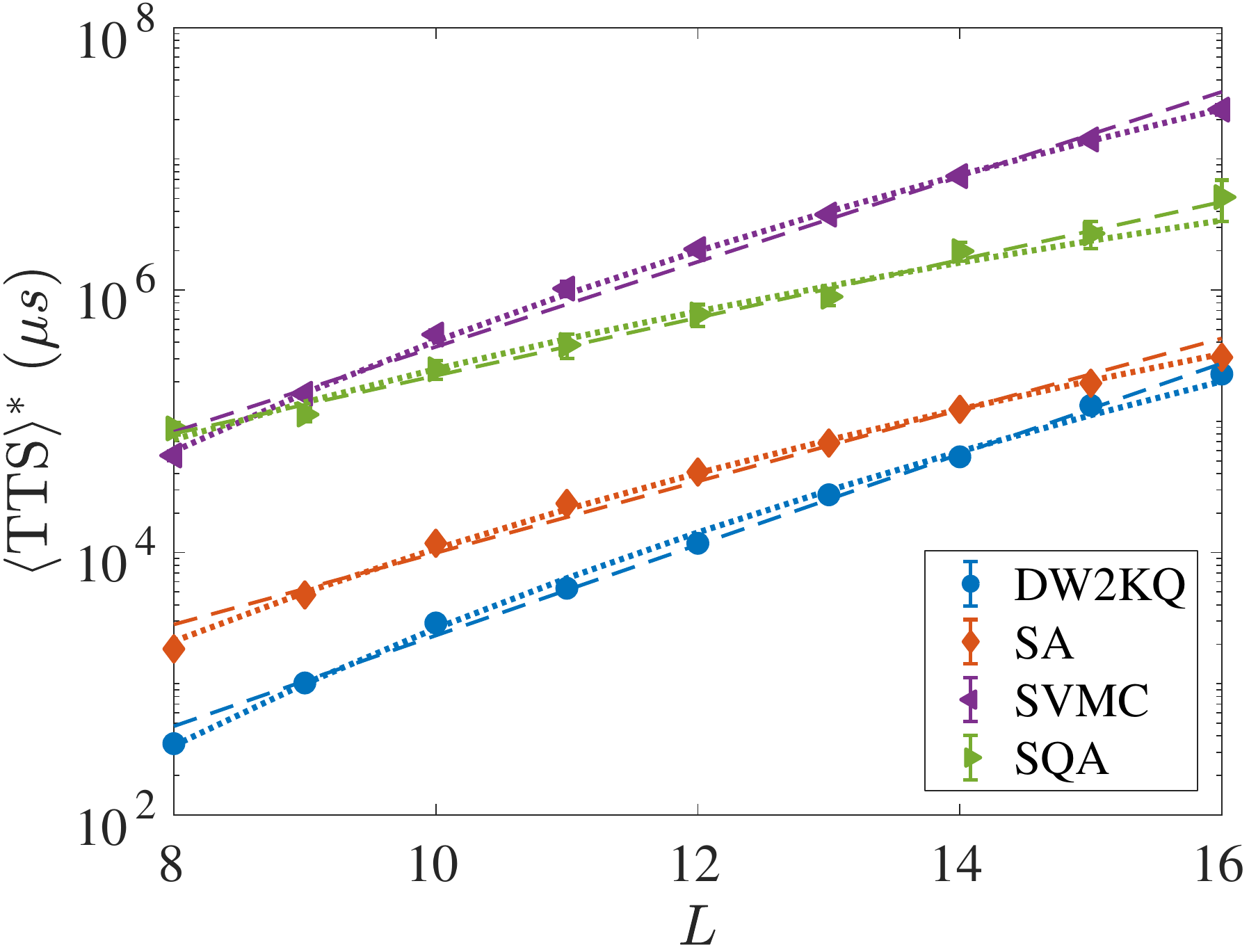} }
\subfigure[\ $q=0.50$]{\includegraphics[width=0.3\textwidth]{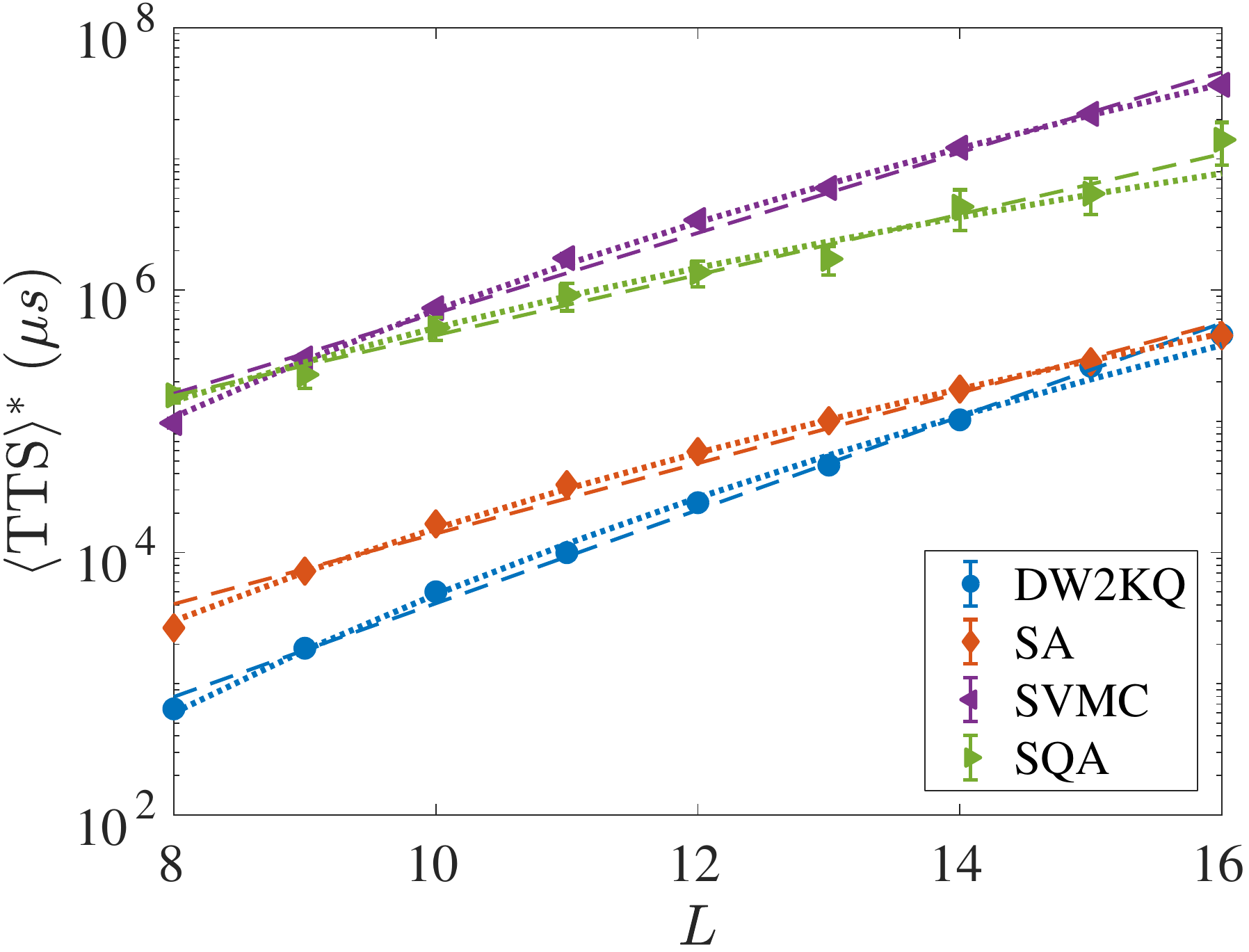} \label{fig:scalingHardware}}
   \subfigure[\ $q=0.75$]{\includegraphics[width=0.3\textwidth]{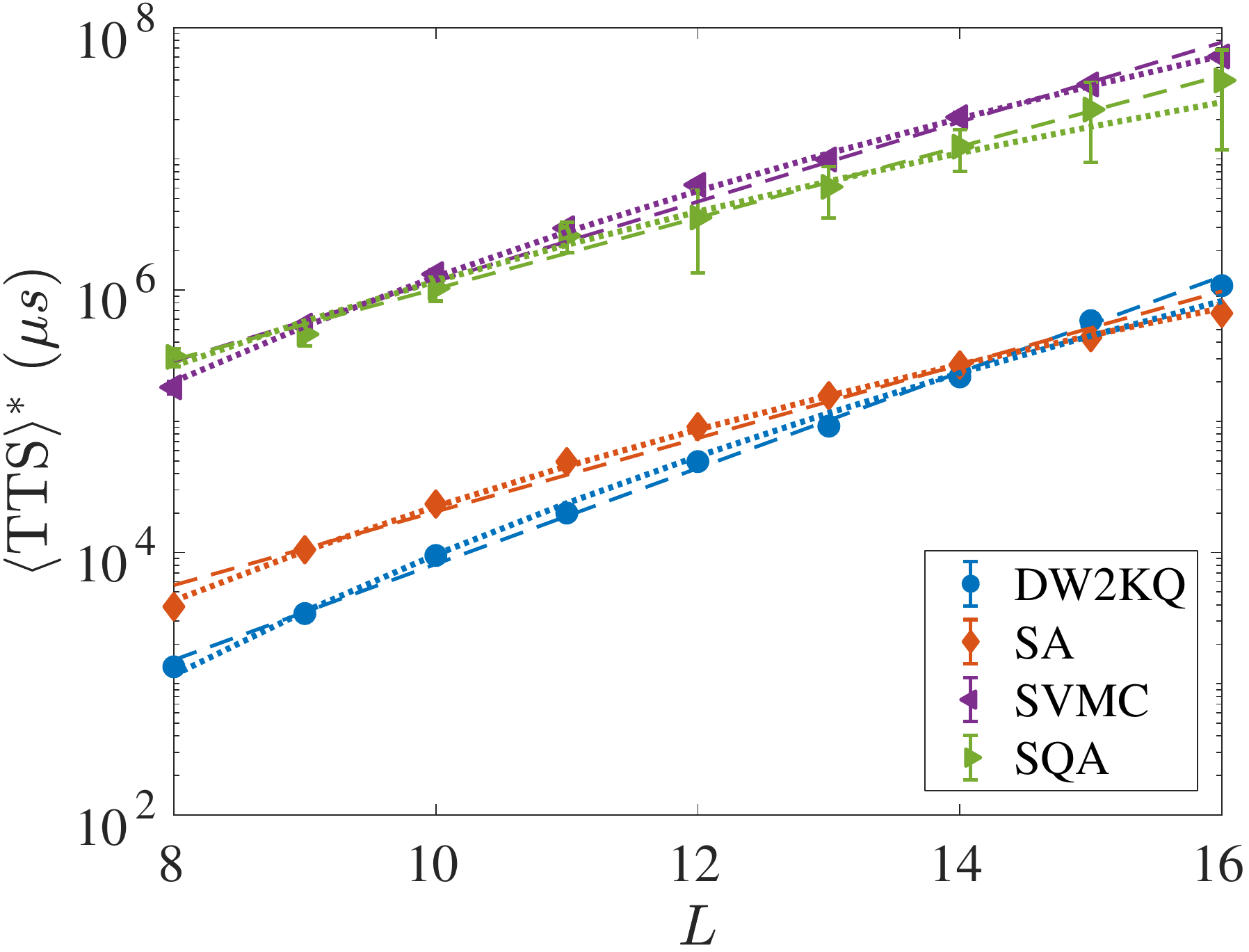} }
\caption{\textbf{Scaling of the optimal TTS with problem size for hardware-planted instances.} The data points represent the DW2KQ (blue circles) and three classical solvers, SA (red diamonds), SVMC (purple left triangle), and SQA (green right triangle). The dashed and dotted curves correspond, respectively, to exponential and polynomial best fits with parameters given in Table~\ref{table:scalingFitsHardware}. Panels (a), (b) and (c) correspond to the $25$th quantile, median, and $75$th quantile, respectively. Simulation parameters for SA, SVMC, and SQA are given in Appendix~\ref{app:SA_SVMC}. The data symbols obscure the error bars, representing the 95\% confidence interval for each optimal TTS data point (computed from the fit of $\ln\langle \mathrm{TTS} \rangle$  to a quadratic function as explained in Appendix~\ref{app:optimumFitting}).} 
   \label{fig:TTSPRC}
\end{figure*}

\begin{table}[t]
\centering
\begin{tabular}{c c c c c c}
\hline\hline
(a) Solver & $q=0.75$ & $q=0.50$ & $q=0.25$  \\ [0.5ex] 
\hline 
DW2KQ &$0.842 \pm 0.01$  &    $0.820 \pm 0.009$ & $0.796 \pm 0.009$\\ 
SA &    $0.645 \pm  0.008 $ & $0.617 \pm  0.006 $ & $0.628 \pm  0.007 $ \\
SQA & $0.594 \pm 0.029$  & $0.510 \pm 0.018$ &   $0.487 \pm 0.015$ \\
SVMC & $0.699 \pm 0.012$ &  $0.705 \pm 0.010$ & $0.746 \pm 0.012$ \\
HFS & & $ 0.796 \pm 0.008$ &  &  \\
SAC & & $0.420 \pm 0.015$ &  & \\
[1ex]
\hline\hline
(b) Solver & $q=0.75$ & $q=0.50$ & $q=0.25$  \\ [0.5ex] 
\hline
DW2KQ &  $9.470 \pm 0.111$ & $9.310 \pm 0.097$&  $9.228 \pm 0.105$  \\ 
SA &   $7.405 \pm  0.089 $ & $7.275 \pm  0.071 $&  $7.277 \pm  0.085 $ \\
SQA & $6.461 \pm 0.315$ &  $5.703 \pm 0.199$ &  $5.383 \pm 0.164$ \\
SVMC &  $8.258 \pm 0.143$ & $8.431 \pm 0.119$   & $8.669 \pm 0.133$ \\
HFS & & $ 9.231 \pm 0.029 $ &  &  \\
SAC & & $5.001 \pm 0.177$ &  & \\
\hline\hline
(c) Solver & $q=0.75$ & $q=0.50$ & $q=0.25$  \\ [0.5ex] 
\hline 
DW2KQ &$0.59 \pm 0.12$  &    $0.12 \pm 0.10$ & $-0.21 \pm 0.11$\\ 
SA &    $14.30 \pm 0.09 $ & $14.19 \pm 0.08 $ & $13.75 \pm 0.09 $ \\
SQA & $16.48 \pm 0.30$  & $16.44 \pm 0.20$ &   $15.97 \pm 0.16$ \\
SVMC & $17.26 \pm 0.15$ &  $16.63 \pm 0.13$ & $15.64 \pm 0.15$ \\
HFS & & $ 4.29 \pm 0.12$ &  &  \\
SAC & & $12.16 \pm 0.18$ &  & \\
[1ex]
\hline\hline
(d) Solver & $q=0.75$ & $q=0.50$ & $q=0.25$  \\ [0.5ex] 
\hline
DW2KQ &  $-12.63 \pm 0.26$ & $-12.96 \pm 0.23$&  $-13.37 \pm 0.25$  \\ 
SA &   $3.79 \pm 0.22 $ & $3.70 \pm 0.18 $&  $3.34 \pm 0.21 $ \\
SQA & $7.65 \pm 0.73$ &  $8.51 \pm 0.47$ &  $8.56 \pm 0.39$ \\
SVMC &  $5.31 \pm 0.36$ & $4.33 \pm 0.30$   & $3.24 \pm 0.33$ \\
HFS & & $ -8.85 \pm 0.07 $ &  &  \\
SAC & & $4.88 \pm 0.43$ &  & \\
[1ex]
\hline\hline
\end{tabular}
\caption{The coefficient $b$ in fits to (a) $\exp( a + b L)$ and (b) $\exp(a) L^b$  and the coefficient $a$ in fits (c) $\exp(a + b L)$ and (d) $\exp(a) L^b$ for the hardware-planted instances using $L \in [8,16]$; $q$ denotes the quantile. Errors are $95\%$ confidence intervals.}
\label{table:scalingFitsHardware}
\end{table}

\begin{table}[ht]
\centering
\begin{tabular}{c c c c c c}
\hline\hline
(a) Solver & $q=0.75$ & $q=0.50$ & $q=0.25$  \\ [0.5ex] 
\hline 
DW2KQ &$0.864 \pm  0.028$ &   $0.760 \pm  0.017$ & $0.701 \pm  0.014$ \\ 
SA &   $1.064 \pm 0.031$ & $0.996 \pm 0.024$ & $0.961 \pm 0.023$\\
SVMC & $0.773 \pm 0.060$ &  $0.500 \pm 0.029$ &$0.441 \pm 0.020$ \\
SQA & $0.450 \pm  0.050$ & $0.365 \pm 0.035 $ & $0.331 \pm 0.024$ \\
HFS & & $0.678 \pm 0.013$ &  &  \\
SAC & & $0.510 \pm 0.018$ &  & \\
[1ex]
\hline
\hline
(b) Solver & $q=0.75$ & $q=0.50$ & $q=0.25$  \\ [0.5ex] 
\hline
DW2KQ &   $11.962 \pm  0.391$ & $10.573 \pm  0.242$ &$9.746 \pm  0.201$  \\ 
SA &   $14.635 \pm 0.433$ & $13.746 \pm 0.331$ & $13.299 \pm 0.316$ \\
SVMC &  $10.735 \pm 0.834$ & $6.890 \pm 0.399$  &  $6.141 \pm 0.273$\\
SQA & $6.221 \pm 0.697$ & $5.047 \pm 0.484 $ & $4.561 \pm 0.331 $ \\
HFS & & $9.455 \pm 0.183$ &  &  \\
SAC & & $7.134 \pm 0.259$ &  & \\
\hline\hline
(c) Solver & $q=0.75$ & $q=0.50$ & $q=0.25$  \\ [0.5ex] 
\hline 
DW2KQ &$-2.85 \pm  0.40$ &   $-2.54 \pm  0.25$ & $-2.53 \pm  0.21$ \\ 
SA &   $12.42 \pm 0.44$ & $12.46 \pm 0.33$ & $12.18 \pm 0.32$\\
SVMC & $16.88 \pm 0.84$ &  $19.26 \pm 0.39$ &$19.46 \pm 0.27$ \\
SQA & $16.60 \pm  0.69$ & $17.13 \pm 0.48 $ & $17.12 \pm 0.33$ \\
HFS & & $4.89 \pm 0.19$ &  &  \\
SAC & & $10.39 \pm 0.26$ &  & \\
[1ex]
\hline
\hline
(d) Solver & $q=0.75$ & $q=0.50$ & $q=0.25$  \\ [0.5ex] 
\hline
DW2KQ &   $-22.25 \pm  1.03$ & $-19.75 \pm  0.64$ &$-18.39 \pm  0.53$  \\ 
SA &   $-11.23 \pm 1.14$ & $-9.80 \pm 0.87$ & $-9.39 \pm 0.83$ \\
SVMC &  $-0.56 \pm 2.18$ & $8.11 \pm 1.03$  &  $9.47 \pm 0.72$\\
SQA & $6.51 \pm 1.82$ & $8.95 \pm 1.27 $ & $9.74 \pm 0.86 $ \\
HFS & & $-10.52 \pm 0.49$ &  &  \\
SAC & & $-1.23 \pm 0.68$ &  & \\
[1ex]
\hline\hline
\end{tabular}
\caption{The coefficient $b$ in fits to (a) $\exp( a + b L)$ and (b) $\exp(a) L^b$  and the coefficient $a$ in fits (c) $\exp(a + b L)$ and (d) $\exp(a) L^b$ for the logical-planted instances using $L \in [12,16]$; $q$ denotes the quantile. Errors are $95\%$ confidence intervals.  Given here are the fits for SVMC and SQA at $\beta = 2.5$ only.}
\label{table:scalingFitsHFS}
\end{table}

In the main text we focused on the annealing time.  There are several other relevant timescales that we present here for completeness.  We used the default initial state preparation time ($t_{\mathrm{initial}}$).  The readout time for the DW2KQ is $t_{\mathrm{readout}} = 124.98 \mu$s.  A complete characterization of the required runtime (the ``wall-clock time") would include the thermalization and readout times in each independent run of the quantum annealer. Furthermore, since we program the same instance multiple times using different gauges,
the programming time of $t_{\mathrm{program}} = 6987.80 \mu$s needs to be accounted for.  In total, the wall-clock TTS would be given by:
\begin{equation}
\TTS_{\mathrm{wallclock}} = G t_{\mathrm{program}} + (t_f + t_{\mathrm{initial}} + t_{\mathrm{readout}}) \frac{R}{ \lfloor \frac{N_{\mathrm{max}}}{N} \rfloor} \ ,
\label{eq:wall-clock}
\end{equation}
where $G$ is the number of gauges, and $R$ is the total number of runs, divided equally among the $G$ gauges.
However, since these timescales can be much larger than the optimal annealing time, they can mask the scaling of the TTS, and hence we focus just on the annealing time, as in previous work \cite{q108,speedup}. In principle, the initial state preparation time can be reduced and optimized along with the annealing time if included as part of the TTS, but we have not explored in this work how this impacts performance.

\section{Simulation Parameters and Timing}  
\label{app:SA_SVMC}

Our implementation of the SA, SQA, and SVMC algorithms is based on the graphics processing unit (GPU) implementation used in Ref.~\cite{DW2000Q} and described in more detail in Ref.~\cite{KingGPU}.  We briefly describe our CUDA implementation of these algorithms here for completeness. In what follows, a sweep is a single Monte Carlo update of all the spins. For all implementations, we use the default cuRAND random number generator (XORWOW).  We compile the CUDA code using the `-use\_fast\_math' flag, which, we note, may not be suitable for Monte Carlo simulations that require accurate calculations of thermal expectation values.

We first discuss our implementation of SA \cite{kirkpatrick_optimization_1983}.  Each GPU thread updates the eight spins in a single unit cell.  Because the Chimera graph is bipartite, each thread updates the four spins in one partition followed by the four spins in the second partition.  A key feature of the implementation is that the eight local fields, $16$ inter-cell couplers, and $16$ intra-cell couplers are stored in the memory registers of the GPU.  Only the spin configuration is stored on local memory.  This minimizes the cost of retrieving data from global memory.  We use the GPU intrinsic math function for the calculation of the Metropolis acceptance probability in order to maximize execution speed. 
As many copies $n_{\mathrm{copies}}$ as allowed by register memory are run in parallel in separate GPU blocks.  Therefore, we have for the timing of SA:
\begin{equation}
\TTS = \tau_{\mathrm{sweep}} n_{\mathrm{sweep}} \frac{R}{n_{\mathrm{copies}}} \ ,
\end{equation}
where $n_{\mathrm{sweep}}$ is the number of sweeps and $\tau_{\mathrm{sweep}}$ is the time required to perform a single sweep.  Because $n_{\mathrm{copies}}$ depends on the total number of threads ($L^2)$ and hence on the problem size, this can be equivalently written as:
\begin{equation} \label{eq:TimingSA}
\TTS =  8 L^2 n_{\mathrm{sweep}} R / f_{\mathrm{SA}}
\end{equation}
where $f_{\mathrm{SA}}$ is the number of total spin updates per unit time performed by the GPU.  For consistency we use the timings reported in Ref.~\cite{KingGPU} for runs performed on an NVIDIA GTX 980, which have $f_{\mathrm{SA}} = 50$ns$^{-1}$.  For SA, we use a temperature annealing schedule that is the DW2X annealing schedule for $B(s)$ (shown in Appendix~\ref{app:DW}) times $\beta = 0.132$ (in units where the maximum Ising coupling strength $|J_{ij}|$ is $1$), such that $\beta B(1) \approx 5$.

The implementation of SVMC follows the same structure as SA, except that the spin configuration is replaced by angles $\left\{\theta_i\right\} \in (0, 2\pi]$ \cite{SSSV}.  The energy potential along the anneal is given by:
\begin{equation}
V(s) = - A(s) \sum_i \sin \theta_i  + B(s) \sum_{i<j} J_{ij} \cos \theta_i \cos \theta_j\ ,
\end{equation}
a special case of Eq.~\eqref{eqt:VTIM} in the main text.
An update involves drawing a random angle $\in (0, 2\pi]$, and it is accepted according to the Metropolis-Hastings rule \cite{HASTINGS01041970,Metropolis} with $\beta = 2.5$ for the logical-planted instances and $\beta = 0.51$ for the hardware-planted instances (in units where the maximum Ising coupling strength $|J_{ij}|$ is $1$).  We use the GPU intrinsic math function for the calculation of the cosine, sine, and Metropolis acceptance probability in order to maximize the speed of the algorithm.  The timing of SVMC is the same as in Eq.~\eqref{eq:TimingSA} but with $f_{\mathrm{SVMC}} = 29$ns$^{-1}$ replacing $f_{\mathrm{SA}}$. We use the DW2X annealing schedule for $A(s)$ and $B(s)$ shown in Appendix~\ref{app:DW}; this schedule keeps $A(s)>0$ longer than that of the DW2KQ, which favors the SVMC algorithm, since once $A(s)=0$ the system becomes the classical Ising model and the most efficient updates use $\theta = 0, \pi$, but SVMC chooses angles randomly. 

The implementation of SQA follows the same structure as SA, and also uses the DW2X schedule for similar reasons as just mentioned for SVMC.  We restrict the Trotter slicing to $64$ in order to fit the spins along the imaginary-time direction into a $64$-bit word.  A sweep involves performing a single Wolff cluster update \cite{PhysRevLett.62.361} along the imaginary time direction for each spin.  Once a cluster of spins is picked, it is flipped according to the Metropolis-Hastings rule using the Ising energy of the cluster with $\beta = 2.5$ for the logical-planted instances and $\beta = 4.25$ for the hardware-planted instances.  We use the GPU intrinsic math function for the calculation of the Metropolis acceptance probability in order to maximize execution speed. 
At the end of the anneal, one of the $64$ slices is picked randomly as the final classical state.  The timing of SQA is the same as in Eq.~\eqref{eq:TimingSA} but with $f_{\mathrm{SQA}} = 5$ns$^{-1}$ replacing $f_{\mathrm{SA}}$. \\

We note that increasing the number of Trotter slices, while decreasing the Trotter error, appears to reduce the final success probability for one of the instances we have checked (see Fig.~\ref{fig:SQATrotter}, where the peak success probability occurs for $64$ slices), an effect noted in Ref.~\cite{Heim:2014jf}.  Studying this effect over the entire set of instances is computationally prohibitive at our $>2000$ qubits scale.

\section{Alternative fits} 
\label{app:AlternativeFits}
In Fig.~\ref{fig:TTSScaling} of the main text
(see also Figs.~\ref{fig:TTSPRC} and ~\ref{fig:TTSScalingHFS} below), 
we present exponential and polynomial fits to the optimal TTS as a function of $L$.  Here we show that a hybrid three-parameter fit, i.e., $\ln \mathrm{TTS} = a + b \ln L + c L$, does not give reasonable fits with good confidence bounds for all solvers.  We restrict our attention to the hardware-planted instances, since in that case we have $9$ sizes for the fit.  Table~\ref{table:scalingFitsHardwareExpPoly} gives the results of the fits for the median; we see that the estimate for the exponential scaling coefficient $c$ of the classical solvers is especially poor, likely due an insufficient number of data points for a three-parameter fit.

\section{SA with a linear schedule} 
\label{app:SA_DifferentSchedule}

We used the DW2X annealing schedule in Fig.~\ref{fig:DW2Xannealing} for the SQA, SVMC, and SA simulations. 
Further optimization of this schedule is likely to improve the overall performance of the algorithms, although it is not evident whether it will substantially change their scaling with problem size.  As an example, we provide results for the median TTS for SA using the DW2X schedule with a different overall temperature $\beta = 0.396$ and a linear schedule in Fig.~\ref{fig:SAschedule}, where we observe that the different schedules only shift the TTS curve but do not change the scaling within the statistical error bars. This indicates that our SA scaling results are robust to minor modifications of the schedule.

\section{Quantile of Ratios}
\label{app:QofR}
The benchmarking analysis we performed in the main text is akin to the `ratio-of-quantiles' comparison performed in Ref.~\cite{speedup}, where an alternative metric for speedups was also defined, called the `quantile-of-ratios'.  For this case, we find the annealing time that minimizes the TTS for each instance \emph{individually}, and the per-instance optimal TTS, denoted $\TTS^\ast_i$, is the minimal TTS for each instance individually.  For each instance, the ratio of $\TTS^\ast_i$ for two different solvers is calculated, and different quantiles over the set of ratios is taken.  We show in Fig.~\ref{fig:QofR} the results for the median ratio using the logical-planted instances. The advantage of the DW2KQ relative to SA continues to hold, and SQA continues to exhibit the best scaling. 

\section{Gadget and Instance construction} 
\label{sec:InstanceConstruction}

The key new ingredient in our instance construction is an eight qubit `gadget' that fits into the unit cell of the D-Wave processors. The gadget has a bipartite $K_{4,4}$ graph connectivity with the following Ising parameters, as also depicted in Fig.~\ref{fig:8gadget} in the main text:
\begin{eqnarray} \label{eqt:WTGParameters}
\vec{h}^T &=& \left(-1,-2/3,2/3,-1,1/3,1,-1,1 \right)  \\
J_{1,5} &=& +1, \ J_{1,6} = -1, \ J_{1,7} = -1, \ J_{1,8} = -1 \nonumber \\
J_{2,5} &=& -1 , \  J_{2,6} = -1,\  J_{2,7} = +1, \ J_{2,8} = -1 \nonumber \\
J_{3,5} &=& -1,\  J_{3,6} = -1, \ J_{3,7} = -1, \ J_{3,8} = -1 \nonumber \\
J_{4,5} &=& -1,\  J_{4,6} = -1, \ J_{4,7} = -1, \ J_{4,8} = -1 \nonumber \ .
\end{eqnarray}

The logical-planted class of instances involves constructing planted-solution instances on the logical graph of the DW2KQ.  The construction of the planted instance is similar to that of Ref.~\cite{DW2000Q}.  We define the logical graph of the DW2KQ as being comprised of vertices corresponding to only the complete unit cells (with no faulty qubits or couplers).  We also included one unit cell that was missing a single intra-cell coupler (unit cell $251$), since having this missing coupler does not change the analysis.  This is a minor difference relative to Ref.~\cite{DW2000Q}, where only complete unit cells were used.  The edges of the logical graph correspond to having all four inter-cell couplers. We did remove the logical edge between unit cells $251$ and $252$.  On an ideal Chimera graph, this would form a square grid.  We constructed an Ising Hamiltonian as a sum of $\lfloor \alpha L^2 \rfloor$ frustrated loops, where 
we picked $\alpha = 0.65$.  We again chose to plant the all-zero state.  We constructed loops as follows.  Choose a random vertex on the graph as the starting vertex, and randomly pick an available edge.  If that edge does not already have $|J|=3$,
add the vertex connecting it to the chain until a loop is formed.  Continue until the chain forms a loop by hitting a member of the chain.  Only the loop and not the tail is kept.  Accept the loop if it includes more than $4$ vertices; this 
means that the minimum loop has $6$ vertices.  This then generates a planted-solution instance on the logical graph.  In order to embed it on the hardware graph, turn on all the available couplings in the unit cell to be ferromagnetic with $J=-3$.
In the notation of Ref.~\cite{DW2000Q}, this amounts to constructing instances with $R = \rho = 3$.  

Finally, we randomly placed our gadget into a fraction $p = 0.1$ of all the connected unit cells in the planted-solution instance, and added these terms to the Ising Hamiltonian. (The gadget on unit cell $251$ has the same ground state even with the one missing coupling.)   The final Hamiltonian now has a maximum range of $6$, i.e., $|J_{ij}|\leq 6$ for all couplers.  Again, the ground state of the final Hamiltonian remains the all-zero state.\\

\subsection{The eight qubit gadget}
\label{sec:Case0989_Changed6}
%
The key ingredient in our study is an eight qubit `gadget' that fits into the unit cell of the D-Wave processors.  Figure~\ref{fig:gadget} compares the results for the $8$-qubit gadget on the two D-Wave processor generations, for different representative unit cells.  The success probability exhibits a single maximum, with the peaks occurring at different annealing times on the two devices.  While there is some variation in the magnitude of the success probability depending on which unit cell is used, the position of the peak remains robust. We note however that the position of the peak differs on the two devices (around $100 \mu s$ on the DW2X and around $300 \mu s$ on the DW2KQ), indicating that the physical characteristics of the two devices are different beyond simply having different connectivity graphs.

\subsection{Hardware-planted instances}\label{app:hardware-instances}

Here we describe a class of instances we call ``hardware-planted" (not discussed in the main text), that also exhibits an optimal annealing time within the accessible range of the DW2KQ, as we demonstrate below. The class is defined by constructing planted-solution instances on the hardware graph of the DW2KQ, shown in Fig.~\ref{fig:hardwaregraph}.  This method builds an Ising Hamiltonian as a sum of $\lfloor \alpha 8L^2 \rfloor$ frustrated loops, where the all-zero state is a ground state of all loops  (somewhat confusingly, the Hamiltonian is thus `frustration-free' in the terminology of Ref.~\cite{Bravyi:2009sp}).  
We picked $\alpha = 0.35$ (this value is approximately where the peak in hardness occurs for the HFS algorithm, described in Appendix~\ref{app:HFS}).  We constructed loops as follows.  Choose a random vertex on the graph as the starting vertex.  From the (at most six) available edges connected to this vertex, randomly pick one.  If this new vertex has not been visited already, it is added to the chain.  Continue until the chain forms a loop by hitting a member of the chain.  Only the loop and not the tail is kept.  The loop is discarded if any of the couplings along the loop already have $|J|=3$, and if the loop does not visit at least two unit cells~\cite{King:2015zr}.  The second condition means that the shortest possible loop includes six vertices (within each unit cell the degree of each vertex is four, but including other unit cells the degree is six, except for unit cells along the boundary of the Chimera graph, where the degree can be five).  Along the loop, choose the couplings to satisfy the planted solution, i.e., set them all to be ferromagnetic.  Then randomly pick a single coupling and flip it.  The couplings along the loop are added to the already-present coupling values on the graph.  This process is repeated until $\lfloor \alpha 8L^2 \rfloor$ loops are generated for the chosen value of $\alpha$.

Finally, we randomly placed our gadget into $p L^2$ complete unit cells (without faulty qubits or couplers), where in this work we set $p = 0.1$, and added these terms to the Ising Hamiltonian.  The final Hamiltonian now has a maximum range of $6$, i.e., $|J_{ij}|\leq 6$ for all couplers. 
The ground state of the final Hamiltonian remains the all-zero state because this state is the ground state of all loop and gadget terms in the Hamiltonian.

We provide in Fig.~\ref{fig:OptimalAnnealingTime2} analogous results to those in Fig.~\ref{fig:OptimalAnnealingTime1} of the main text for the hardware-planted instances.  In Fig.~\ref{fig:ExampleOptimalAnnealingTime2}, we show a representative instance at $L=16$ that exhibits an  optimal annealing time above $500 \mu$s.  In Fig.~\ref{fig:GroupTTS2}, we show that the median TTS exhibits a clear minimum for sizes $L \in [8,16]$ (no minimum was observed for $L<8$), which moves to higher annealing time values with increasing problem size.  This is reflected in the distribution of instance optimal annealing times, as shown in Fig.~\ref{fig:HardwareDistribution}.  The steady increase in the hardness of the instances with problem size is reflected in the upward shift of the minimum TTS in Fig.~\ref{fig:GroupTTS2}.

Apart from the obvious difference of the existence of optimal annealing times at smaller sizes ($L\geq 8$ compared to $L\geq 12$), the optimal annealing time is significantly higher for the hardware-planted instances than for the logical-planted instance class, as summarized in Fig.~\ref{fig:toptScaling}. The optimal annealing time is seen to increase with problem size in all cases, rising faster for the DW2KQ than for the DW2X, but eventually flattening for both types of problem instances. The increase is consistent with both the possibility of benefit from a longer adiabatic evolution time or from a longer thermal relaxation time at larger problem sizes.

In Fig.~\ref{fig:TTSPRC} we present the scaling results for the hardware-planted instances at three different quantiles, in analogy to Fig.~\ref{fig:TTSScaling} in the main text.  The simulation parameters for the solvers are identical except that we use colder temperatures for SA and SQA.  For SA we use $\beta = 0.396$ (this corresponds to $\beta B(1) \approx 15$), while for SQA we use $\beta = 4.25$.  The scaling coefficients are summarized in Table~\ref{table:scalingFitsHardware}. We again find that SQA has the smallest scaling coefficient. The scaling coefficient of the DW2KQ is larger than that of all the classical solvers we tested, so for this class of instances we can definitively rule out the possibility of scaling advantage against the solvers we tested.

\subsection{Correlating SQA and SVMC for logical-planted instances}
\label{app:SQA-vs-SVMC}

\begin{figure}[t] 
   \centering
   \includegraphics[width=0.8\columnwidth]{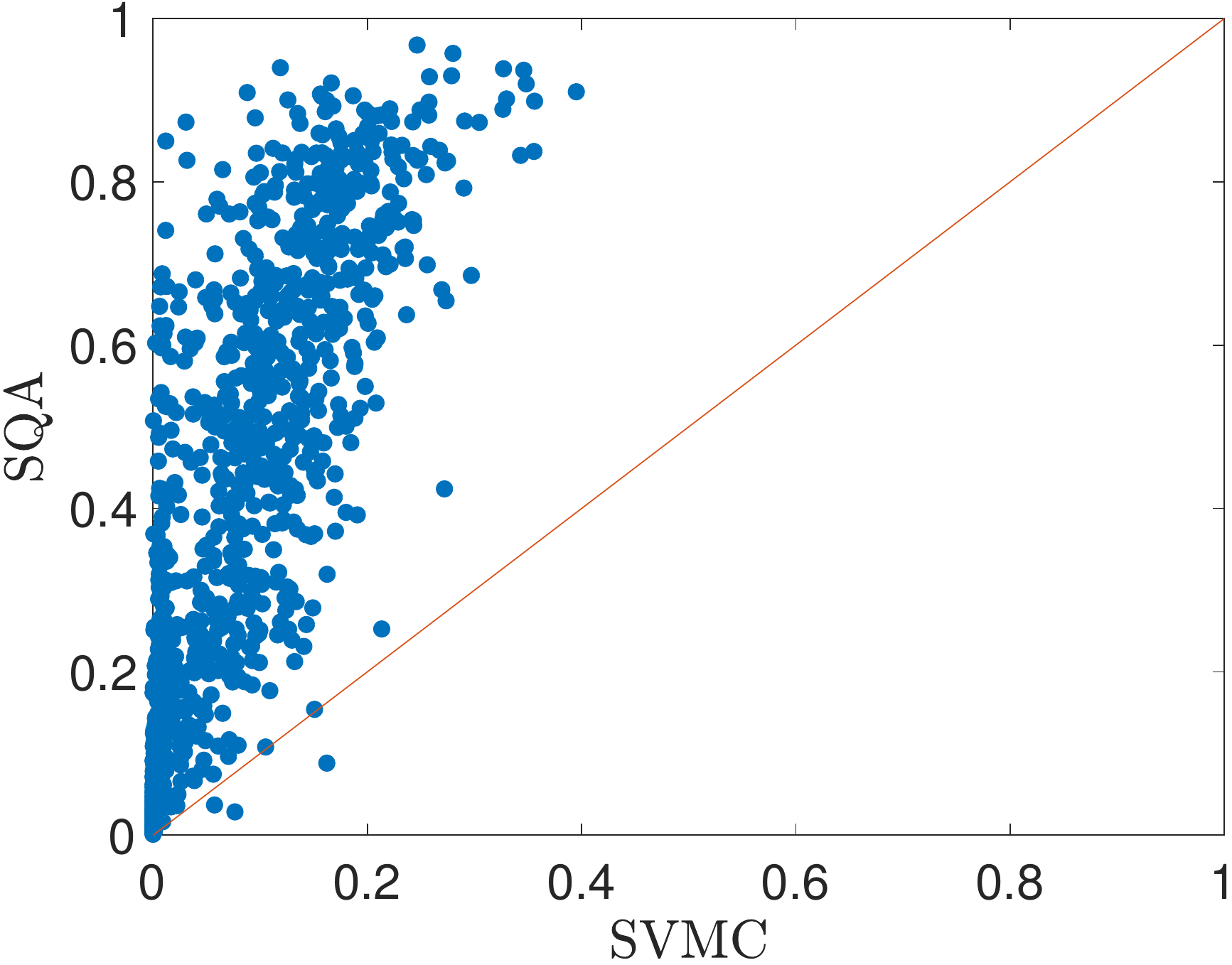} 
   \caption{\textbf{Correlating the SQA and SVMC success probabilities.} Shown is a scatter plot correlating the highest ground state probability for SQA (up to $2$M sweeps)  and SVMC (up to $8$M sweeps).  For the results shown here both algorithms use the DW2X annealing schedule with an inverse-temperature of $\beta = 2.5$.}
   \label{fig:ScatterSQASVMC}
\end{figure}

While a detailed analysis for each instance such as shown in Fig.~\ref{fig:L=16SQASVMC} in the main text is prohibitive, we correlate in Fig.~\ref{fig:ScatterSQASVMC} the performance of SQA and SVMC at $\beta = 2.5$ and a relatively large number of sweeps. We observe that for almost all the instances, SQA finds the ground state with a significantly higher success probability and substantially fewer spin updates.  Furthermore, a significant (but not overwhelming) number of instances hug the vertical axis of the scatter plot, corresponding to instances where SVMC completely fails to find the ground state but SQA succeeds with a non-vanishing probability. 

\begin{figure}[t]
\centering
\subfigure[]{\includegraphics[width=0.725\columnwidth]{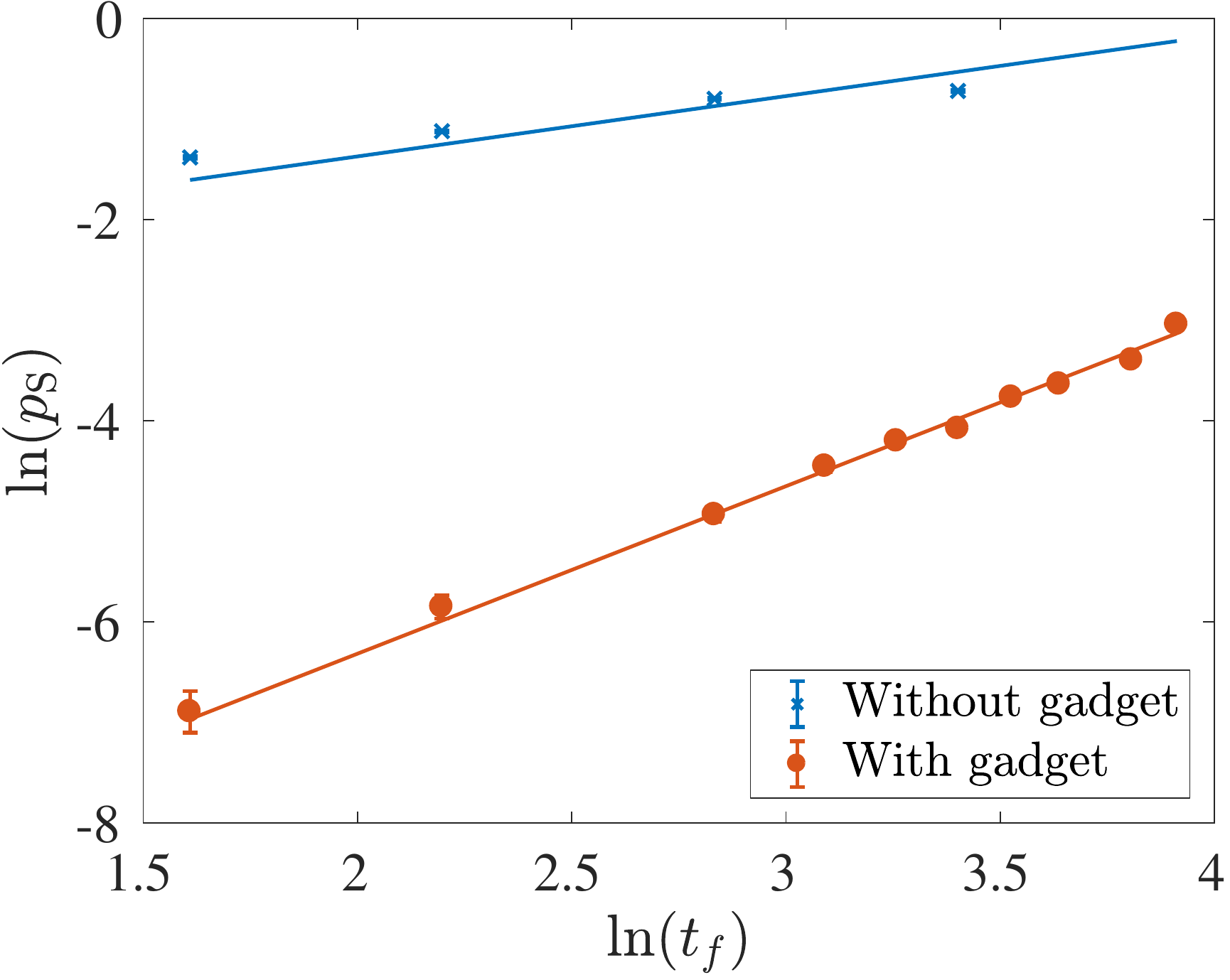} \label{fig:pGSScalingInstance}}
\subfigure[]{\includegraphics[width=0.8\columnwidth]{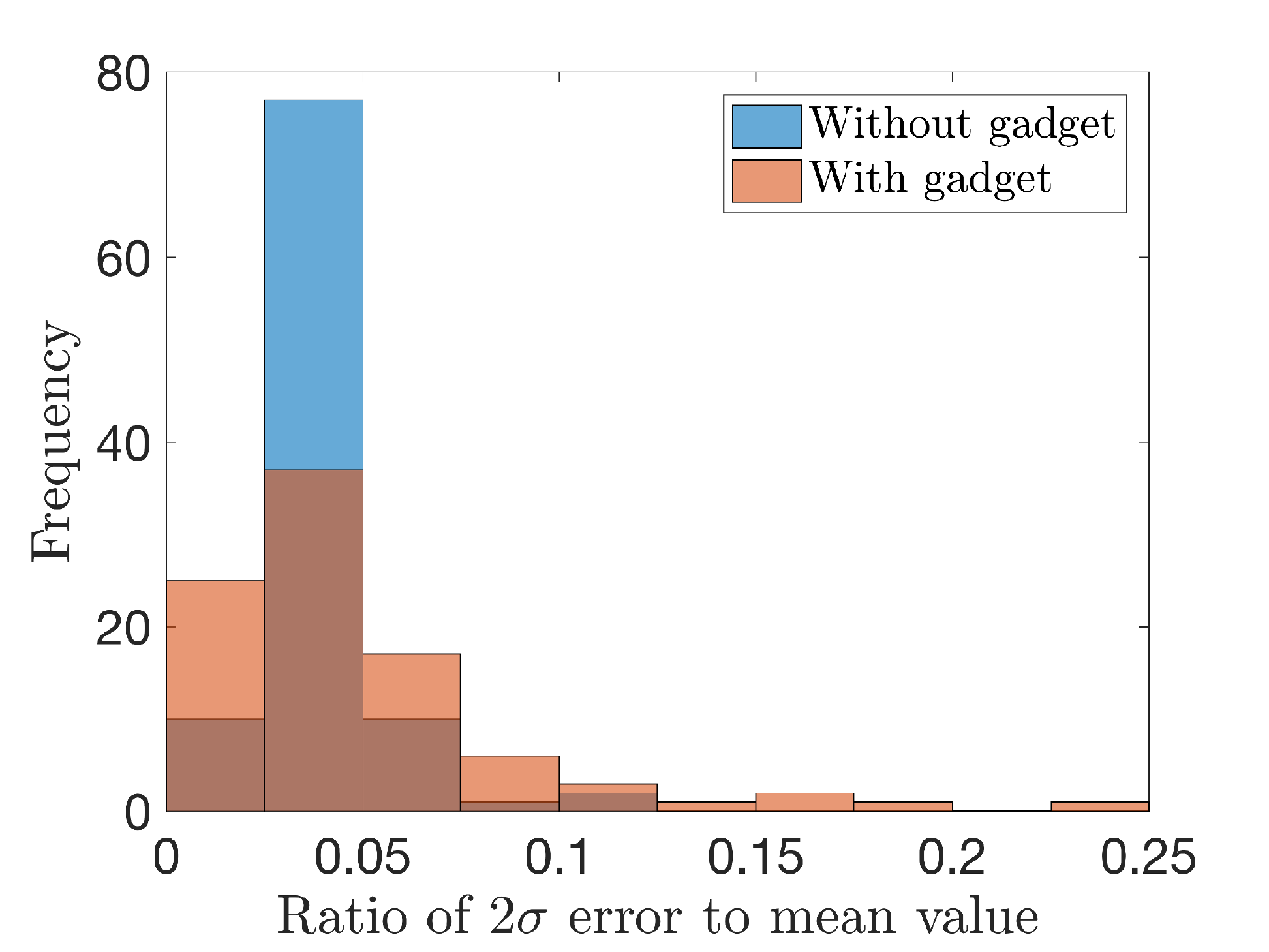} \label{fig:pGSScalingError}}
\caption{\textbf{Fits of the success probability to a power law.} (a) The data points represent the DW2KQ $\ln(p_\mathrm{S})$ results for the logical-planted instances with/without the gadget (red circles/blue crosses) over the range $t_f \in [5, 50] \mu s$. The solid lines correspond to the linear best fits $a \ln(t_f) + b$, where $a =1.546 \pm 0.015, b =-8.348 \pm 0.052$ (with the gadget), and $a = 0.367 \pm 0.007 , b = -1.918 \pm 0.019$ (without the gadget). The 2$\sigma$ error bars are not visible as they are smaller than the data marker size.  (b) The ratio of the $2 \sigma$ error to the best-fit value of the linear  coefficient (shown in Fig.~\ref{fig:GroupOptimum} of the main text) for $100$ instances of the logical-planted instances without (blue) and with (red) the gadget.} 
\label{fig:pGSScalingFit2}
\end{figure}
%

\section{Success probability scaling} 
\label{app:pGSscaling}
In Fig.~\ref{fig:GroupOptimum} of the main text, we showed the power law scaling coefficient of the success probability extracted for $t_f \leq 50 \mu s$.  Here we provide supplemental data to support the quality of these fits.  First, we show the data and fit in Fig.~\ref{fig:pGSScalingInstance} for the instance depicted in Fig.~\ref{fig:ExampleOptimalAnnealingTime1} of the main text as well as its counterpart without the gadget.  The data for this instance nicely agrees with a polynomial fit.  In Fig.~\ref{fig:pGSScalingError}, we show that the uncertainty in the power law scaling coefficient for the majority of the instances is below $10\%$, indicating that the polynomial fits to the data points are reasonable.

\section{Calculating the normalized overlap of instances.}  
\label{app:Overlap}

In Fig.~\ref{fig:Overlap} of the main text we showed the overlap of the logical-planted instances below the median between the classical solvers and the DW2KQ.  In order to calculate this quantity, we first fit the $\ln (\TTS)$ of each instance to the function $a ( \ln t_f - b)^2 + c$, and we evaluate the function at the optimal annealing time for the median TTS.  This gives us a mean value $\overline{\TTS}_i(t^\ast)$ and its associated 1$\sigma$ error $\Delta \TTS_i$ for the $i$-th instance.  We then perform $1000$ bootstraps over $400$ instances, where for each bootstrap we generate two sets of $100$ normally distributed random numbers $\eta_{i,(1,2)}$ in order to calculate two TTS realizations for each instance, i.e. $\TTS_{i,(1,2)} = \overline{\TTS}_i(t^\ast) + \eta_{i,(1,2)} \Delta \TTS_i$.  For the two sets of TTS realizations, we calculate the median TTS and find which instances have a TTS below the median.  We calculate the overlap fraction of instances between two solvers $S$ and $S'$ for realizations $\alpha$ and $\beta$ respectively, which we denote by $f_{S_\alpha,S_\beta'}$.  The normalized fraction $\bar{f}_{C,\mathrm{DW2KQ}}$ between a solver $C$ and the DW2KQ is then given by:
\begin{equation}
\bar{f}_{C,\mathrm{DW2KQ}} = f_{C_1,\mathrm{DW2KQ}_2} / \sqrt{f_{C_1,C_2} f_{\mathrm{DW2KQ}_1,\mathrm{DW2KQ}_2}}
\end{equation}
The normalization ensures that even with the noisy realization of the TTS, the overlap of instances between a solver and itself is one.

\section{Comparison to other algorithms}
\label{app:HFS}

In this section we present results from testing a number of other algorithms. Of course, for practical reasons we cannot consider all other relevant algorithms (e.g., we do not consider the iso-energetic cluster updates algorithm~\cite{PhysRevLett.115.077201}). Instead, we aimed to find other algorithms in addition to SQA that have a better scaling than the DW2KQ for the logical-planted instances. SQA remains the best-scaling algorithm among those we tested.

\begin{figure}[t]
\centering
\subfigure[\ hardware-planted]{\includegraphics[width=0.8\columnwidth]{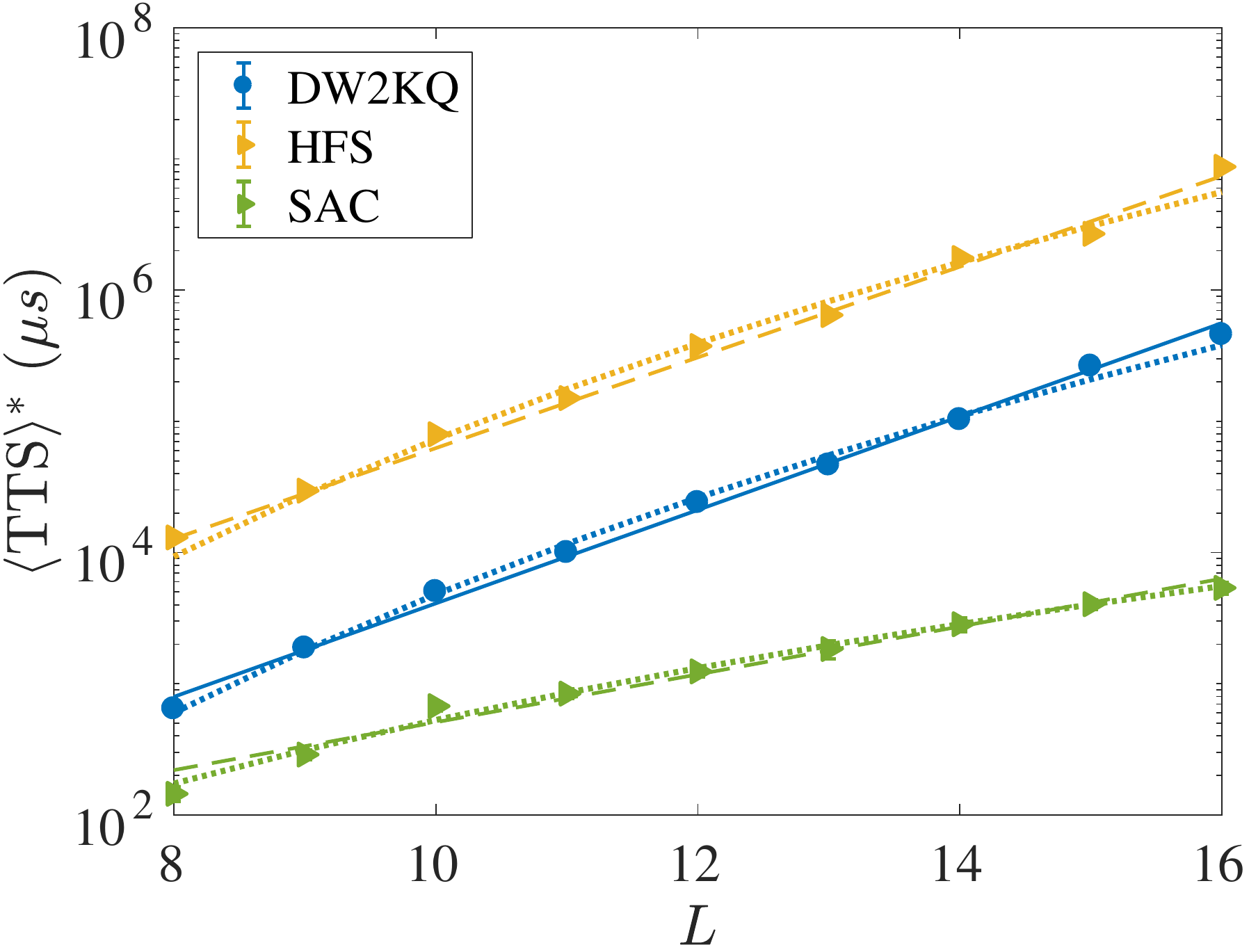} \label{fig:scalingHardware-app}}
\subfigure[\ logical-planted]{\includegraphics[width=0.8\columnwidth]{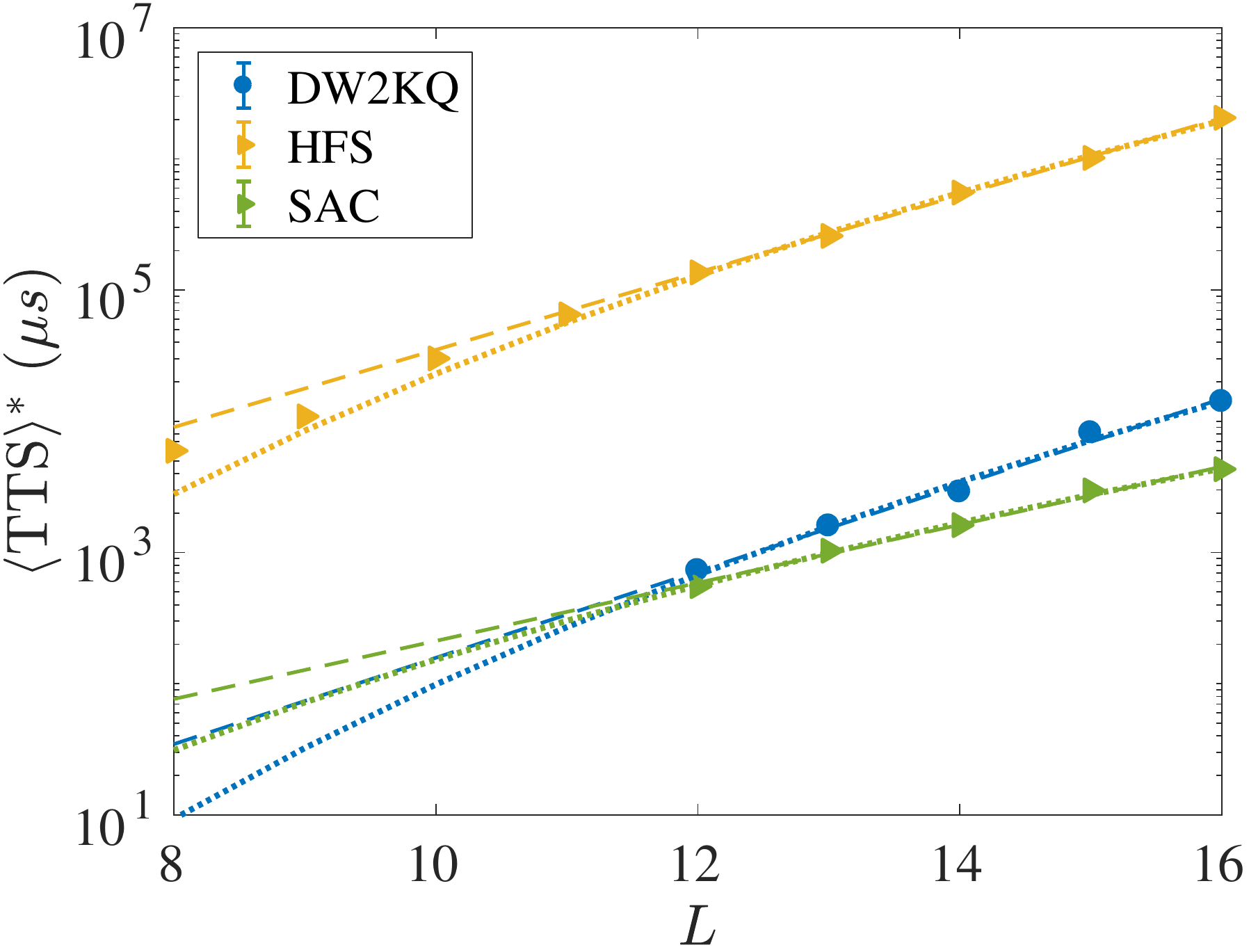} \label{fig:scalingLogical-app}}
\caption{\textbf{Scaling of the median optimal TTS with problem size for DW2KQ \textit{vs} HFS and SAC.} The data points represent the DW2KQ (blue circles), HFS (yellow left triangle), and SAC (green right triangle). The dashed and dotted curves correspond, respectively, to exponential and polynomial best fits with parameters given in Table~\ref{table:scalingFitsHFS}. (a) Hardware-planted instances. (b) Logical-planted instances. The data symbols obscure the error bars, representing the 95\% confidence interval for each optimal TTS data point (computed from the fit of $\ln\langle \mathrm{TTS} \rangle$  to a quadratic function as explained in  in Appendix~\ref{app:optimumFitting}).} 
\label{fig:TTSScalingHFS}
\end{figure}
%

\subsection{HFS}
In the main text, we did not make comparisons to the HFS algorithm because it does not implement the same algorithmic approach as the other annealing algorithms.  Nevertheless, because it is an algorithm tailored to solve spin-glass problems on the Chimera architecture, it is instructive to compare its performance.  For HFS, we use the implementation provided by Ref.~\cite{selby:13a} (which does not utilize a GPU), and we ran it in mode ``-S3'', meaning that maximal induced trees (treewidth $1$ in this case) were used.  The TTS is given by \cite{Hen:2015rt}:
\begin{equation}
\mathrm{TTS}_{\mathrm{HFS}} = \tau_{\mathrm{HFS}} L \left( \frac{5}{4} L  + 2 \right) n_{\mathrm{trees}} R(n_{\mathrm{trees}})
\end{equation}
where $\tau_{\mathrm{HFS}}  = 0.3 \mu$s is the time for a single update. $R(n_{\mathrm{trees}})$ is the number of repetitions with $n_{\mathrm{trees}}$ tree updates. In principle, the optimal TTS is found by finding the value of $ n_{\mathrm{trees}}$ that minimizes the TTS, but the implementation of Ref.~\cite{selby:13a} continues to increase $n_{\mathrm{trees}}$ until an exit criterion is reached.  Specifically, the algorithm exits when the same lowest energy is found consecutively after $n_{\mathrm{exit}}=4$ tree updates.  We have found that this can be highly non-optimal, especially for the hardware-planted instances.   Therefore, in all our scaling plots, we have optimized the value of $n_{\mathrm{trees}}$. This is an important distinction from all previous work using the HFS algorithm, which to the best of our knowledge did not optimize $n_{\mathrm{trees}}$, and hence the scaling of the HFS algorithm in previous work is likely to be an underestimate of the true scaling, in the very same sense that the D-Wave scaling reported previously underestimates the true scaling.

The behavior of the optimal TTS with problem size is shown in Fig.~\ref{fig:TTSScalingHFS}, with the scaling parameter fits given in Table~\ref{table:scalingFitsHFS}.  We find that for the logical-planted instances, HFS scales better than the DW2KQ, while for the hardware-planted instances the scaling of the two is statistically indistinguishable. 

For the HFS algorithm, we find that a quadratic fit does not capture the general features of the TTS curve as a function of number of tree updates.  Instead, we find that a function of the form $\ln \langle \TTS \rangle= a (\ln n_{\mathrm{tree}})^{-3} + b (\ln n_{\mathrm{tree}}) - \frac{4}{3^{3/4}}(a^3 b)^{1/4} + c$ captures the data well.  The value of $c$ gives the value of $\ln\TTSopt{}$.  We give the fit values and their confidence intervals in Tables~\ref{table:logicalHFS} and \ref{table:hardwareHFS}.

\subsection{SAC}
We can also consider simulated annealing with both single and multi-spin updates (SAC), with the latter being simultaneous updates of all the spins comprising a unit cell (super-spin approximation~\cite{2016arXiv160401746M}). This requires the algorithm to know about the underlying hardware graph. The implementation of SAC is identical to that of SA, except each sweep of single spin updates is followed by a sweep of unit cell updates.  The eight spins in the unit cell are flipped, and the move is accepted according to the Metropolis-Hastings rule.  Because the unit cell graph is bipartite, unit cells in the first partition are updated first, followed by the unit cells in the second partition.  This algorithm can be implemented as efficiently on GPU's as the single spin SA algorithm since it does not require storing any more data in memory.  Because SAC effectively involves updating twice as many spins as SA in a single sweep, the timing of SAC is the same as in Eq.~\eqref{eq:TimingSA} in Appendix~\ref{app:SA_SVMC} but with $f_{\mathrm{SAC}} = 25$ns$^{-1}$.  For consistency, we use the same annealing schedule in $B(s) \beta$ for SAC as we did for SA, with $B(s)$ as in Fig.~\ref{fig:DW2Xannealing} (in units where the maximum Ising coupling strength $|J_{ij}|$ is $1$). We use $\beta = 0.132$ (this corresponds to $\beta B(1) \approx 5$) for the logical-planted instances and $\beta = 0.396$ (this corresponds to $\beta B(1) \approx 15$) for the hardware-planted instances. We give the fit values and their confidence intervals in Tables~\ref{table:logicalSAC} and \ref{table:hardwareSAC}, and the behavior of the optimal TTS with problem size is shown in Fig.~\ref{fig:TTSScalingHFS}, with the scaling parameter fits given in Table~\ref{table:scalingFitsHardware}. We see that HFS and DW2KQ are statistically indistinguishable, and SAC is the top-scaling algorithm  for the hardware-planted instances, outperforming even SQA. Results for the logical-planted instances are given in Table~\ref{table:scalingFitsHFS}, which for convenience reproduces data from Fig~\ref{fig:scalingFitsLogical} in the main text. Here we see that SAC outperforms HFS, which in turn outperforms DW2KQ, while SQA is the top-scaling algorithm, outperforming SAC.  

\subsection{Minimum-weight perfect-matching}
\begin{figure}[t] 
   \centering
   \includegraphics[width=0.75\columnwidth]{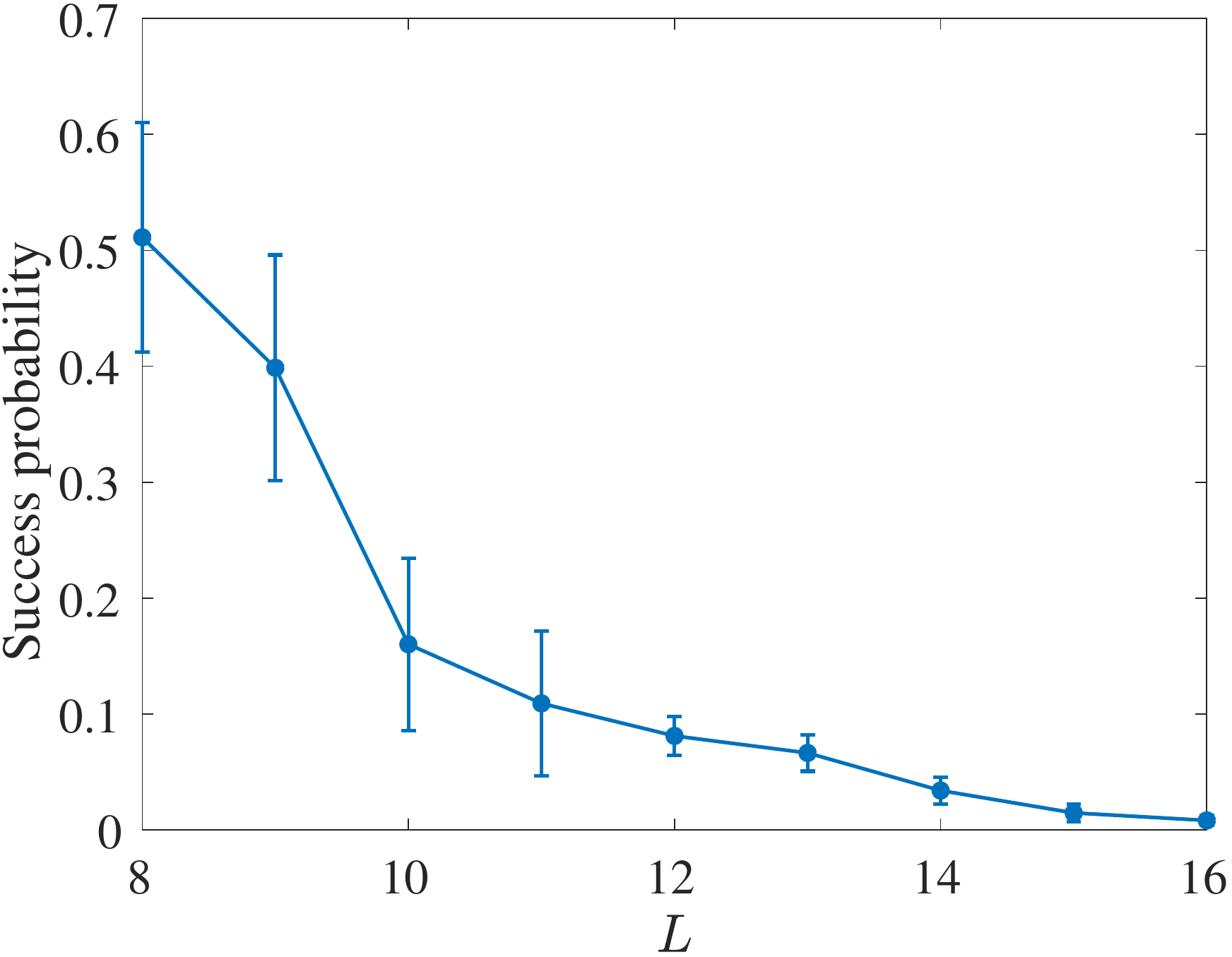} 
   \caption{{\bf Success probability for MWPM on the logical-planted instances.}  The data symbols correspond to the average success probability calculated using $1000$ bootstraps of $100$ instances for $L<12$ and of 1000 instances for $L\geq 12$, and the error bars represent the $95\%$ confidence intervals ($2\sigma$) calculated from the same bootstrap.}
   \label{fig:MWPM}
\end{figure}

Before the mapping onto Chimera and the introduction of the gadget, the logical-planted solution instances are defined on a two dimensional square grid.  Here we check a polynomial-time algorithm for solving the minimum-weight perfect-matching (MWPM) problem~\cite{Edmonds1965a,Kolmogorov2009} and show that it cannot be used to efficiently determine the ground state of the logical-planted instances defined on Chimera with the gadget.  In order to do so, we map the the Ising Hamiltonian on the square grid to a MWPM problem \cite{Bieche:1980}, run the Blossom V algorithm~\cite{Kolmogorov2009} to determine the solution to the MWPM problem, and finally map the solution of the MWPM problem to a ground state of the Ising Hamiltonian.  While the MWPM algorithm does find the ground state of planted solution instance defined on the square grid, it does not necessarily find the ground state of the associated Chimera instance with the gadget.  The reason for this is that the gadget reduces the degeneracy of the ground state by selecting only those states for which the unit cell on which the gadget is placed points up.  Without knowing this and because of the large ground state degeneracy, the MWPM predominantly selects the wrong ground state.  We show in Fig.~\ref{fig:MWPM} how the success probability of finding the ground state decreases with increasing problem size.  While at $L=8$, MWPM finds the ground state for approximately half the instances, at $L=16$, it finds the ground state of only 16 instances out of 1000 instances.  This means MWPM is not competitive with SAC, HFS, or SQA.

%

\section{Determining the optimal annealing time and optimal TTS}
\label{app:optimumFitting}

In order to determine the position of the optimal annealing time $t^\ast$ and its associated $\TTSopt{}$ (we drop the quantile notation $q$ for simplicity) at a given size $L$ for the DW2KQ, DW2X, SA, SQA, and SVMC results, we fit the $\ln \langle \TTS \rangle$ data for different annealing times to a function of the form $a (\ln t_f - b)^2 + c$.  The value of $b$ gives the value of $t^\ast$, and the value of $c$ is the associated $\ln\TTSopt{}$.  

The fit values and their confidence intervals are given in Tables~\ref{table:logicalDW2KQ} (logical-planted) and~\ref{table:hardwareDW2KQ} (hardware-planted) for the DW2KQ, in Table~\ref{table:hardwareDW2X} for the DW2X (hardware-planted only, since logical-planted instances did not exhibit an accessible optimal annealing time on the DW2X), in Tables~\ref{table:logicalSAA} and~\ref{table:hardwareSAA} for SA, in Tables~\ref{table:logicalSQA} and~\ref{table:hardwareSQA} for SQA, and in Tables~\ref{table:logicalSVMC} and~\ref{table:hardwareSVMC} for SVMC. 

\begin{figure}[t] 
   \centering
 \includegraphics[width=0.8\columnwidth]
{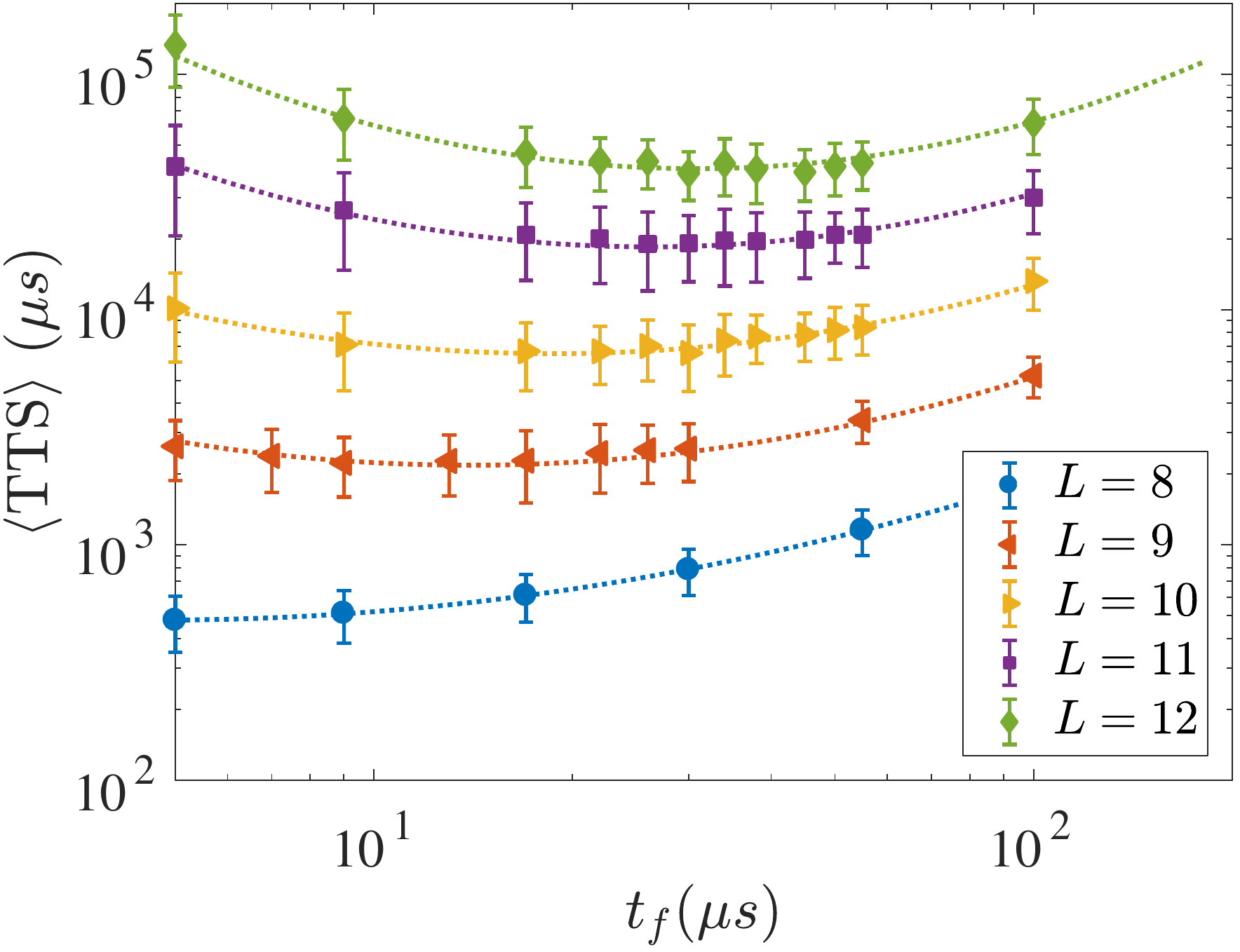}
   \caption{{\bf Median TTS as a function of annealing time for hardware-planted instances on the DW2X.}  Unlike the DW2KQ, the optimal annealing time for $L=8$ appears to be smaller than $5\mu$s; for $L\geq 9$ the optimal annealing time $t^*$ lies within the range of achievable annealing times for the DW2X device. Furthermore, for the same problem size, the DW2X optimal annealing times are consistently smaller than those of the DW2KQ [compare to Fig.~\ref{fig:GroupTTS2} and recall Fig.~\ref{fig:toptScaling}].  While the difference in the hardware graph may play a role in this, it is likely that the intrinsic differences between the two devices is responsible (see Appendix~\ref{sec:Case0989_Changed6} for additional comparisons).  
  }
   \label{fig:hardwareplantedDW2X}
\end{figure}

Results of this fitting procedure for the DW2KQ results on the logical-planted instances are shown in Fig.~\ref{fig:GroupTTS1} of the main text and on the hardware-planted instances in Fig.~\ref{fig:GroupTTS2}.  The fits for the DW2X on the hardware-planted instances are shown in Fig.~\ref{fig:hardwareplantedDW2X}.
Because the largest problem size we can program on the DW2X is at $L=12$, we studied only hardware-planted instances on this device.  Note that because the hardware graph of the DW2X differs from that of the DW2KQ [see Figs.~\ref{fig:hardwaregraph} and \ref{fig:DW2Xhardwaregraph}], we should not assume that the instances defined on both are necessarily from the same class.  

\begin{table*}[t]
\centering
\begin{tabular}{c c c c c c}
\hline
$L$ & Min trees & Max trees & $a$ & $b$ & $c$\\ [0.5ex] 
\hline \hline
$8$ & $3$ & $30$ &   $0.841 \pm 0.082$ & $2.221 \pm  0.453$ & $9.897 \pm 0.063$ \\
$9$ & $3$ & $30$ &   $0.675 \pm 0.078$ & $2.458 \pm  0.362$ & $10.502 \pm 0.052$ \\
$10$ & $4$ & $30$ & $0.747 \pm 0.116$ & $4.088 \pm  0.941$ & $11.518 \pm 0.055$ \\
$11$ & $4$ & $30$ & $0.633 \pm 0.117$ & $5.253 \pm  0.989$ & $12.292 \pm 0.053$ \\
$12$ & $5$ & $30$ & $0.669 \pm 0.087$ & $7.320 \pm  0.957$ & $13.029 \pm 0.028$ \\
$13$ & $5$ & $30$ & $0.707 \pm 0.104$ & $12.266 \pm  1.951$ & $13.668 \pm 0.029$ \\
$14$ & $5$ & $30$ & $0.666 \pm 0.103$ & $13.070 \pm  2.049$ & $14.430 \pm 0.024$ \\
$15$ & $6$ & $30$ & $0.593 \pm 0.095$ & $14.247 \pm  1.674$ & $15.038 \pm 0.022$ \\
$16$ & $5$ & $30$ & $0.674 \pm 0.086$ & $19.071 \pm  1.855$ & $15.742 \pm 0.024$  \\
 [1ex]
\hline
\end{tabular}
\caption{Fit to $a x^{-3} +b x + c - 4 (a^3 b)^{1/4}/3^{3/4}$ for the median results of the logical-planted instances using HFS with $y = \log \TTS$ and $ x= \log n_{\mathrm{trees}}$.  The factor $c$ does not include $\tau_{\mathrm{HFS}}$.}
\label{table:logicalHFS}
\end{table*}
\begin{table*}[t]
\centering
\begin{tabular}{c c c c c c}
\hline
$L$ & Min trees & Max trees & $a$ & $b$ & $c$\\ [0.5ex] 
\hline \hline
$8$ & $5$ & $30$ 		& $0.868 \pm 0.068$ & $7.838 \pm  0.643$ & $10.679 \pm 0.020$ \\
$9$ & $7$ & $30$ 		& $0.489 \pm 0.102$ & $8.908 \pm  1.392$ & $11.500 \pm 0.015$ \\
$10$ & $5$ & $50$		& $0.670 \pm 0.041$ & $13.950 \pm  0.854$ & $12.491 \pm 0.018$ \\
$11$ & $5$ & $50$ 		& $0.432 \pm 0.040$ & $15.410 \pm  0.671$ & $13.125 \pm 0.011$ \\
$12$ & $10$ & $30$ 	& $0.672 \pm 0.263$ & $23.631 \pm  5.858$ & $14.038 \pm 0.022$ \\
$13$ & $30$ & $100$	& $1.061 \pm 0.154$ & $76.298 \pm  12.888$ & $14.582 \pm 0.012$ \\
$14$ & $10$ & $100$	& $0.486 \pm 0.086$ & $35.793 \pm  3.1606$ & $15.583 \pm 0.026$ \\
$15$ & $35$ & $100$	&  $0.943 \pm 0.239$ & $96.115 \pm  22.116$ & $16.009 \pm 0.014$ \\
$16$ & $35$ & $100$ 	& $0.878 \pm 0.245$ & $69.559 \pm  22.095$ & $17.185 \pm 0.014$ \\
 [1ex]
\hline
\end{tabular}
\caption{Fit to $a x^{-3} +b x + c - 4 (a^3 b)^{1/4}/3^{3/4}$ for the median results of the hardware-planted instances using HFS  with $y = \log \TTS$ and $ x= \log n_{\mathrm{trees}}$.  The factor $c$ does not include $\tau_{\mathrm{HFS}}$.}
\label{table:hardwareHFS}
\end{table*}

\begin{table*}[t]
\centering
\begin{tabular}{c c c c c c}
\hline
$L$ & Min $t_f$ & Max $t_f$ & $a$ & $b$ & $c$\\ [0.5ex] 
\hline \hline
$12$ & $10$ & $2154 $ & $0.101 \pm 0.010$ & $4.495 \pm 0.099$ & $16.446 \pm 0.039$ \\
$13$ & $10$ & $2154 $ & $0.122 \pm 0.009$ & $5.104 \pm 0.061$ & $17.067 \pm 0.037$ \\
$14$ & $10^2$ & $2154 $ & $0.123 \pm 0.026$ & $5.504 \pm 0.171$ & $17.515 \pm 0.031$ \\
$15$ & $10^2$ & $2154 $ & $0.143 \pm 0.034$ & $5.876 \pm 0.130$ & $18.124 \pm 0.046$ \\
$16$ & $10^2$ & $3593 $ & $0.145 \pm 0.020$ & $6.247 \pm 0.079$ & $18.497 \pm 0.034$  \\ [1ex]
\hline
\end{tabular}
\caption{Fit to $y = a (x-b)^2 + c$ for the median results of the logical-planted instances using SAC with $y = \log \TTS$ and $ x= \log t_f$.  The factor $c$ does not include $f_{\mathrm{SAC}}$.}
\label{table:logicalSAC}
\end{table*}
\begin{table*}[t]
\centering
\begin{tabular}{c c c c c c}
\hline
$L$ & Min sweeps & Max sweeps & $a$ & $b$ & $c$\\ [0.5ex] 
\hline \hline
$8$ & $46$ & $10^3 $ & $0.262 \pm 0.102$ & $5.911 \pm 0.227$ & $15.098 \pm 0.125$ \\
$9$ & $46$ & $4641 $ & $0.217 \pm 0.038$ & $6.656 \pm 0.110$ & $15.792 \pm 0.082$ \\
$10$ & $166$ & $5994 $ & $0.207 \pm 0.048$ & $6.907 \pm 0.129$ & $16.641 \pm 0.078$ \\
$11$ & $166$ & $5994 $ & $0.327 \pm 0.059$ & $6.903 \pm 0.101$ & $16.859 \pm 0.079$ \\
$12$ & $166$ & $10^4 $ & $0.293 \pm 0.042$ & $7.480 \pm 0.100$ & $17.250 \pm 0.089$ \\
$13$ & $215$ & $10^4 $ & $0.277 \pm 0.078$ & $7.743 \pm 0.146$ & $17.643 \pm 0.165$ \\
$14$ & $278$ & $10^4 $ & $0.295 \pm 0.075$ & $8.016 \pm 0.180$ & $18.075 \pm 0.118$ \\
$15$ & $278$ & $10^4 $ & $0.326 \pm 0.057$ & $8.160 \pm 0.123$ & $18.427 \pm 0.083$ \\
$16$ & $464$ & $10^4 $ & $0.325 \pm 0.091$ & $8.201 \pm 0.166$ & $18.712 \pm 0.104$  \\
 [1ex]
\hline
\end{tabular}
\caption{Fit to $y = a (x-b)^2 + c$ for the median results of the hardware-planted instances using SAC with $y = \log \TTS$ and $ x= \log n_{\mathrm{sweeps}}$.  The factor $c$ does not include $f_{\mathrm{SAC}}$.}
\label{table:hardwareSAC}
\end{table*}
\begin{table*}[h]
\centering
\begin{tabular}{c c c c c c}
\hline
$L$ & Min $t_f$ & Max $t_f$ & $a$ & $b$ & $c$\\ [0.5ex] 
\hline \hline
$12$ & $5$ & $100$ & $0.124 \pm 0.040$ & $2.233 \pm  0.317$ & $6.588 \pm 0.036$ \\
$13$ & $5$ & $100$ & $0.141 \pm 0.044$ & $2.732 \pm  0.196$ & $7.371 \pm 0.040$ \\
$14$ & $5$ & $100$ & $0.144 \pm 0.040$ & $3.341 \pm  0.160$ & $7.968 \pm 0.032$ \\
$15$ & $5$ & $100$ & $0.182 \pm 0.050$ & $3.672 \pm  0.224$ & $9.007 \pm 0.032$ \\
$16$ & $5$ & $180$ & $0.221 \pm 0.031$ & $3.798 \pm  0.104$ & $9.557 \pm 0.032$  \\ [1ex]
\hline
\end{tabular}
\caption{Fit to $y = a (x-b)^2 + c$ for the median results of the logical-planted instances using the DW2KQ with $y = \log \TTS$ and $ x= \log t_f$. 
}
\label{table:logicalDW2KQ}
\end{table*}
\begin{table*}[h]
\centering
\begin{tabular}{c c c c c c}
\hline
$L$ & Min $t_f$ & Max $t_f$ & $a$ & $b$ & $c$\\ [0.5ex] 
\hline \hline
$8$ & $5$ & $330$      & $0.115 \pm 0.026$ & $3.089 \pm  0.266$ & $6.465 \pm 0.046$ \\
$9$ & $5$ & $330$      & $0.148 \pm 0.026$ & $3.997 \pm  0.122$ & $7.531 \pm 0.051$ \\
$10$ & $5$ & $610$    & $0.203 \pm 0.025$ & $4.476 \pm  0.092$ & $8.521 \pm 0.049$ \\
$11$ & $5$ & $2000$  & $0.268 \pm 0.019$ & $5.059 \pm  0.070$ & $9.209 \pm 0.055$ \\
$12$ & $9$ & $2000$  & $0.281 \pm 0.029$ & $5.390 \pm  0.092$ & $10.084 \pm 0.060$ \\
$13$ & $17$ & $2000$ & $0.305 \pm 0.029$ & $5.583 \pm  0.070$ & $10.745 \pm 0.050$  \\
$14$ & $30$ & $2000$ & $0.338 \pm 0.044$ & $5.834 \pm  0.093$ & $11.540 \pm 0.064$ \\
$15$ & $30$ & $2000$ & $0.330 \pm 0.038$ & $5.795 \pm  0.080$ & $12.479 \pm 0.063$ \\
$16$ & $55$ & $2000$ & $0.342 \pm 0.055$ & $5.853 \pm  0.107$ & $13.033 \pm 0.068$  \\
 [1ex]
\hline
\end{tabular}
\caption{Fit to $y = a (x-b)^2 + c$ for the median results of the hardware-planted instances using the DW2KQ with $y = \log \TTS$ and $ x= \log t_f$.}
\label{table:hardwareDW2KQ}
\end{table*}
\begin{table*}[h]
\centering
\begin{tabular}{c c c c c c}
\hline
$L$ & Min $t_f$ & Max $t_f$ & $a$ & $b$ & $c$\\ [0.5ex] 
\hline \hline
$9$ & $5$ & $2000$   & $0.223 \pm 0.026$ & $2.636 \pm  0.244$ & $7.688 \pm 0.103$ \\
$10$ & $5$ & $2000$ & $0.240 \pm 0.034$ & $2.926 \pm  0.270$ & $8.780 \pm 0.103$ \\
$11$ & $5$ & $2000$ & $0.291 \pm 0.037$ & $3.265 \pm  0.226$ & $9.825 \pm 0.112$ \\
$12$ & $5$ & $2000$ & $0.333 \pm 0.030$ & $3.428 \pm  0.137$ & $10.591 \pm 0.083$ \\
 [1ex]
\hline
\end{tabular}
\caption{Fit to $y = a (x-b)^2 + c$ for the median results of the hardware-planted instances using the DW2X with $y = \log \TTS$ and $ x= \log t_f$.}
\label{table:hardwareDW2X}
\end{table*}
\begin{table*}[h]
\centering
\begin{tabular}{c c c c c c}
\hline
$L$ & Min sweeps & Max sweeps & $a$ & $b$ & $c$\\ [0.5ex] 
\hline \hline
$8$ & $784$ & $14384 $ & $0.130 \pm 0.029$ & $7.843 \pm 0.118$ & $19.272 \pm 0.034$ \\
$9$ & $1000$ & $37926 $ & $0.139 \pm 0.022$ & $8.58 \pm 0.086$ & $20.355 \pm 0.040$ \\
$10$ & $3359$ & $10^5 $ & $0.124 \pm 0.025$ & $9.408 \pm 0.118$ & $22.105 \pm 0.037$ \\
$11$ & $5455$ & $2\times 10^5 $ & $0.125 \pm 0.026$ & $10.196 \pm 0.118$ & $23.29 \pm 0.035$ \\
$12$ & $5000$ & $10^6 $ & $0.122 \pm 0.012$ & $10.970 \pm 0.075$ & $24.423 \pm 0.041$ \\
$13$ & $10000$ & $10^6 $ & $0.124 \pm 0.020$ & $11.485 \pm 0.109$ & $25.445 \pm 0.054$ \\
$14$ & $22000$ & $2\times 10^6 $ & $0.122 \pm 0.019$ & $12.057 \pm 0.114$ & $26.301 \pm 0.044$ \\
$15$ & $35000$ & $2\times 10^6 $ & $0.132 \pm 0.028$ & $12.601 \pm 0.125$ & $27.568 \pm 0.059$ \\
$16$ & $60000$ & $8\times 10^6  $ & $0.109 \pm 0.017$ & $13.103 \pm 0.139$ & $28.366 \pm 0.052$  \\
 [1ex]
\hline
\end{tabular}
\caption{Fit to $y = a (x-b)^2 + c$ for the median results of the logical-planted instances using SA with $y = \log \TTS$ and $ x= \log n_{\mathrm{sweeps}}$.  The factor $c$ does not include $f_{\mathrm{SA}}$.}
\label{table:logicalSAA}
\end{table*}
\begin{table*}[h]
\centering
\begin{tabular}{c c c c c c}
\hline
$L$ & Min sweeps & Max sweeps & $a$ & $b$ & $c$\\ [0.5ex] 
\hline \hline
$8$ & $297$ & $23357 $ & $0.127 \pm 0.015$ & $7.971 \pm 0.072$ & $18.708 \pm 0.040$ \\
$9$ & $483$ & $37926 $ & $0.139 \pm 0.018$ & $8.772 \pm 0.080$ & $19.703 \pm 0.049$ \\
$10$ & $1274$ & $10^5 $ & $0.124 \pm 0.014$ & $9.373 \pm 0.073$ & $20.533 \pm 0.032$ \\
$11$ & $2069$ & $1.2\times10^5 $ & $0.145 \pm 0.020$ & $10.046 \pm 0.081$ & $21.221 \pm 0.050$ \\
$12$ & $5455$ & $1.2\times10^5 $ & $0.154 \pm 0.034$ & $10.45 \pm 0.111$ & $21.802 \pm 0.038$\\
$13$ & $5455$ & $3\times10^5 $ & $0.153 \pm 0.024$ & $11.037 \pm 0.115$ & $22.344 \pm 0.051$\\
$14$ & $5455$ & $3\times10^5 $ & $0.188 \pm 0.025$ & $11.429 \pm 0.115$ & $22.898 \pm 0.051$ \\
$15$ & $8858$ & $5.6\times10^5 $ & $0.179 \pm 0.017$ & $11.887 \pm 0.077$ & $23.378 \pm 0.035$\\
$16$ & $14384$ & $10^6 $ & $0.172 \pm 0.018$ & $12.293 \pm 0.073$ & $23.837 \pm 0.039$  \\
 [1ex]
\hline
\end{tabular}
\caption{Fit to $y = a (x-b)^2 + c$ for the median results of the hardware-planted instances using SAA with $y = \log \TTS$ and $ x= \log n_{\mathrm{sweeps}}$.  The factor $c$ does not include $f_{\mathrm{SA}}$.}
\label{table:hardwareSAA}
\end{table*}
\begin{table*}[h]
\centering
\begin{tabular}{c c c c c c}
\hline
$L$ & Min sweeps & Max sweeps & $a$ & $b$ & $c$\\ [0.5ex] 
\hline \hline
$8$ & $27825$ & $359381 $ & $0.339 \pm 0.039$ & $11.475 \pm 0.046$ & $19.476 \pm 0.036$ \\
$9$ & $27825$ & $599484 $ & $0.325 \pm 0.028$ & $11.792 \pm 0.037$ & $19.811 \pm 0.037$ \\
$10$ & $46415$ & $10^6 $ & $0.324 \pm 0.033$ & $12.068 \pm 0.048$ & $20.577 \pm 0.039$ \\
$11$ & $46415$ & $599484 $ & $0.451 \pm 0.053$ & $12.17 \pm 0.050$ & $20.977 \pm 0.05$\\
$12$ & $77426$ & $10^6 $ & $0.322 \pm 0.064$ & $12.296 \pm 0.082$ & $21.505 \pm 0.061$\\
$13$ & $77426$ & $10^6 $ & $0.414 \pm 0.075$ & $12.391 \pm 0.070$ & $21.905 \pm 0.073$ \\
$14$ & $77426$ & $10^6 $ & $0.469 \pm 0.078$ & $12.477 \pm 0.065$ & $22.171 \pm 0.069$ \\
$15$ & $77426$ & $10^6 $ & $0.575 \pm 0.095$ & $12.589 \pm 0.071$ & $22.678 \pm 0.089$ \\
$16$ & $77426$ & $10^6 $ & $0.606 \pm 0.094$ & $12.653 \pm 0.071$ & $22.955 \pm 0.073$  \\
 [1ex]
\hline
\end{tabular}
\caption{Fit to $y = a (x-b)^2 + c$ for the median results of the logical-planted instances using SQA with $y = \log \TTS$ and $ x= \log n_{\mathrm{sweeps}}$.  The factor $c$ does not include $f_{\mathrm{SQA}}$.}
\label{table:logicalSQA}
\end{table*}
\begin{table*}[h]
\centering
\begin{tabular}{c c c c c c}
\hline
$L$ & Min sweeps & Max sweeps & $a$ & $b$ & $c$\\ [0.5ex] 
\hline \hline
$8$ & $7742$ & $599484 $ & $0.373 \pm 0.036$ & $12.005 \pm 0.085$ & $20.424 \pm 0.082$ \\
$9$ & $12915$ & $599484 $ & $0.504 \pm 0.050$ & $12.162 \pm 0.089$ & $20.985 \pm 0.086$\\
$10$ & $16681$ & $599484 $ & $0.567 \pm 0.052$ & $12.258 \pm 0.075$ & $21.700 \pm 0.086$ \\
$11$ & $27825$ & $599484 $ & $0.704 \pm 0.092$ & $12.382 \pm 0.095$ & $22.057 \pm 0.117$\\
$12$ & $46415$ & $599484 $ & $0.701 \pm 0.164$ & $12.497 \pm 0.130$ & $22.632 \pm 0.142$\\
$13$ & $46415$ & $10^6 $ & $0.613 \pm 0.089$ & $12.742 \pm 0.089$ & $23.043 \pm 0.108$\\
$14$ & $46415$ & $10^6 $ & $0.756 \pm 0.119$ & $12.791 \pm 0.098$ & $23.597 \pm 0.142$ \\
$15$ & $77426$ & $10^6 $ & $0.662 \pm 0.173$ & $12.928 \pm 0.129$ & $24.069 \pm 0.156$ \\
$16$ & $77426$ & $10^6 $ & $0.710 \pm 0.181$ & $12.945 \pm 0.139$ & $24.434 \pm 0.166$  \\
 [1ex]
\hline
\end{tabular}
\caption{Fit to $y=a (x-b)^2 + c$ for the median results of the hardware-planted instances using SQA with $y = \log \TTS$ and $ x= \log n_{\mathrm{sweeps}}$.  The factor $c$ does not include $f_{\mathrm{SQA}}$.}
\label{table:hardwareSQA}
\end{table*}
\begin{table*}[h]
\centering
\begin{tabular}{c c c c c c}
\hline
$L$ & Min sweeps & Max sweeps & $a$ & $b$ & $c$\\ [0.5ex] 
\hline \hline
$8$ & $4641$ & $10^5 $ & $0.183 \pm 0.032$ & $10.166 \pm 0.075$ & $21.545 \pm 0.041$\\
$9$ & $7742$ & $4\times10^5 $ & $0.135 \pm 0.017$ & $10.919 \pm 0.079$ & $22.485 \pm 0.037$\\
$10$ & $27825$ & $10^6 $ & $0.141 \pm 0.023$ & $11.705 \pm 0.103$ & $23.515 \pm 0.040$\\
$11$ & $27825$ & $4\times10^6 $ & $0.140 \pm 0.011$ & $12.546 \pm 0.059$ & $23.797 \pm 0.036$ \\
$12$ & $27825$ & $4\times10^6 $ & $0.168 \pm 0.012$ & $13.003 \pm 0.049$ & $25.184 \pm 0.042$\\
$13$ & $46415$ & $4\times10^6 $ & $0.174 \pm 0.023$ & $13.362 \pm 0.076$ & $25.864 \pm 0.056$ \\
$14$ & $77426$ & $3\times10^6 $ & $0.217 \pm 0.041$ & $13.700 \pm 0.105$ & $26.457 \pm 0.069$ \\
$15$ & $77426$ & $8\times10^6 $ & $0.220 \pm 0.023$ & $14.159 \pm 0.110$ & $26.582 \pm 0.077$\\
$16$ & $10^5$ & $8\times10^6 $ & $0.235 \pm 0.024$ & $14.434 \pm 0.126$ & $27.208 \pm 0.072$ \\
 [1ex]
\hline
\end{tabular}
\caption{Fit to $y = a (x-b)^2 + c$ for the median results of the logical-planted instances using SVMC with $y = \log \TTS$ and $ x= \log n_{\mathrm{sweeps}}$.  The factor $c$ does not include $f_{\mathrm{SVMC}}$.}
\label{table:logicalSVMC}
\end{table*}
\begin{table*}[h]
\centering
\begin{tabular}{c c c c c c}
\hline
$L$ & Min sweeps & Max sweeps & $a$ & $b$ & $c$\\ [0.5ex] 
\hline \hline
$8$ & $4641$ & $10^5 $ & $0.087 \pm 0.066$ & $10.229 \pm 0.406$ & $21.754 \pm 0.082$ \\
$9$ & $2782$ & $4\times10^5 $ & $0.106 \pm 0.026$ & $11.350 \pm 0.215$ & $22.890 \pm 0.075$ \\
$10$ & $4641$ & $4\times10^5 $ & $0.132 \pm 0.030$ & $11.961 \pm 0.255$ & $23.774 \pm 0.061$ \\
$11$ & $10^5$ & $10^6 $ & $0.148 \pm 0.042$ & $12.875 \pm 0.285$ & $24.651 \pm 0.079$ \\
$12$ & $46415$ & $4 \times 10^6 $ & $0.143 \pm 0.023$ & $13.492 \pm 0.111$ & $25.321 \pm 0.060$ \\
$13$ & $46415$ & $4 \times 10^6  $ & $0.158 \pm 0.023$ & $13.887 \pm 0.107$ & $25.884 \pm 0.050$ \\
$14$ & $77426$ & $4 \times 10^6  $ & $0.176 \pm 0.048$ & $14.394 \pm 0.264$ & $26.592 \pm 0.075$ \\
$15$ & $129154$ & $8 \times 10^6  $ & $0.179 \pm 0.034$ & $14.919 \pm 0.167$ & $27.183 \pm 0.057$ \\
$16$ & $215443$ & $8 \times 10^6  $ & $0.182 \pm 0.043$ & $15.215 \pm 0.241$ & $27.693 \pm 0.054$ \\
 [1ex]
\hline
\end{tabular}
\caption{Fit to $y = a (x-b)^2 + c$ for the median results of the hardware-planted instances using SVMC with $y = \log \TTS$ and $ x= \log n_{\mathrm{sweeps}}$.  The factor $c$ does not include $f_{\mathrm{SVMC}}$.}
\label{table:hardwareSVMC}
\end{table*}
%
%


\end{document}